\documentclass[a4paper,10pt,twoside,openright]{report}%
\usepackage[german,english]{babel}
\usepackage[ansinew]{inputenc}
\usepackage{geometry}
\usepackage{graphicx}
\usepackage[bf]{caption2}
\usepackage{amsmath}
\usepackage{amsthm}
\usepackage{bm}
\usepackage{layout}
\usepackage{float}
\usepackage{fancyhdr}
\usepackage{amsfonts}
\usepackage{amssymb}%
\normalsize

\newcommand\openone{\leavevmode\hbox{\small1\kern-3.3pt\normalsize1}}



\begin{document}
\begin{titlepage}
\begin{figure}[t]
\vspace{-1.75cm}
\begin{flushright}
\includegraphics{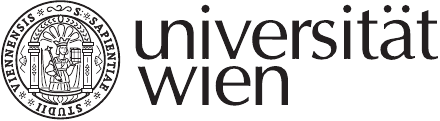}
\end{flushright}
\vspace{1.75cm}
\end{figure}
\vspace{4cm} \centering{\Huge{\bfseries{\textsf{DISSERTATION}}}}
\vspace{1.75cm}
\\
\centering{\large{\textsf{Titel der Dissertation}}} \vspace{0.5cm}
\\
\centering{\huge{\textsf{Quantum violation of macroscopic
realism}}}\vspace{0.1cm}
\\
\centering{\huge{\textsf{and the transition to classical physics}}}
\vspace{2.75cm}
\\
\centering{\large{\textsf{Verfasser}}} \vspace{0.5cm}
\\
\centering{\LARGE{\textsf{Johannes Kofler}}}\vspace{1.75cm}
\\
\centering{\large{\textsf{angestrebter akademischer Grad}}}
\vspace{0.5cm}
\\
\centering{\LARGE{\textsf{Doktor der
Naturwissenschaften}}}\vspace{1.5cm}
\\
\flushleft{\large{\textsf{Wien, im Juni 2008}}}\vspace{0.75cm}
\\
\flushleft{\textsf{Studienkennzahl laut
Studienblatt:}}\hspace{1cm}\textsf{A 091 411}
\\
\flushleft{\textsf{Dissertationsgebiet laut
Studienblatt:}}\hspace{0.6cm}\textsf{Physik}
\\
\flushleft{\textsf{Betreuer:}}\hspace{4.72cm}\textsf{ao.\
Univ.-Prof.\ \v{C}aslav Brukner}
\end{titlepage}

\newpage

\thispagestyle{empty}

$\ $

\newpage

\thispagestyle{empty}

\hspace{0cm}
\begin{tabular}
[c]{p{13cm}}%
\vspace{6.25cm} \textit{Ich weiß ehrlich nicht, was die Leute meinen, wenn sie
von der Freiheit des menschlichen Willens sprechen. Ich habe zum Beispiel das
Gefühl, daß ich irgend etwas will; aber was das mit Freiheit zu tun hat, kann
ich überhaupt nicht verstehen. Ich spüre, daß ich meine Pfeife anzünden will
und tue das auch; aber wie kann ich das mit der Idee der Freiheit verbinden?
Was liegt hinter dem Willensakt, daß ich meine Pfeife anzünden will? Ein
anderer Willensakt? Schopenhauer hat einmal gesagt: ``Der Mensch kann tun was
er will; er kann aber nicht wollen was er will.''}\\
\\
\textit{[Honestly I cannot understand what people mean when they talk about
the freedom of the human will. I have a feeling, for instance, that I will
something or other; but what relation this has with freedom I cannot
understand at all. I feel that I will to light my pipe and I do it; but how
can I connect that up with the idea of freedom? What is behind the act of
willing to light the pipe? Another act of willing? Schopenhauer once said:
``Man can do what he wills but he cannot will what he wills.'']}\\
\flushright{Albert Einstein}
\end{tabular}

\newpage

\thispagestyle{empty} ~\newpage

\pagenumbering{arabic} \setcounter{page}{5}%

\tableofcontents

\chapter*{\vspace{-3cm}Abstract}

\addcontentsline{toc}{chapter}{Abstract}

\vspace{0.25cm}The descriptions of the quantum realm and the macroscopic
classical world differ significantly not only in their mathematical
formulations but also in their foundational concepts and philosophical
consequences. The assumptions of a genuine classical world---local realism and
macroscopic realism (macrorealism)---are at variance with quantum mechanical
predictions as characterized by the violation of the Bell and the Leggett-Garg
inequality, respectively. When and how physical systems stop to behave
quantumly and begin to behave classically is still heavily debated in the
physics community and subject to theoretical and experimental research.

The first chapter of this dissertation puts forward a novel approach to the
quantum-to-classical transition fully within quantum theory and conceptually
different from already existing models. It neither needs to refer to the
uncontrollable environment of a system (decoherence) nor to change the quantum
laws itself (collapse models), but puts the stress on the limits of
observability of quantum phenomena due to imprecisions of our measurement
apparatuses. Naively, one would say that the predictions of quantum mechanics
reduce to those of classical physics merely by going to large quantum numbers.
Using a quantum spin as a model object, we first demonstrate that for
unrestricted measurement accuracy the system's time evolution cannot be
described classically and is conflict with macrorealism through violation of
the Leggett-Garg inequality, no matter how large the spin is. How then does
the classical world arise? Under realistic conditions in every-day life, we
are only able to perform coarse-grained measurements and do not resolve
individual quantum levels of the macroscopic system. We show that for some
\textquotedblleft classical\textquotedblright\ Hamiltonians it is this mere
restriction to fuzzy measurements which is sufficient to see the natural
emergence of macrorealism and the classical Newtonian laws out of the full
quantum formalism. This resolves the apparent impossibility of how classical
realism and deterministic laws can emerge out of fundamentally random quantum events.

In the second chapter, we find that the restriction of coarse-grained
measurements usually allows to describe the time evolution of any quantum spin
state by a time evolution of a statistical mixture. However, we demonstrate
that there exist \textquotedblleft non-classical\textquotedblright%
\ Hamiltonians for which the time evolution of this mixture cannot be
understood classically, leading to a violation of macrorealism. We derive the
necessary condition for these non-classical time evolutions and illustrate it
with the example of an oscillating Schrödinger cat-like state. Constant
interaction of the system with an environment establishes macrorealism but
cannot account for a continuous spatiotemporal description of the system's
non-classical time evolution in terms of classical laws of motion. We argue
that non-classical Hamiltonians are unlikely to appear in nature because they
require interactions between a large number of particles or are of high
computational complexity.

The third chapter investigates entanglement between collective operators in
two specific physical systems, namely in a linear chain of harmonic
oscillators and in ensembles of spin-$\tfrac{1}{2}$ particles. We show that
under certain conditions entanglement between macroscopic observables can
persist for large system sizes. However, since this analysis uses sharp
measurements, it is not in disagreement with our quantum-to-classical approach.

The last chapter addresses the question of the origin of quantum randomness
and proposes a link with mathematical undecidability. We demonstrate that the
states of elementary quantum systems are capable of encoding a set of
mathematical axioms. Quantum measurements reveal whether a given proposition
is decidable or undecidable within this set. We theoretically find and
experimentally confirm that whenever a mathematical proposition is undecidable
within the axiomatic set encoded in the state, the measurement associated to
the proposition has random outcomes. This supports the view that quantum
randomness is irreducible and a manifestation of mathematical
undecidability.\newpage

\chapter*{\vspace{-3cm}Zusammenfassung}

\addcontentsline{toc}{chapter}{Zusammenfassung}

\selectlanguage{german} \vspace{0.25cm}Die Beschreibungen der
quantenmechanischen und der klassischen Welt unterscheiden sich nicht nur
signifikant in ihren mathematischen Formulierungen sondern auch in ihren
grundsätzlichen Konzepten und philosophischen Implikationen. Die Annahmen
einer klassischen Welt -- lokaler und makroskopischer Realismus
(Makrorealismus) -- widersprechen den Vorhersagen der Quantenphysik, was durch
die Verletzung der Bell- und der Leggett-Garg-Ungleichung charakterisiert
wird. Die Frage, wann und wie physikalische Systeme aufhören sich
quantenmechanisch und anfangen sich klassisch zu verhalten, wird in der
wissenschaftlichen Gemeinschaft noch immer heftig diskutiert und ist
Gegenstand experimenteller und theoretischer Forschung.

Das erste Kapitel der vorliegenden Dissertation entwickelt einen neuen Zugang
zum Übergang der Quanten- zur klassischen Physik, und zwar vollkommen
innerhalb der Quantentheorie und konzeptionell verschieden von bereits
bestehenden Modellen. Dieser Zugang muss sich weder auf die unkontrollierbare
Umgebung von Systemen beziehen (Dekohärenz) noch die Gesetze der
Quantenmechanik selbst abändern (Kollaps-Modelle). Er fokussiert sich vielmehr
auf die Limitierung der Beobachtbarkeit von Quantenphänomen aufgrund der
Ungenauigkeit unserer Messapparate. Naiverweise würde man annehmen, dass sich
die Vorhersagen der Quantenmechanik auf jene der klassischen Physik allein
dadurch reduzieren, indem man zu großen Quantenzahlen geht. Wir verwenden
einen Quantenspin als Modellobjekt und zeigen zunächst, dass bei
uneingeschränkter Messgenauigkeit die Zeitevolution des Systems nicht
klassisch verstanden werden kann und aufgrund der Verletzung der
Leggett-Garg-Ungleichung im Widerspruch zu Makrorealismus steht, selbst wenn
der Spin beliebig groß ist. Wie entsteht dann die klassische Welt? Unter
realistischen alltäglichen Bedingungen sind wir nur in der Lage, grobkörnige
Messungen durchzuführen, die die einzelnen Quantenniveaus des makroskopischen
Systems nicht auflösen können. Wir zeigen, dass für bestimmte
\textquotedblleft klassische\textquotedblright\ Hamilton-Operatoren diese
bloße Einschränkung zu unscharfen Messungen ausreicht, um die natürliche
Emergenz von Makrorealismus und klassischer Newtonscher Gesetze aus dem vollen
quantenmechanischen Formalismus zu sehen. Dies löst die scheinbare
Unmöglichkeit auf, wie klassischer Realismus und deterministische Gesetze aus
fundamental zufälligen Quantenereignissen entstehen können.

Im zweiten Kapitel zeigen wir, dass die Einschränkung grobkörniger Messungen
es üblicher\-weise erlaubt, die zeitliche Evolution jedes beliebigen
quantenmechanischen Spinzustands durch die zeitliche Evolution einer
statistischen Mischung zu beschreiben. Ungeachtet dessen demonstrieren wir,
dass es \textquotedblleft nicht-klassische\textquotedblright%
\ Hamilton-Operatoren gibt, für die die Zeitentwicklung dieser Mischung nicht
klassisch verstanden werden kann. Wir leiten die allgemeine Bedingung für
solche nicht-klassischen Zeitentwicklungen her und veranschaulichen sie anhand
des Beispiels einer oszillierenden Schrödinger-Katze. Andauernde Interaktion
des Systems mit einer Umgebung etabliert Makrorealismus, kann aber keine
kontinuierliche raumzeitliche Beschreibung der nicht-klassischen Zeitevolution
des Systems durch klassische Bewegungsgleichungen liefern. Wir argumentieren,
dass nicht-klassische Hamilton-Operatoren in der Natur wahrscheinlich nicht
vorkommen, weil sie Vielteilchen-Wechselwirkungen benötigen oder von hoher
Komplexität sind.

Das dritte Kapitel untersucht Verschränkung zwischen kollektiven Operatoren in
zwei spezifischen physikalischen Systemen, nämlich in einer linearen Kette von
harmonischen Oszillatoren und in Ensembles von Spin-$\tfrac{1}{2}$-Teilchen.
Wir zeigen, dass Verschränkung zwischen makroskopischen Observablen für große
Systeme bestehen bleiben kann. Zumal diese Analyse scharfe Messungen
verwendet, steht sie nicht im Widerspruch zu unserem Zugang zum Übergang von
der Quanten- zur klassischen Physik.

Das letzte Kapitel beschäftigt sich mit der Frage nach dem Ursprung von
quantenmechanischem Zufall und schlägt eine Verbindung mit mathematischer
Unentscheidbarkeit vor. Wir demonstrieren, dass Zustände elementarer
Quantensysteme einen Satz von mathematischen Axiomen kodieren können.
Quantenmechanische Messungen bringen zum Vorschein, ob eine gegebene
Proposition innerhalb dieses Satzes entscheidbar oder unentscheidbar ist. Wir
finden theoretisch und bestätigen experimentell, dass Messungen, die mit
innerhalb des vom Quantenzustand kodierten Axiomensatzes unentscheidbaren
mathematischen Propositionen assoziiert sind, zu zufälligen Resultaten führen.
Dies unterstützt die Sichtweise, dass quantenmechanischer Zufall irreduzibel
und eine Manifestation von mathematischer Unentscheidbarkeit ist.\newpage

\chapter*{\vspace{-3cm}Acknowledgement}

\addcontentsline{toc}{chapter}{Acknowledgement}

\selectlanguage{english}\vspace{0.25cm}When I was writing my diploma thesis at
the University of Linz in summer 2004, I applied for a PhD position in Vienna
because of a strong interest in quantum physics and foundational questions. In
retrospect, this was likely one of the best decisions that I could have
made---both from a professional and a private perspective.

\noindent A few days after my arrival in Vienna in January 2005, I was the
first person to move into the building of the Institute for Quantum Optics and
Quantum Information of the Austrian Academy of Sciences. I did not yet know
how great the pleasure would be to learn and work in this extremely inspiring
and challenging environment during the following years. Now is the time to
thank:\vspace{0.5cm}

\noindent First and foremost \v{C}aslav Brukner for his perfect scientific and
personal support. I am sincerely grateful for all the time and effort he has
put into our joint research, and without exaggeration, I could not have wished
for a better supervision. His vast knowledge and insight into quantum theory
as well as the true joy and interest with which he is doing science have been
an enduring inspiration through my whole doctoral studies.\vspace{0.25cm}

\noindent Nikita Arnold and Urbaan M.~Titulaer for their objective advice on
my interest to go to Vienna after my graduate studies.\vspace{0.25cm}

\noindent Anton Zeilinger for advocating my application for a theoretical PhD
position, for his open-minded way of leading the institute, and for
integrating me into experimental activities.\vspace{0.25cm}

\noindent Markus Aspelmeyer, Tomasz Paterek, and Rupert Ursin for many
fascinating discussions, not only on quantum mechanics.\vspace{0.25cm}

\noindent Simon Gröblacher and Robert Prevedel for forcing me every now and
then to have a glass of wine. Katharina Gugler for sharing with me the
disbelieve in free will.\vspace{0.25cm}

\noindent The whole Vienna quantum group---Academy and University---for
creating a very pleasant atmosphere.\vspace{0.25cm}

\noindent The Austrian Academy of Sciences for a doctoral fellowship that
allowed me to pursue the second half of my research in a fully independent
way.\vspace{0.25cm}

\noindent And last but not least my family for their unconditional support and
interest in my work throughout the years, and my friends for all the hiking
trips, cinema evenings, and discussions on Life, the Universe, and
Everything.\vspace{0.75cm}

Johannes Kofler

\vspace{0.25cm}\hspace{-0.18cm}Vienna, June 2008

\newpage~\newpage

\chapter*{\vspace{-3cm}List of publications}

\addcontentsline{toc}{chapter}{List of publications}

\vspace{0.25cm}The titles of publications that are directly relevant for this
dissertation are written in bold face.\vspace{0.25cm}

\subsection*{Articles in refereed journals}

\vspace{0.25cm}

\begin{itemize}
\item J. Kofler and \v{C}. Brukner\newline\textit{\textbf{The conditions for
quantum violation of macroscopic realism}}\newline
Phys.~Rev.~Lett.~(accepted); arXiv:0706.0668 [quant-ph].

\item J. Kofler and \v{C}. Brukner\newline\textit{\textbf{Classical world
arising out of quantum physics under the restriction of coarse-grained
measurements}}\newline Phys.~Rev.~Lett.~\textbf{99}, 180403 (2007).

\item J. Kofler and \v{C}. Brukner\newline\textit{\textbf{Entanglement
distribution revealed by macroscopic observations}}\newline Phys.~Rev.~A
\textbf{74}, 050304(R) (2006).

\item M. Lindenthal and J. Kofler\newline\textit{Measuring the absolute photo
detection efficiency using photon number correlations}\newline
Appl.~Opt.~\textbf{45}, 6059 (2006).

\item J. Kofler and N. Arnold\newline\textit{Axially symmetric focusing as a
cuspoid diffraction catastrophe: Scalar and vector cases and comparison with
the theory of Mie}\newline Phys.~Rev.~B \textbf{73}, 235401 (2006).

\item J. Kofler, V. Vedral, M. S. Kim, and \v{C}. Brukner\newline%
\textit{\textbf{Entanglement between collective operators in a linear harmonic
chain}}\newline Phys.~Rev.~A \textbf{73}, 052107 (2006).

\item J. Kofler, T. Paterek, and \v{C}. Brukner\newline\textit{Experimenter's
freedom in Bell's theorem and quantum cryptography}\newline Phys.~Rev.~A
\textbf{73}, 022104 (2006).
\end{itemize}

\subsection*{Submitted or in preparation}

\vspace{0.25cm}

\begin{itemize}
\item J. Kofler and \v{C}. Brukner\newline\textit{\textbf{The conditions for
quantum violation of macroscopic realism}}\newline ArXiv:0706.0668 [quant-ph] (submitted).

\item X. Ma, A. Qarry, J. Kofler, T. Jennewein, and A. Zeilinger\newline%
\textit{Experimental violation of a Bell inequality with two different degrees
of freedom}\newline In preparation (2008).

\item X. Ma, A. Qarry, N. Tetik, J. Kofler, T. Jennewein, and A.
Zeilinger\newline\textit{Entanglement-assisted delayed-choice experiment}%
\newline In preparation (2008).

\item J. Kofler and \v{C}. Brukner\newline\textit{Fundamental limits on
observing quantum phenomena from within quantum theory}\newline In preparation (2008).
\end{itemize}

\subsection*{Contributions in books}

\vspace{0.25cm}

\begin{itemize}
\item J. Kofler and \v{C}. Brukner\newline\textit{\textbf{A coarse-grained
Schrödinger cat}}\newline In:~\textit{Quantum Communication and Security},
ed.~M. \.{Z}ukowski, S. Kilin, and J. Kowalik (IOS Press 2007).

\item J. Kofler and N. Arnold\newline\textit{Axially symmetric focusing of
light in dry laser cleaning and nanopatterning}\newline In:~\textit{Laser
Cleaning II}, ed.~D.~M.~Kane (World Scientific Publishing, 2006).

\item D. Bäuerle, T. Gumpenberger, D. Brodoceanu, G. Langer, J. Kofler, J.
Heitz, and K. Piglmayer\newline\textit{Laser cleaning and surface
modifications: Applications in nano- and biotechnology}\newline
In:~\textit{Laser Cleaning II}, ed.~D.~M.~Kane (World Scientific Publishing, 2006).
\end{itemize}

\subsection*{Contributions in proceedings}

\vspace{0.25cm}

\begin{itemize}
\item R. Ursin, T. Jennewein, J. Kofler, J. Perdigues, L. Cacciapuoti, C. J.
de Matos, M. Aspelmeyer, A. Valencia, T. Scheidl, A. Fedrizzi, A. Acin, C.
Barbieri, G. Bianco, \v{C}. Brukner, J. Capmany, S. Cova, D. Giggenbach, W.
Leeb, R. H. Hadfield, R. Laflamme, N. Lütkenhaus, G. Milburn, M. Peev, T.
Ralph, J. Rarity, R. Renner, E. Samain, N. Solomos, W. Tittel, J. P. Torres,
M. Toyoshima, A. Ortigosa-Blanch, V. Pruneri, P. Villoresi, I. Walmsley, G.
Weihs, H. Weinfurter,\ M. \.{Z}ukowski, and A. Zeilinger\newline%
\textit{Space-QUEST:~Experiments with quantum entanglement in space}\newline
Accepted for the 59th International Astronautical Congress (2008);
arXiv:0806.0945v1 [quant-ph].

\item D. Bäuerle, L. Landström, J. Kofler, N. Arnold, and K. Piglmayer\newline%
\textit{Laser-processing with colloid monolayers}\newline Proc.~SPIE 5339,
\textbf{20} (2004).
\end{itemize}

\subsection*{Articles in popular journals}

\vspace{0.25cm}

\begin{itemize}
\item A. Zeilinger and J. Kofler\newline\textit{La dissolution du
paradoxe}\newline Sciences et Avenir Hors-Série \textbf{148}, 54 (2006).
\end{itemize}

\subsection*{Theses}

\vspace{0.25cm}

\begin{itemize}
\item J. Kofler\newline\textit{Focusing of light in axially symmetric systems
within the wave optics approximation}\newline Diploma Thesis, Johannes Kepler
University Linz, Austria (2004).
\end{itemize}

\newpage

~\newpage

\setcounter{chapter}{-1}

\chapter{Introduction}

\pagestyle{fancy} \fancyhf{} \fancyhead[LE,RO]{\thepage}
\fancyhead[LO]{Introduction} \fancyhead[RE]{Introduction}
\renewcommand{\headrulewidth}{0.25pt} \renewcommand{\footrulewidth}{0pt}
\addtolength{\headheight}{0.5pt} \fancypagestyle{plain}{\fancyhead{}
\renewcommand{\headrulewidth}{0pt}}

\begin{figure}[t]
\begin{center}
\includegraphics[width=.45\textwidth]{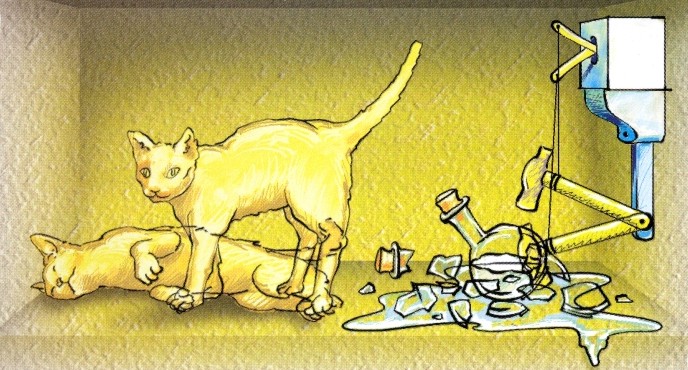}
\end{center}
\par
\vspace{-0.25cm}\caption{In 1935, Schrödinger put forward the
\textquotedblleft burlesque\textquotedblright\ \textit{Gedankenexperiment} of
a \textquotedblleft hell machine\textquotedblright\ to illustrate that,
according to quantum physics, it is possible to prepare a cat in a
superposition of `dead'\ and `alive'~\cite{Schr1935}. [Picture taken from
Sciences et Avenir Hors-Série \textbf{148}, 54 (2006).]}%
\label{Figure_Schroedinger_cat}%
\end{figure}

Since the birth of quantum theory in the 1920s, \textit{quantum entanglement}
and \textit{quantum superposition} have been used to highlight a number of
counter-intuitive phenomena. They lie at the heart of the
Einstein-Podolsky-Rosen~\cite{Eins1935} and the Schrödinger cat
paradox~\cite{Schr1935} (Figure~\ref{Figure_Schroedinger_cat}). The
corresponding conflicts between quantum mechanics on one side and a classical
world---\textit{local realism} and \textit{macroscopic realism}
(macrorealism)---on the other, are quantitatively expressed by the violation
of Bell's~\cite{Bell1964} and the Leggett-Garg inequality~\cite{Legg1985},
respectively. The quantum violation of local realism shows that the view is
untenable that space-like separated events do not influence each other
\textit{and} that objects have their properties prior to and independent of
measurement. The quantum violation of macrorealism means that it is wrong to
believe that a macroscopic object has definite properties at any time
\textit{and} that it can be measured without effecting them or their
subsequent dynamics.

The importance of this incongruousness today exceeds the realm of the
foundations of quantum physics and has become an important conceptual tool for
developing new quantum information technology. Entanglement and superposition
allow to perform certain computation and communication tasks such as quantum
cryptography~\cite{Benn1984,Gisi2002}, teleportation~\cite{Benn1993,Bouw1997}
or quantum computation~\cite{Deut1985,Niel2000}, which are not possible
classically. Experiments in the near future will be realized with increasingly
complex objects, either by entangling more and more systems with each other,
or by entangling systems with a very large number of degrees of freedom.
Eventually, all these developments will push the realm of quantum physics well
into the macroscopic world. Moreover, implications on society in a cultural
sense may manifest themselves, for the characteristics and peculiarities of
the quantum world---in particular quantum entanglement and quantum
superposition---could eventually become part of the every-day experience.

However, the macroscopic classical world that we perceive around us does not
show any characteristics of the quantum realm. The question \textquotedblleft
Why do classical systems stop to show quantum features?\textquotedblright\ is
still answered in radically different ways within the physics community. Since
classical apparatuses are needed for performing measurements on quantum
systems, this question is also related to the so called \textquotedblleft
measurement problem\textquotedblright\ and the various interpretations of
quantum mechanics, ranging from the Copenhagen over the Bohmian to the
many-worlds interpretation~\cite{Jamm1974}.

On the one hand, there exist a number of so-called collapse
models~\cite{Ghir1986,Penr1996} which try to explain the discrepancy between
the quantum and the classical world by introducing a fundamental breakdown of
quantum superpositions at some quantum-classical border. On the other, the
decoherence program~\cite{Zure1991,Zure2003} demonstrates that the states of
complex systems interacting with an environment, which cannot be accessed and
controlled in detail, rapidly evolve into statistical mixtures and lose their
quantum character.

While neither of these approaches can give a definite or already
experimentally settled answer, the understanding of the quantum-to-classical
transition is not only of prior importance for the future development towards
macroscopic superpositions and entanglement but also necessary for a
consistent description of the physical world. Collapse models make assumptions
about inherently non-quantum mechanical background fields or gravitational
mechanisms which are still to be tested experimentally. The decoherence
program is inherently quantum mechanical and can give good explanations for
many observations though it has to rely on the assumption of a preferred
pointer basis. However, the effects of decoherence can in principle be always
reduced and the experimental progress of the last years has already
demonstrated quantum interference of (Schrödinger-cat like) macroscopic
superpositions, e.g., interference fringes with large molecules of
$\sim\!10^{3}$ atomic mass units~\cite{Arnd1999}, entanglement between clouds
of $\sim\!10^{12}$ atoms~\cite{Juls2001}, or superpositions of macroscopically
distinct flux states in superconducting rings corresponding to $\sim\!10^{-6}$
amperes of current flowing clock- or anticlockwise~\cite{Frie2000}. These
experiments circumvent the problem of decoherence but did not yet come into
the region where they could exclude collapse models.

Until today there exists no definite answer to the problem of the
quantum-to-classical transition. Hence, certainly one of the most fundamental
and interesting questions in modern physics still remains unanswered:

\begin{quote}
\textit{How does the classical physical world emerge out of the quantum
realm?}
\end{quote}

\textbf{Chapter 1} of this dissertation addresses this question from a novel
perspective and develops an approach to the \textit{quantum-to-classical
transition} fully within quantum theory and conceptually different from
already existing models. It neither needs to refer to the environment of a
system (decoherence) nor to change the quantum laws itself (collapse models)
but puts the stress on the limits of observability of quantum phenomena due to
our measurement apparatuses. Using a quantum spin as a model object, we first
demonstrate that for unrestricted measurement accuracy the system's time
evolution cannot be described classically and is in conflict with macrorealism
through violation of the Leggett-Garg inequality. This conflict remains even
if the spin is arbitrarily large and macroscopic.

\begin{figure}[t]
\begin{center}
\includegraphics[width=.9\textwidth]{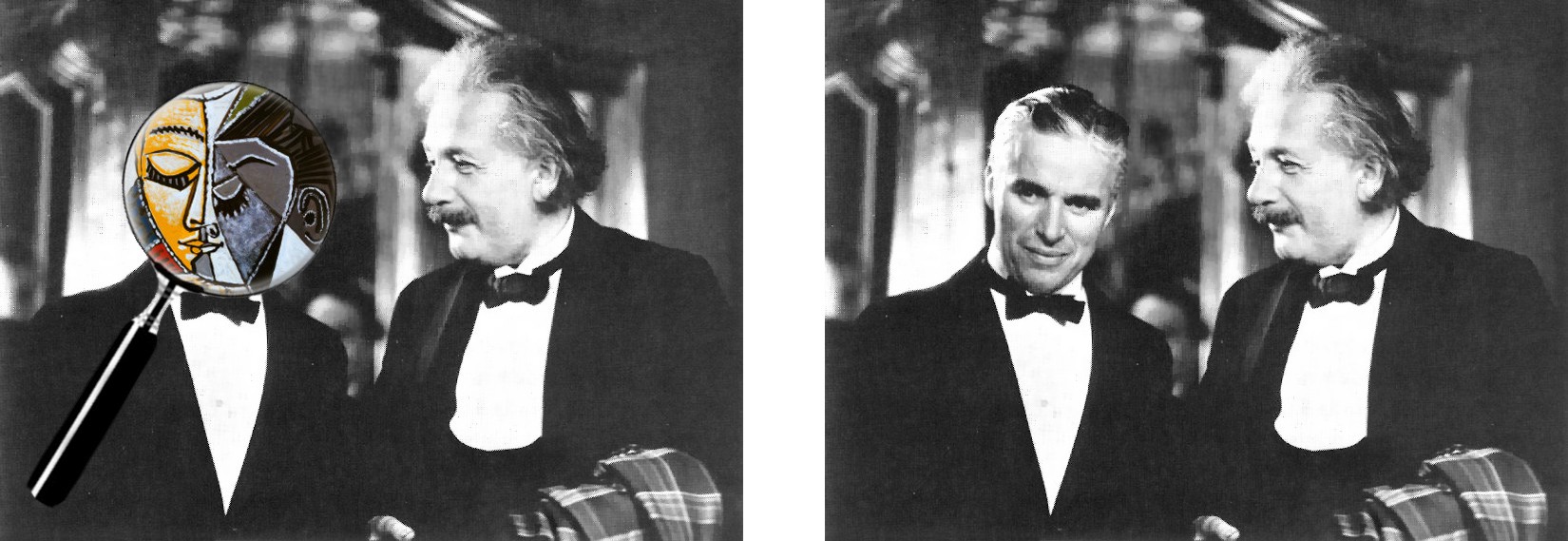}
\end{center}
\par
\vspace{-0.25cm}\caption{Under the magnifying glass of sharp measurements
Albert Einstein sees a strange and colorful quantum picture of the face next
to him. Its abstractness is symbolized by Pablo Picasso's \textquotedblleft
Head of a Reading Woman\textquotedblright. Under an every-day coarse-grained
view the classical appearance of Charlie Chaplin emerges.}%
\label{Figure_Einstein_and_Chaplin}%
\end{figure}

Under realistic conditions in every-day life, however, we are only able to
perform \textit{coarse-grained measurements} and do not resolve individual
quantum levels of the macroscopic system. As we show, it is this mere
restriction to fuzzy measurements which is sufficient to see the natural
emergence of macrorealism and the classical Newtonian laws out of the full
quantum formalism: the system's time evolution governed by the Schrödinger
equation and the state projection induced by measurements. This resolves the
apparent impossibility of how classical realism and deterministic laws can
emerge out of fundamentally random quantum events.
Figure~\ref{Figure_Einstein_and_Chaplin} presents an illustration of this
approach.\bigskip

\textbf{Chapter 2} first shows that a violation of the Leggett-Garg inequality
itself is possible for arbitrary Hamiltonians given the ability to perform
sharp quantum measurements. Apparatus decoherence or the restriction of
coarse-grained measurements usually allow to describe the time evolution of
any quantum spin state by a time evolution of a statistical mixture. However,
we demonstrate that there are \textit{\textquotedblleft
non-classical\textquotedblright\ Hamiltonians} for which the time evolution of
this mixture cannot be understood classically, leading to a violation of
macrorealism. We derive the necessary condition for these non-classical time
evolutions and illustrate it with the example of an oscillating Schrödinger
cat-like state. System decoherence, i.e.\ the continuous monitoring of the
system by an environment, leads to macrorealism but a dynamical description of
non-classical time evolutions in terms of classical laws of motion remains impossible.

In the last part we argue that non-classical Hamiltonians either require
interactions between a large number of particles or are of high
\textit{computational complexity}. This might be understood as the reason why
they are unlikely to appear in nature.\bigskip

\textbf{Chapter 3} investigates \textit{entanglement between collective
operators} in two specific physical systems, namely in a linear chain of
harmonic oscillators and in ensembles of spin-$\tfrac{1}{2}$ particles. We
demonstrate that under certain conditions entanglement between macroscopic
observables can indeed persist for large system sizes. However, since this
analysis uses sharp measurements, it is not in disagreement with our
quantum-to-classical approach.\bigskip

\textbf{Chapter 4} addresses the question of the \textit{origin of quantum
randomness}. In our view, classical physics emerges out of the quantum world
but the randomness in the classical mixture is still irreducible and of
quantum nature. We propose to link quantum randomness with
\textit{mathematical undecidability} in the sense of Chaitin's version of
Gödel's theorem. It states that given a set of axioms that contains a certain
amount of information, it is impossible to deduce the truth value of a
proposition which, together with the axioms, contains more information than
the set of axioms itself.

First, we demonstrate that the states of elementary quantum systems are
capable of encoding mathematical axioms. Quantum mechanics imposes an upper
limit on how much information can be carried by a quantum state, thus limiting
the information content of the set of axioms. Then, we show that quantum
measurements are capable of revealing whether a given proposition is decidable
or undecidable within this set. This allows for an experimental test of
mathematical undecidability by realizing in the laboratory the actual quantum
states and operations required. We theoretically find and experimentally
confirm that whenever a mathematical proposition is undecidable within the
system of axioms encoded in the state, the measurement associated to the
proposition gives random outcomes. Our results support the view that quantum
randomness is irreducible and a manifestation of mathematical undecidability.

\newpage

\pagestyle{fancy} \fancyhf{}
\renewcommand{\chaptermark}[1]{\markboth{\thechapter\ #1}{}}
\renewcommand{\sectionmark}[1]{\markright{\thesection\ #1}}
\fancyhead[LE,RO]{\thepage} \fancyhead[LO]{\rightmark}
\fancyhead[RE]{\leftmark} \renewcommand{\headrulewidth}{0.25pt}
\renewcommand{\footrulewidth}{0pt} \addtolength{\headheight}{0.5pt} \fancypagestyle{plain}{\fancyhead{}\renewcommand{\headrulewidth}{0pt}}

\chapter{Classical world emerging from quantum physics}

\textbf{Summary:}\bigskip

Inspired by the thoughts of Peres on the classical limit~\cite{Pere1995}---we
present a \textit{novel theoretical approach to macroscopic realism and
classical physics within quantum theory}. While our approach is not at
variance with the decoherence program~\cite{Zure1991,Zure2003}, it differs
conceptually from it. It is not dynamical and puts the stress on the limits of
observability of quantum effects of macroscopic objects, i.e.\ on the required
precision of our measurement apparatuses such that quantum phenomena can still
be observed. The term \textquotedblleft macroscopic\textquotedblright\ is used
here to denote a system with a high dimensionality rather than a
low-dimensional system with a large parameter such as mass or size.
Furthermore, there is no need to change the quantum laws itself like in
collapse models~\cite{Ghir1986,Penr1996}.

Using a quantum spin as a model system, we first show that, if consecutive
eigenvalues of a spin component can be experimentally resolved in sharp
quantum measurements, the Leggett-Garg inequality is violated for arbitrary
spin lengths and the violation persists even in the limit of infinitely large
spins. This contradicts the naive assumption that the predictions of quantum
mechanics reduce to those of classical physics merely due to the fact that a
system becomes \textquotedblleft large\textquotedblright. For local realism
this persistence of quantum features was demonstrated by Garg and
Mermin~\cite{Garg1982}, and the violation even increases with the systems'
dimensionality~\cite{Kasz2000,Coll2002}.

In every-day life, however, one not only encounters very high-dimensional
systems but is experimentally restricted to \textit{coarse-grained
measurements}. They only distinguish between eigenvalues which are separated
by much more than the intrinsic quantum uncertainty. We show for arbitrary
spin states that, given a certain time evolution, the macroscopically distinct
outcomes obey the classical Newtonian laws which emerge out of the Schrödinger
equation and the projection postulate.

This suggests that \textit{classical physics can be seen as implied by quantum
mechanics under the restriction of fuzzy measurements} and resolves the
apparent impossibility of how classical realism and deterministic laws can
emerge out of fundamentally random quantum events.\bigskip

\noindent This chapter mainly bases on and also uses parts of
Reference~\cite{Kofl2007b}:

\begin{itemize}
\item J. Kofler and \v{C}. Brukner\newline\textit{Classical world arising out
of quantum physics under the restriction of coarse-grained measurements}%
\newline Phys.~Rev.~Lett.~\textbf{99}, 180403 (2007).\newpage
\end{itemize}

\section{The Leggett-Garg inequality}

In this section we introduce the concept of \textit{macroscopic realism}
(macrorealism) and show how to derive a Leggett-Garg inequality, which can be
used as a tool to indicate whether or not a system's time evolution can be
understood in classical terms. In agreement with this, we then briefly
demonstrate explicitly that the inequality can always be violated for genuine
quantum systems but always satisfied for classical objects.

Macrorealism is defined by the conjunction of the following three
postulates~\cite{Legg2002}:

\begin{enumerate}
\item[(1)] \textit{Macrorealism per se}. A macroscopic object which has
available to it two or more macroscopically distinct states is at any given
time in a definite one of those states.\vspace{-0.2cm}

\item[(2)] \textit{Non-invasive measurability}. It is possible in principle to
determine which of these states the system is in without any effect on the
state itself or on the subsequent system dynamics.\vspace{-0.2cm}

\item[(3)] \textit{Induction}. The properties of ensembles are determined
exclusively by initial conditions (and in particular not by final conditions).
\end{enumerate}

The last two postulates can be phrased into the single assumption that the
object's state is independent of past and future measurements~\cite{Legg1995}.
Classical (Newtonian) physics belongs to the class of macrorealistic theories.

\begin{figure}[b]
\begin{center}
\includegraphics{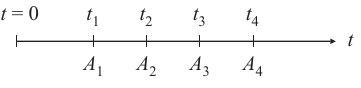}
\end{center}
\par
\vspace{-0.5cm}\caption{In a macrorealistic theory \textit{macrorealism per
se} implies that a time-dependent quantity $A(t)$ of a macroscopic object has
a well defined unambiguous value $A_{i}=A(t_{i})$ at every time $t_{i}$.
\textit{Non-invasive measurability} (together with \textit{induction}) is
reflected by the fact that the $A_{i}$'s do not depend on whether the system
was or was not measured at earlier times.}%
\label{Figure_Leggett_Garg}%
\end{figure}

Now consider a macroscopic physical system and a dichotomic quantity $A$,
which whenever measured is found to take one of the values $\pm1$ only.
Further consider a series of runs starting from identical initial conditions
at time $t=0$ such that on the first set of runs $A$ is measured only at times
$t_{1}$ and $t_{2}$, only at $t_{2}$ and $t_{3}$ on the second, at $t_{3}$ and
$t_{4}$ on the third, and at $t_{1}$ and $t_{4}$ on the fourth $(0\leq
t_{1}<t_{2}<t_{3}<t_{4})$. Let $A_{i}$ denote the value of $A$ at time $t_{i}$
(see Figure~\ref{Figure_Leggett_Garg}). Consider the algebraic combination of
the Clauser-Horne-Shimony-Holt (CHSH) type \cite{Clau1969}:%
\begin{equation}
A_{1}\,(A_{2}-A_{4})+A_{3}\,(A_{2}+A_{4})=\pm2\,. \label{eq Ais}%
\end{equation}
It can only have the values $+2$ or $-2$ for one of the two
brackets has to vanish and the other is $+2$ or $-2$ and then multiplied with
$+1$ or $-1$. \textit{Macrorealism per se} is reflected by the objective
existence of unambiguous values of $A_{i}$ at all times, and
\textit{non-invasive measurability} together with \textit{induction} is
reflected by the fact that the $A_{i}$'s are the same, independent in which
combination they appear. E.g., $A_{4}$ is independent of previous
measurements, i.e.\ whether it appears in a run together with $A_{1}$ or
$A_{3}$. Repeating the experimental runs many times, we introduce the temporal
correlation functions%
\begin{equation}
C_{ij}\equiv\langle A_{i}\,A_{j}\rangle\,.
\end{equation}
By averaging (\ref{eq Ais}) it follows that any macrorealistic theory has to
satisfy the \textit{Leggett-Garg inequality}~\cite{Legg1985}%
\begin{equation}
\fbox{$\;\;K\equiv C_{12}+C_{23}+C_{34}-C_{14}\leq2\,.\;\;$}
\label{eq Leggett}%
\end{equation}
Its violation implies that the object's time evolution cannot be understood classically.

\begin{figure}[t]
\begin{center}
\includegraphics{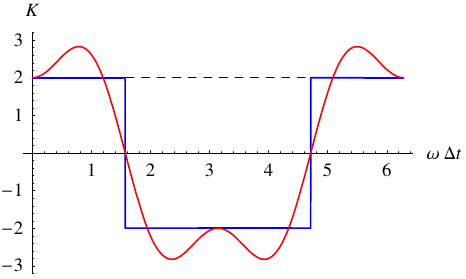}
\end{center}
\par
\vspace{-0.25cm}\caption{Violation of the Leggett-Garg inequality
(\ref{eq Leggett micro}) for a rotating spin-$\frac{1}{2}$ particle with
precession frequency $\omega$. The left-hand side of the inequality, $K$, is
shown by a red line, while the classical limit, $K=2$, is indicated by a
dashed line. If the distance $\Delta t$ between the four possible equidistant
measurement times is chosen as $\Delta t=\frac{\pi}{4\omega}$, then the
inequality is maximally violated with $K=2\,\sqrt{2}$. The blue line shows the
left-hand side of the Leggett-Garg inequality for a classical rotating spin
vector. Its time evolution can be understood classically and does not violate
the inequality.}%
\label{Figure_Violation_c}%
\end{figure}

Let us briefly analyze the quantum evolution of a \textit{microscopic} quantum
object, say the precession of a spin-$\frac{1}{2}$ particle with the
Hamiltonian $\hat{H}=\frac{1}{2}\,\omega\,\hat{\sigma}_{x}$, where $\omega$ is
the angular precession frequency and $\hat{\sigma}_{x}$ is the Pauli
$x$-matrix.\footnote{Throughout this and the subsequent chapter we use units
in which the reduced Planck constant is $\hbar=1$.} If we measure the spin
along the $z$-direction, then we obtain the temporal correlations
$C_{ij}=\langle\hat{\sigma}_{z}(t_{i})\,\hat{\sigma}_{z}(t_{j})\rangle
=\cos[\omega(t_{j}\!-\!t_{i})]$. Choosing the four possible measurement times
as equidistant, with time distance $\Delta t=t_{2}-t_{1}=t_{3}-t_{2}%
=t_{4}-t_{3}$, the Leggett-Garg inequality becomes%
\begin{equation}
K=3\cos(\omega\Delta t)-\cos(3\omega\Delta t)\leq2\,. \label{eq Leggett micro}%
\end{equation}
This is maximally violated for the time distance $\Delta t=\frac{\pi}{4\omega
}$ for which $K=2\,\sqrt{2}$ (see red line in Figure~\ref{Figure_Violation_c}%
). The violation is not surprising as a spin-$\frac{1}{2}$ particle is a
genuine quantum system and cannot have the objective properties tentatively
attributed to macroscopic objects prior to and independent of measurements. In
contrast, we consider an arbitrarily sized uniformly rotating classical spin
vector, again precessing around $x$ and pointing along $z$ at time $t_{1}$. As
dichotomic observable quantity we use $A(t_{i})=\;$sgn$(\cos\omega t_{i})$
such that $A=+1$ ($-1$) if the spin is pointing upwards (downwards) along $z$.
As expected, the inequality (\ref{eq Leggett}) is always satisfied (see blue
line in Figure~\ref{Figure_Violation_c}).

\section{Violation of the Leggett-Garg inequality for arbitrarily large spins}

\begin{figure}[t]
\begin{center}
\includegraphics{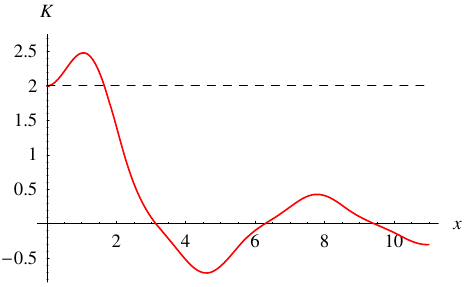}
\end{center}
\par
\vspace{-0.25cm}\caption{Violation of the Leggett-Garg inequality
(\ref{eq Leggett K(x)}) for \textit{arbitrarily large} spin size $j$. The left
hand side of the inequality, $K$, is indicated by a red line and the classical
limit is $K=2$ (dashed line). The initial quantum state is totally mixed, eq.
(\ref{eq mixed state}), and the Hamiltonian (\ref{eq Hamiltonian}) produces a
precession around the $x$-axis with frequency $\omega$. The dichotomic
observable is the parity in a measurement (\ref{eq parity}) of the spin's
$z$-component. For all spin sizes $j$ and precession frequencies $\omega$, one
can always find a time distance $\Delta t$ between the four possible
equidistant measurement times such that the quantity $x\equiv(2j+1)\,\omega
\,\Delta t$ is around the value $1$ and the classical limit is violated. The
maximal violation $K\approx2.481$ is achieved for $x\approx1.054$.}%
\label{Figure_Violation_K}%
\end{figure}

In this section, we demonstrate that the Leggett-Garg inequality
(\ref{eq Leggett}) is violated for arbitrarily large (macroscopic) spin
lengths $j$ as long as accurate measurements can be performed. In any run, the
first of the two measurements acts as a preparation of the state for the
subsequent measurement. Therefore, the initial state of the spin is not
decisive and it is sufficient for us to consider as initial state the
maximally mixed one:%
\begin{equation}
\hat{\rho}(0)\equiv\dfrac{1}{2j+1}\,%
{\displaystyle\sum\nolimits_{m}}
\left\vert m\right\rangle \!\left\langle m\right\vert =\dfrac
{\leavevmode\hbox{\small1\kern-3.3pt\normalsize1}}{2j+1}\,.
\label{eq mixed state}%
\end{equation}
Here, $\leavevmode\hbox{\small1\kern-3.3pt\normalsize1}$ is the identity
operator and $\left\vert m\right\rangle $ are the $\hat{J}_{z}$ (spin
$z$-component) eigenstates with the possible eigenvalues $m=-j,-j+1,...,+j$.
We consider the Hamiltonian%
\begin{equation}
\hat{H}=\frac{\hat{\mathbf{J}}^{2}}{2I}+\omega\,\hat{J}_{x}\,,
\label{eq Hamiltonian}%
\end{equation}
where $\hat{\mathbf{J}}$ is the rotor's total spin vector operator, $\hat
{J}_{x}$ its $x$-component, $I$ the moment of inertia and $\omega$ the angular
precession frequency. The constant of motion $\frac{\hat{\mathbf{J}}^{2}}{2I}$
can be ignored since $\hat{\mathbf{J}}^{2}$ commutes with the individual spin
components and does not contribute to their time evolution. The solution of
the Schrödinger equation produces a rotation about the $x$-axis, represented
by the time evolution operator%
\begin{equation}
\hat{U}_{t}\equiv\text{e}^{-\text{i}\omega t\hat{J}_{x}}.
\label{eq time evolution}%
\end{equation}
We assume that individual eigenstates $\left\vert m\right\rangle $ can be
experimentally resolved and use the parity measurement%
\begin{equation}
\hat{A}\equiv%
{\displaystyle\sum\nolimits_{m}}
(-1)^{j-m}\left\vert m\right\rangle \!\left\langle m\right\vert =\text{e}%
^{\text{i}\pi(j-\hat{J}_{z})} \label{eq parity}%
\end{equation}
with the possible dichotomic outcomes $\pm1$ (identifying $\pm\equiv\pm1$).
The temporal correlation function between results of the parity measurement
$\hat{A}$ at different (arbitrary) times $t_{i}$ and $t_{j}$ ($t_{j}>t_{i}$)
is%
\begin{equation}
C_{ij}=p_{i+}\,q_{j+|i+}+p_{i-}\,q_{j-|i-}-p_{i+}\,q_{j-|i+}-p_{i-}%
\,q_{j+|i-}\,, \label{eq C}%
\end{equation}
where $p_{i+}$ ($p_{i-}$) is the probability for measuring $+$ ($-$) at
$t_{i}$ and $q_{jl|ik}$ is the probability for measuring $l$ at $t_{j}$ given
that $k$ was measured at $t_{i}$ ($k,l=+,-$). Furthermore,%
\begin{align}
p_{i+}  &  =1-p_{i-}=\tfrac{1}{2}\,(\langle\hat{A}_{t_{i}}\rangle+1)\,,\\
q_{j+|i\pm}  &  =1-q_{j-|i\pm}=\tfrac{1}{2}\,(\langle\hat{A}_{t_{j}}%
\rangle_{\pm}+1)\,.
\end{align}
Here $\langle\hat{A}_{t_{i}}\rangle$ is the expectation value of $\hat{A}$ at
$t_{i}$ and $\langle\hat{A}_{t_{j}}\rangle_{\pm}$ is the expectation value of
$\hat{A}$ at $t_{j}$ given that $\pm$ was the outcome at $t_{i}$. The totally
mixed state is not changed until the first measurement: $\hat{\rho}%
(t_{i})=\hat{U}_{t_{i}}\,\hat{\rho}(0)\,\hat{U}_{t_{i}}^{\dagger}=\hat{\rho
}(0)$ and we find%
\begin{equation}
\langle\hat{A}_{t_{i}}\rangle=\text{Tr}[\hat{\rho}(t_{i})\,\hat{A}]=\dfrac
{1}{2j+1}\,%
{\displaystyle\sum\nolimits_{m}}
(-1)^{j-m}\approx0\,. \label{eq expectation P_t1}%
\end{equation}
The approximate sign is accurate for half integer $j$ as well as in the
macroscopic limit $j\gg1$, which is assumed from now on. Hence, we have
$p_{i+}=p_{i-}=\tfrac{1}{2}$, which is self-evident for a totally mixed state.
Depending on the measurement result $\pm$ at $t_{i}$, the state is reduced to%
\begin{equation}
\hat{\rho}_{\pm}(t_{i})=\dfrac{\hat{P}_{\pm}\,\hat{\rho}(t_{i})\,\hat{P}_{\pm
}}{\text{Tr}[\hat{P}_{\pm}\,\hat{\rho}(t_{i})\,\hat{P}_{\pm}]}=\dfrac
{\leavevmode\hbox{\small1\kern-3.3pt\normalsize1}\pm\hat{A}}{2j+1}\,,
\label{eq reduced}%
\end{equation}
with $\hat{P}_{\pm}\equiv\tfrac{1}{2}%
\,(\leavevmode\hbox{\small1\kern-3.3pt\normalsize1}\pm\hat{A})$ the projection
operator onto positive (negative) parity states. Denoting $\theta\equiv
\omega\,(t_{j}-t_{i})$, the remaining expectation value $\langle\hat{A}%
_{t_{j}}\rangle_{\pm}=\;$Tr$[\hat{U}_{t_{j}-t_{i}}\,\hat{\rho}_{\pm}%
(t_{i})\,\hat{U}_{t_{j}-t_{i}}^{\dagger}\,\hat{A}]$ becomes%
\begin{align}
\langle\hat{A}_{t_{j}}\rangle_{\pm}  &  =\pm\dfrac{1}{2j+1}\,\text{Tr}%
[\text{e}^{-\text{i}\theta\hat{J}_{x}}\,\text{e}^{\text{i}\pi(j-\hat{J}_{z}%
)}\,\text{e}^{\text{i}\theta\hat{J}_{x}}\,\text{e}^{\text{i}\pi(j-\hat{J}%
_{z})}]\nonumber\\
&  =\pm\frac{\text{Tr}[\text{e}^{2\text{i}\theta\hat{J}_{x}}]}{2j+1}=\pm
\dfrac{\sin[(2j+1)\,\omega\,(t_{j}-t_{i})]}{(2j+1)\sin[\omega\,(t_{j}-t_{i}%
)]}\,.
\end{align}
Here we used the geometrical meaning of the rotations in the first line. From
$\langle\hat{A}_{t_{j}}\rangle_{+}=-\langle\hat{A}_{t_{j}}\rangle_{-}$ it
follows $q_{+|+}+q_{+|-}=1$. Using this and $p_{i+}=\tfrac{1}{2}$ from above,
the temporal correlation function becomes%
\begin{equation}
C_{ij}=\langle\hat{A}_{t_{j}}\rangle_{+}\,.
\end{equation}
Having four possible equidistant measurements with time distance $\Delta t$,
and using the abbreviation%
\begin{equation}
x\equiv(2j+1)\,\omega\,\Delta t
\end{equation}
the Leggett--Garg inequality (\ref{eq Leggett}) now reads%
\begin{equation}
\fbox{$\;\;K\approx\dfrac{3\sin x}{x}-\dfrac{\sin3x}{3x}\leq2\,.\;\;$}
\label{eq Leggett K(x)}%
\end{equation}
We approximated the sine function in the denominator, assuming $\tfrac
{x}{2j+1}\ll1$. Inequality~(\ref{eq Leggett K(x)}) is violated for all
positive $x\lesssim1.656$ and maximally violated for $x\approx1.054$ where
$K\approx2.481$ (compare with Reference~\cite{Pere1995} for the violation of
local realism) as can be seen in Figure~\ref{Figure_Violation_K}. For every
spin size $j$, and given a precession frequency $\omega$, it is always
possible to choose the time distance $\Delta t$ such that $K>2$%
.

\begin{quote}
\textit{We can conclude that a violation of the Leggett-Garg inequality is
possible for arbitrarily high-dimensional systems and even for initially
totally mixed states.}
\end{quote}

Note, however, that the temporal precision of our measurement apparatuses,
which is required for seeing the violation, increases with $j$, as
$\omega\,\Delta t$ has to scale with $j^{-1}$ in order to keep $x\approx1$.
Moreover, due to the nature of the parity measurement, consecutive values of
$m$ have to be resolved.

\section{The quantum-to-classical transition for a spin-coherent state}

In this section we will show that coarse-grained measurements not only lead to
the validity of macrorealism but even to the \textit{emergence of classical
physics} for a certain class of quantum states. The generalization to
arbitrary states will be done in the next section.

Let us start with a preliminary remark about distinguishability of states in
quantum theory. Any two different eigenvalues $m_{1}$ and $m_{2}$ in a
measurement of a spin's $z$-component correspond to orthogonal states
\textit{without any concept of closeness or distance}. In Hilbert space the
vectors $|m\rangle$ and $|m\!+\!1\rangle$ are as orthogonal as $|m\rangle$ and
$|m\!+\!10^{10}\rangle$:%
\begin{align}
\langle m\!+\!1|m\rangle &  =0\,,\\
\langle m\!+\!10^{10}|m\rangle &  =0\,.
\end{align}

\begin{quote}
\textit{The terms \textquotedblleft close\textquotedblright\ or
\textquotedblleft distant\textquotedblright\ only make sense in a classical
context, where those eigenvalues are treated as close which correspond to
neighboring outcomes in the real configuration space.}
\end{quote}

For example, the \textquotedblleft eigenvalue labels\textquotedblright\ $m$
and $m+1$ of a spin component observable correspond to neighboring outcomes in
a Stern-Gerlach experiment. (Such observables are sometimes called classical
or reasonable~\cite{Yaff1982,Kay1983,Pere1995}.) If our measurement accuracy
is limited, it is those neighboring eigenvalues which we conflate to
coarse-grained observables. It seems thus unavoidable that certain features of
classicality have to be assumed beforehand to give the Hilbert space some
structure which it does not have a priori.

In what follows, we will first consider the special case of a single spin
coherent state and then generalize the transition to classicality for
arbitrary states. \textit{Spin-}$j$\textit{ coherent states }$\left\vert
\Omega\right\rangle \equiv\left\vert \vartheta,\varphi\right\rangle
$~\cite{Radc1971,Atki1971} are the eigenstates with maximal eigenvalue of a
spin operator pointing into the direction $\Omega\equiv(\vartheta,\varphi)$,
where $\vartheta$ and $\varphi$ are the polar and azimuthal angle in spherical
coordinates, respectively:%
\begin{equation}
\fbox{$\;\;\hat{\mathbf{J}}_{\vartheta,\varphi}\left\vert \vartheta
,\varphi\right\rangle =j\left\vert \vartheta,\varphi\right\rangle .\;\;$}%
\end{equation}
Let us consider the initial spin coherent state at time $t=0$ pointing into
the direction ($\vartheta_{0},\varphi_{0}$). In the basis of $\hat{J}_{z}$
eigenstates it reads%
\begin{equation}
|\vartheta_{0},\varphi_{0}\rangle=%
{\displaystyle\sum\nolimits_{m}}
\!\left(  \!%
\genfrac{}{}{0pt}{1}{2\,j}{j+m}%
\!\right)  ^{\!1/2}\cos^{j+m}\!\tfrac{\vartheta_{0}}{2}\,\sin^{j-m}%
\!\tfrac{\vartheta_{0}}{2}\;\text{e}^{-\text{i}m\varphi_{0}}\,|m\rangle\,.
\end{equation}
Under time evolution $\hat{U}_{t}=\;$e$^{-\text{i}\omega t\hat{J}_{x}}$,
eq.\ (\ref{eq time evolution}), the probability that a $\hat{J}_{z}$
measurement at some later time $t$ has the particular outcome $m$ is given by
the binomial distribution%
\begin{equation}
p(m,t)=|\langle m|\vartheta_{t},\varphi_{t}\rangle|^{2} \label{eq p(m,t)}%
\end{equation}
with $\cos\vartheta_{t}=\sin\omega t\sin\vartheta_{0}\sin\varphi_{0}%
+\cos\omega t\cos\vartheta_{0}$, $\tan\varphi_{t}=\cos\omega t\tan\varphi
_{0}-\sin\omega t\tan^{-1}\!\vartheta_{0}\cos^{-1}\!\varphi_{0}$, where
$\vartheta_{t}$ and $\varphi_{t}$ are the polar and azimuthal angle of the
(rotated) spin coherent state $\left\vert \vartheta_{t},\varphi_{t}%
\right\rangle =\hat{U}_{t}\left\vert \vartheta_{0},\varphi_{0}\right\rangle $
at time $t$. In the macroscopic limit, $j\gg1$, the binomial distribution
(\ref{eq p(m,t)}) can be very well approximated by a Gaussian distribution%
\begin{equation}
p(m,t)\approx\dfrac{1}{\sqrt{2\pi}\,\sigma}\;\text{e}^{-\frac{(m-\mu)^{2}%
}{2\sigma^{2}}} \label{eq p(m1,t1)}%
\end{equation}
with $\sigma\equiv\sqrt{j/2}\,\sin\vartheta_{t}$ the width (standard
deviation) and $\mu\equiv j\cos\vartheta_{t}$ the mean value.

\begin{figure}[t]
\begin{center}
\includegraphics{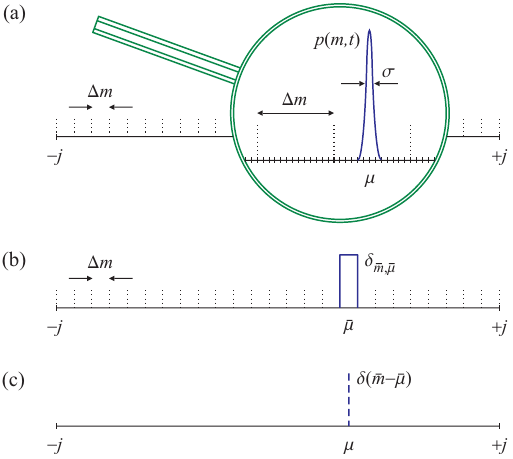}
\end{center}
\par
\vspace{-0.25cm}\caption{An initial spin-$j$ coherent state $\left\vert
\vartheta_{0},\varphi_{0}\right\rangle $ precesses into the coherent state
$\left\vert \vartheta_{t},\varphi_{t}\right\rangle $ at time $t$ under a
quantum time evolution. (a)~The probability $p(m,t)$ for the outcome $m$ in a
measurement of the spin's $z$-component is given by a Gaussian distribution
with width $\sigma$ and mean $\mu$, which can be seen under the magnifying
glass of sharp measurements. (b)~The measurement resolution $\Delta m$ is
finite and subdivides the $2j+1$ possible outcomes into a smaller number of
coarse-grained \textquotedblleft slots\textquotedblright. If the measurement
accuracy is much poorer than the width $\sigma$, i.e., $\Delta m\gg\!\sqrt{j}%
$, the sharply peaked Gaussian cannot be distinguished anymore from the
discrete Kronecker delta $\delta_{\bar{m},\bar{\mu}}$ where $\bar{m}$ is
numbering the slots and $\bar{\mu}$ is the slot in which the center $\mu$ of
the Gaussian lies. (c)~In the limit $j\rightarrow\infty$ the slots
\textit{seem} to become infinitely narrow and $\delta_{\bar{m},\bar{\mu}}$
becomes the Dirac delta function $\delta(\bar{m}\!-\!\bar{\mu})$.}%
\label{Figure_Magnifying_glass}%
\end{figure}

Under the \textquotedblleft magnifying glass\textquotedblright\ of sharp
measurements individual eigenvalues $m$ can be distinguished and the Gaussian
probability distribution $p(m,t)$ can be resolved, as shown in
Figure~\ref{Figure_Magnifying_glass}(a), allowing a violation of the
Leggett-Garg inequality. Let us now assume that, as in every-day life, the
resolution of the measurement apparatus, $\Delta m$, is restricted and that it
subdivides the $2j+1$ possible different outcomes $m$ into a much smaller
number of $\frac{2j+1}{\Delta m}$ coarse-grained \textquotedblleft
slots\textquotedblright\ $\bar{m}$. If the slot size is much larger than the
standard deviation $\sigma\sim\!\sqrt{j}$, i.e.%
\begin{equation}
\fbox{$\;\;\Delta m\gg\sqrt{j}\,,\;\;$}%
\end{equation}
the sharply peaked Gaussian cannot be distinguished anymore from the discrete
Kronecker delta,%
\begin{equation}
\Delta m\gg\!\sqrt{j}:\quad p(m,t)\rightarrow\delta_{\bar{m},\bar{\mu}}\,.
\label{eq Kronecker}%
\end{equation}
Here, $\bar{m}$ is numbering the slots (from $-j+\frac{\Delta m}{2}$ to
$j-\frac{\Delta m}{2}$ in steps $\Delta m$) and $\bar{\mu}$ is the number of
the slot in which the center $\mu$ of the Gaussian lies. This is indicated in
Figure~\ref{Figure_Magnifying_glass}(b).

In the \textit{classical limit}, $j\rightarrow\infty$, one can distinguish two
cases: (1) If the inaccuracy $\Delta m$ scales linearly with $j$,
i.e.\ $\Delta m=O(j)$, the discreteness remains. (2) If $\Delta m$ scales
slower than $j$, i.e.\ $\Delta m=o(j)$ but still $\Delta m\gg\!\sqrt{j}$, then
the slots seem to become infinitely narrow. Pictorially, the \textit{real
space length} of the eigenvalue axis, representing the $2j+1$ possible
outcomes $m$, is limited in any laboratory, e.g., by the size of the
observation screen after a Stern-Gerlach magnet, whereas the number of slots
grows as $j/\Delta m$. Then, in the limit $j\rightarrow\infty$, the discrete
Kronecker delta becomes the Dirac delta function,%
\begin{equation}
\Delta m\gg\!\sqrt{j}\;\;\&\;\;j\rightarrow\infty:\quad p(m,t)\rightarrow
\delta(\bar{m}\!-\!\bar{\mu})\,, \label{eq delta}%
\end{equation}
which is shown in Figure~\ref{Figure_Magnifying_glass}(c).

Now we have to focus on the question in which sense coarse-grained von Neumann
measurements disturb the spin coherent state. Let%
\begin{equation}
\hat{P}_{\bar{m}}\equiv%
{\displaystyle\sum\nolimits_{m\in\{\bar{m}\}}}
\left\vert m\right\rangle \!\left\langle m\right\vert \label{eq Pmbar}%
\end{equation}
denote the projector onto the slot $\bar{m}$ with $\{\bar{m}\}$ the set of all
$m$ belonging to $\bar{m}$. Then $\hat{P}_{\bar{m}}\,|\vartheta,\varphi
\rangle$ is \textit{almost} $|\vartheta,\varphi\rangle$ (the zero vector
$\mathbf{0}$) for all coherent states lying inside (outside) the slot,
respectively:%
\begin{equation}
\hat{P}_{\bar{m}}\,|\vartheta,\varphi\rangle\approx\left\{
\begin{array}
[c]{ll}%
|\vartheta,\varphi\rangle & \text{for }\bar{\mu}\text{ inside }\bar{m}%
\text{,}\\
\mathbf{0} & \text{for }\bar{\mu}\text{ outside }\bar{m}\text{.}%
\end{array}
\right.  \label{eq Projector}%
\end{equation}
This means that the reduced (projected) state is essentially the state before
the measurement or projected away. If $\left\vert \vartheta,\varphi
\right\rangle $ is centered well inside the slot, the above relation holds
with merely exponentially small deviation. Only in the cases where $\left\vert
\vartheta,\varphi\right\rangle $ is close to the border between two slots, the
measurement is invasive and disturbs the state. Presuming that the measurement
times and/or slot positions chosen by the observer are statistically
independent of the (initial) position of the coherent state, a significant
disturbance happens merely in the fraction $\sigma/\Delta m\ll1$ of all
measurements. This is equivalent to the already assumed condition $\!\sqrt
{j}\ll\Delta m$. Therefore, fuzzy measurements of a spin coherent state are
almost always non-invasive such as in any macrorealistic theory, in particular
classical Newtonian physics. Small errors may accumulate over many
measurements and eventually there might appear deviations from the classical
time evolution. This, however, is unavoidable in any explanation of
classicality \textit{gradually} emerging out of quantum theory.\footnote{For
the general trade-off between measurement accuracy and state disturbance for
the more realistic smoothed positive operator value measure (POVM) and for
related approaches to classicality, see
References~\cite{Busc1995,Poul2005,Gell2007}. A natural way of implementing
coarse-grained measurements as POVM is presented in the next chapter. In
contrast to von Neumann measurements, they do not allow to distinguish
perfectly between two states at two sides of a slot border. But under all
circumstances it is in general unavoidable that quantum measurements are
invasive to some extent. Classicality arises in the sense that the effects of
these deviations become negligibly small.}

Hence, at the coarse-grained level the physics of the (quantum) spin system
can completely be described by a \textquotedblleft new\textquotedblright%
\ formalism, utilizing an initial (classical) \textit{spin vector}
$\mathbf{J}$ at time $t=0$, pointing in the ($\vartheta_{0},\varphi_{0}%
$)-direction with length $J\equiv|\mathbf{J}|=\!\sqrt{j(j\!+\!1)}\approx j$,
where $j\gg1$, and a (Hamilton) function%
\begin{equation}
H=\frac{\mathbf{J}^{2}}{2I}+\omega\,J_{x}\,. \label{eq Hamilton}%
\end{equation}
At any time the probability that the spin vector's $z$-component
$J\cos\vartheta_{t}\approx j\cos\vartheta_{t}$ is in slot $\bar{m}$ is given
by $\delta_{\bar{m},\bar{\mu}}$, eq.~(\ref{eq Kronecker}), \textit{as if} the
time evolution of the spin components $J_{i}$ ($i=x,y,z$) is given by the
Poisson brackets,%
\begin{equation}
\dot{J}_{i}=[J_{i},H]_{\text{PB}}\,, \label{eq Poisson}%
\end{equation}
and measurements are non-invasive. Only the term $\omega\,J_{x}$ in
eq.~(\ref{eq Hamilton}) governs the time evolution and the solutions
correspond to a rotation around the $x$-axis. The spin vector at time $t$
points in the ($\vartheta_{t},\varphi_{t}$)-direction where $\vartheta_{t}$
and $\varphi_{t}$ are the same as for the spin coherent state and the
probability of measurement outcomes is given by $\delta(\bar{m}\!-\!\bar{\mu
})$, eq.~(\ref{eq delta}).

\begin{quote}
\textit{This is classical (Newtonian) mechanics of a single spin emerging from
quantum physics}.
\end{quote}

\section{The quantum-to-classical transition for an arbitrary spin state}

Now we demonstrate that the time evolution of \textit{any} spin-$j$ quantum
state becomes classical under the restriction of coarse-grained measurements.
At all times any (pure or mixed) spin-$j$ density matrix can be written in the
quasi-diagonal form~\cite{Arec1972}%
\begin{equation}
\hat{\rho}=%
{\displaystyle\iint}
\,P(\Omega)\,|\Omega\rangle\langle\Omega|\,\text{d}^{2}\Omega\label{eq rho}%
\end{equation}
with d$^{2}\Omega\equiv\sin\vartheta\,$d$\vartheta\,$d$\varphi$ the
infinitesimal solid angle element and $P(\Omega)$ a \textit{not necessarily
positive} real function with the normalization $%
{\textstyle\iint}
P(\Omega)\,$d$^{2}\Omega=1$.

\begin{figure}[t]
\begin{center}
\includegraphics{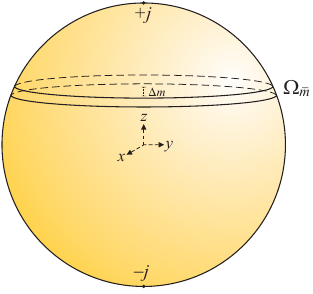}
\end{center}
\par
\vspace{-0.25cm}\caption{Given a coarse-grained measurement of the spin's
$z$-component ($\Delta m\gg\!\sqrt{j}$), one can compute the probability
$w_{\bar{m}}$ that the outcome will be a certain slot $\bar{m}$ via
integration of the quantum state's $P$-function over the region $\Omega
_{\bar{m}}$. This is the region between two circles of latitude corresponding
to the slot $\bar{m}$ of $\hat{J}_{z}$ eigenstates. The coarse-graining
condition $\Delta m\gg\!\sqrt{j}$ ensures that alternatively the integration
can be carried out over the state's $Q$-function, representing the probability
distribution of an ensemble of classical spins.}%
\label{Figure_Sphere}%
\end{figure}

The probability for an outcome $m$ in a $\hat{J}_{z}$ measurement in the state
(\ref{eq rho}) is given by%
\begin{equation}
w(m)=\text{Tr}[\hat{\rho}\left\vert m\right\rangle \!\left\langle m\right\vert
]=%
{\displaystyle\iint}
\,P(\Omega)\,p(m)\,\text{d}^{2}\Omega\,,
\end{equation}
where $p(m)$ is written in eq.~(\ref{eq p(m,t)}). At the coarse-grained level
of classical physics only the probability for a slot outcome $\bar{m}$ can be
measured, i.e.%
\begin{equation}
w_{\bar{m}}\equiv\text{Tr}[\hat{\rho}\,\hat{P}_{\bar{m}}]=%
{\displaystyle\sum\nolimits_{m\in\{\bar{m}\}}}
w(m)
\end{equation}
with $\hat{P}_{\bar{m}}$ from eq.\ (\ref{eq Pmbar}). Inserting
eq.~(\ref{eq rho}), we get $w_{\bar{m}}=%
{\textstyle\iint}
P(\Omega)\,$Tr$[\hat{P}_{\bar{m}}|\Omega\rangle\langle\Omega|]\,$d$^{2}\Omega
$. Using eq.~(\ref{eq Projector}), $\Delta m\gg\!\sqrt{j}$, and Tr$[|\Omega
\rangle\langle\Omega|]=1$, this can be well approximated by%
\begin{equation}
w_{\bar{m}}\approx%
{\displaystyle\iint\nolimits_{\Omega_{\bar{m}}}}
P(\Omega)\,\text{d}^{2}\Omega\,, \label{eq P(m) f}%
\end{equation}
where $\Omega_{\bar{m}}$ is the region between two circles of latitude at
polar angles $\vartheta_{1}(\bar{m})$ and $\vartheta_{2}(\bar{m})$
corresponding to the slot $\bar{m}$ (Figure~\ref{Figure_Sphere}). We will show
that $w_{\bar{m}}$ can be obtained from a \textit{positive probability
distribution of classical spin vectors}. Consider the well known
$Q$-function~\cite{Agar1981,Agar1993}%
\begin{equation}
Q(\Omega)\equiv\frac{2j+1}{4\,\pi}\,%
{\displaystyle\iint}
\,P(\Omega^{\prime})\,\cos^{4j}\!\tfrac{\Theta}{2}\,\,\text{d}^{2}%
\Omega^{\prime} \label{eq g}%
\end{equation}
with d$^{2}\Omega^{\prime}\equiv\sin\vartheta^{\prime}%
\,$d$\vartheta^{\prime}\,$d$\varphi^{\prime}$ and $\Theta=2\arccos\{\tfrac
{1}{2}\,[1+\cos\vartheta\cos\vartheta^{\prime}+\sin\vartheta\sin
\vartheta^{\prime}\cos(\varphi-\varphi^{\prime})]\}^{1/2}$ the angle between
the directions $\Omega\equiv(\vartheta,\varphi)$ and $\Omega^{\prime}%
\equiv(\vartheta^{\prime},\varphi^{\prime})$. In the case of large spins the
factor $\cos^{4j}\!\tfrac{\Theta}{2}\sim\exp(-j\Theta^{2})$ in the integrand
is sharply peaked around vanishing relative angle $\Theta$ and significant
contributions arise only from regions where $\Theta\lesssim1/\!\sqrt{j}$. The
normalization factor $\frac{2j+1}{4\pi}$ in eq.~(\ref{eq g}) is the inverse
size of the solid angle element for which the integrand contributes
significantly and makes $Q$ normalized: $%
{\textstyle\iint\nolimits_{\Omega}}
Q(\Omega)\,$d$^{2}\Omega=1$.

The distribution $Q$ is \textit{positive} because it is, up to a normalization
factor, the expectation value of the state $|\Omega\rangle$:%
\begin{equation}
\fbox{$\;\;Q(\Omega)\equiv\dfrac{2j+1}{4\pi}\,\langle\Omega|\hat{\rho}%
|\Omega\rangle\,.\;\;$}%
\end{equation}
For fuzzy measurements with (angular) inaccuracy $\Delta\Theta\sim
\vartheta_{2}(\bar{m})-\vartheta_{1}(\bar{m})\gg1/\!\sqrt{j}$, which is
equivalent to $\Delta m\gg\!\sqrt{j}$, the probability for having an outcome
$\bar{m}$ can now be expressed only in terms of the positive distribution $Q$:%
\begin{equation}
w_{\bar{m}}\approx%
{\displaystyle\iint\nolimits_{\Omega_{\bar{m}}}}
Q(\Omega)\,\text{d}^{2}\Omega\,. \label{eq P(m) g}%
\end{equation}
Figure~\ref{Figure_Sphere} shows the integration region $\Omega_{\bar{m}}$
over which $P$ and $Q$ have to be integrated. The approximate equivalence of
eqs.~(\ref{eq P(m) f}) and (\ref{eq P(m) g}) is verified by substituting
eq.~(\ref{eq g}) into (\ref{eq P(m) g}) and is not accurately fulfilled for
quantum states $\rho$ directly at a slot border.\footnote{This issue is
related to the fact that our von Neumann coarse-grained measurements have
sharp slot borders and will be resolved in the subsequent chapter.}

Note that $Q$ is a mere mathematical tool and not experimentally accessible in
coarse-grained measurements. Operationally, because of $\Delta m\gg\!\sqrt{j}$
an averaged version of $Q$, denoted as $R$, is used by the experimenter to
describe the system. Mathematically, this function $R$ is obtained by
integrating $Q$ over solid angle elements corresponding to the actual
measurement inaccuracy. In the classical limit, without the \textquotedblleft
magnifying glass\textquotedblright, the regions given by the experimenter's
resolution become \textquotedblleft points\textquotedblright\ on the sphere
where $R$ is defined.

\begin{quote}
\textit{Thus, under coarse-grained measurements, a full description of an
arbitrary quantum spin state is provided by an ensemble of classical spins
with a positive probability distribution.}
\end{quote}

In other words, there exists a hidden variable description. The \textit{time
evolution} of the general state (\ref{eq rho}) is determined by
(\ref{eq Hamiltonian}). In the classical limit it can be described by an
ensemble of classical spins characterized by the initial distribution $Q$
($R$), where each spin is rotating according to the Hamilton function
(\ref{eq Hamilton}). From eq.~(\ref{eq P(m) g}) one can see that for the
non-invasiveness at the classical level \textit{it is the change of the
distribution }$Q$\textit{ (}$R$)\textit{ which is important and not the change
of the quantum state or equivalently }$P$\textit{ itself}. In fact, upon a
fuzzy $\hat{J}_{z}$ measurement the state $\hat{\rho}$ \textit{is} reduced to
one particular state depending on the outcome $\bar{m}$,%
\begin{equation}
\hat{\rho}_{\bar{m}}=\dfrac{\hat{P}_{\bar{m}}\,\hat{\rho}\,\hat{P}_{\bar{m}}%
}{w_{\bar{m}}}\,,
\end{equation}
with the corresponding (normalized) functions $P_{\bar{m}}$, $Q_{\bar{m}}$ and
$R_{\bar{m}}$. The reduction to $\hat{\rho}_{\bar{m}}$ happens with
probability $w_{\bar{m}}$, which is given by eq.~(\ref{eq P(m) f}) or
(\ref{eq P(m) g}). Whereas the $P$-function can change dramatically upon
reduction, $Q_{\bar{m}}$ is (up to normalization) approximately the same as
the original $Q$ in the region $\Omega_{\bar{m}}$. Thus,%
\begin{equation}
Q_{\bar{m}}(\Omega)\propto\langle\Omega|\,\hat{\rho}_{\bar{m}}\,|\Omega
\rangle\propto\langle\Omega|\,\hat{P}_{\bar{m}}\,\hat{\rho}\,\hat{P}_{\bar{m}%
}\,|\Omega\rangle\approx\langle\Omega|\,\hat{\rho}\,|\Omega\rangle
\propto\left\{
\begin{array}
[c]{ll}%
Q(\Omega) & \!\!\text{for }\Omega\text{ inside }\Omega_{\bar{m}}\text{,}\\
0 & \!\!\text{for }\Omega\text{ outside }\Omega_{\bar{m}}\text{.}%
\end{array}
\right.
\end{equation}
Therefore, at the coarse-grained level the distribution $Q_{\bar{m}}$
($R_{\bar{m}}$) of the reduced state after the measurement can always be
understood approximately as a subensemble of the (classical) distribution $Q$
before the measurement.

\begin{quote}
\textit{Effectively, the measurement only reveals already existing properties
in the mixture and does not alter the subsequent rotation of the individual
classical spins.}
\end{quote}

The disturbance at the slot borders at that level is quantified by how much
$Q_{\bar{m}}$ differs from a function which is (up to normalization) $Q$
within $\Omega_{\bar{m}}$ and zero outside. One may think of dividing all
quantum states and their $Q$-distributions into two extreme classes: The ones
which show narrow pronounced regions of size comparable to individual coherent
states and the ones which change smoothly over regions larger or comparable to
the slot size. The former can be highly disturbed but in an extremely rare
fraction of all measurements. The latter is disturbed in general in a single
measurement but to very small extent, as the weight---in terms of the
$Q$-distribution---on the slot borders ($\propto\!\!\sqrt{j}$) is small
compared to the weight well inside the slot ($\propto\!\Delta m$). (In the
intermediate cases one has a trade-off between these two scenarios.) The
typical fraction of these weights is $\!\sqrt{j}/\Delta m\ll1$. In any case,
classicality arises with overwhelming statistical weight. In the next chapter
we will introduce measurements with smooth borders (POVM) and therefore reduce
the disturbance dramatically, even for states near a slot border.

Finally, we want to point out explicitly: The angular resolution which is
necessary to see the quantumness, i.e.\ the superposition character, of a
given quantum state is of the order of $1/\!\sqrt{j}$. In other words, it is
necessary to be able to distinguish at least of the order of $\!\sqrt{j}$
different measurement outcomes. For a macroscopic object, $j\sim10^{20}$, it
would be necessary to resolve $\sim\!10^{10}$ different measurement outcomes.
If this precision cannot be met, macrorealism emerges out of quantum physics
for the rotation Hamiltonian.

\section{An alternative derivation}

For the sake of completeness we now present an alternative way to derive that
classicality emerges under coarse-grained measurements. Again, we consider the
totally mixed state (\ref{eq mixed state}) and the time evolution
(\ref{eq time evolution}). We remind that this allows to violate the
Leggett-Garg inequality if sharp measurements can be performed. Now we are
interested in the probability for obtaining the results $m_{1}$ at time
$t_{1}$ and $m_{2}$ at $t_{2}$ in measurements of the spin operator's
$z$-component $\hat{J}_{z}$---in analogy to~\cite{Pere1995}, where a
generalized singlet state and correlations in space are considered. This
probability can be written as%
\begin{equation}
p(m_{1},t_{1};m_{2},t_{2})=p(m_{1},t_{1})\,p(m_{2},t_{2})_{m_{1},t_{1}}\,,
\end{equation}
i.e.\ as the probability that $m_{1}$ is obtained at $t_{1}$ times the
probability that $m_{2}$ is the result at $t_{2}$ given $m_{1}$ at $t_{1}$.
Let $\hat{P}_{m}\equiv\left\vert m\right\rangle \!\left\langle m\right\vert $
denote the projector onto $\left\vert m\right\rangle $. Using that $\hat{\rho
}(t_{1})=\hat{\rho}(0)$, we have $p(m_{1},t_{1})=\;$Tr$[\hat{\rho}%
(t_{1})\,\hat{P}_{m_{1}}]=\tfrac{1}{2j+1}$, reflecting the fact that all of
the $2j+1$ eigenstates are equally probable in a maximally mixed state. If
$m_{1}$ was obtained at $t_{1}$, the state is reduced to $\hat{\rho}_{m_{1}%
}(t_{1})=\hat{P}_{m_{1}}=\left\vert m_{1}\right\rangle \!\left\langle
m_{1}\right\vert $. Then, with $\Delta t\equiv t_{2}-t_{1}$ and $\theta
\equiv\omega\,\Delta t$, we have $p(m_{2},t_{2})_{m_{1},t_{1}}=\;$Tr$[\hat
{U}_{t_{2}-t_{1}}\hat{P}_{m_{1}}\,\hat{U}_{t_{2}-t_{1}}^{\dagger}\hat
{P}_{m_{2}}]=|\!\left\langle m_{2}\right\vert $e$^{-\text{i}\theta\hat{J}_{x}%
}\left\vert m_{1}\right\rangle \!|^{2}$. Therefore,%
\begin{equation}
p(m_{1},t_{1};m_{2},t_{2})=\dfrac{1}{2j+1}\,|\!\left\langle m_{2}\right\vert
\text{e}^{-\text{i}\theta\hat{J}_{x}}\left\vert m_{1}\right\rangle \!|^{2}\,.
\end{equation}
To continue we perform a discrete Fourier transform with \textquotedblleft
frequencies\textquotedblright\ $\xi$ and $\eta$:%
\begin{align}
\tilde{p}(\xi,\eta)  &  =%
{\displaystyle\sum\nolimits_{m_{1},m_{2}}}
\text{e}^{\text{i}(\xi m_{1}+\eta m_{2})}\,p(m_{1},t_{1};m_{2},t_{2}%
)\nonumber\\
&  =\dfrac{\text{Tr}[\text{e}^{\text{i}\eta\hat{J}_{z}}\,\text{e}%
^{-\text{i}\theta\hat{J}_{x}}\,\text{e}^{\text{i}\xi\hat{J}_{z}}%
\,\text{e}^{\text{i}\theta\hat{J}_{x}}]}{2j+1}=\frac{\text{Tr}[\text{e}%
^{-\text{i}\,\bm{\kappa}\cdot\mathbf{\hat{J}}}]}{2j+1}=\dfrac{\sin
[(2j+1)\,\tfrac{\kappa}{2}]}{(2j+1)\sin\tfrac{\kappa}{2}}\,,
\label{eq quantum corr}%
\end{align}
where the four rotations were expressed as a single rotation e$^{-\text{i}%
\,\bm{\kappa}\cdot\mathbf{\hat{J}}}$ around a vector $\bm{\kappa}$ about an
angle of the size $\kappa=|\bm{\kappa}|$ and the trace was evaluated in the
basis where $\bm{\kappa}\cdot\mathbf{\hat{J}}$ is diagonal. Following firmly
Reference~\cite{Pere1995}, it is enough to treat the rotation matrices as
$2\times2$ as if $j=\tfrac{1}{2}$, since $j$ does not affect the geometrical
meaning. If one equates the product of the four matrices in
(\ref{eq quantum corr}) to e$^{-\text{i}\,\bm{\kappa
}\cdot\hat{\bm{\sigma}}/2}$ (with $\hat{\bm{\sigma}}$ the vector of Pauli
matrices), then one obtains the dependence of $\kappa$ on $\xi$, $\eta$ and
$\theta$:%
\begin{equation}
\cos\tfrac{\kappa}{2}=\cos\tfrac{\xi}{2}\,\cos\tfrac{\eta}{2}-\sin\tfrac{\xi
}{2}\,\sin\tfrac{\eta}{2}\,\cos\theta\,. \label{eq cos kappa}%
\end{equation}
Due to \textit{noise or coarse-graining of measurements}, respectively, the
high frequency components ($\xi,\eta\sim1$) are not experimentally observable.
Because of $\xi,\eta\sim\tfrac{1}{\Delta m}$ high frequency components
correspond to a sharp measurement resolution $\Delta m$. The low frequency
limit ($\xi,\eta\ll1$) of the expression for $\kappa$ in the first non-trivial
order reads%
\begin{equation}
\kappa^{2}\approx\xi^{2}+\eta^{2}+2\,\xi\,\eta\cos\theta\,.
\label{eq low frequency}%
\end{equation}

Now consider a \textit{classical spin vector} $\mathbf{J}$ at time $t=0$ with
length $J=|\mathbf{J}|=\sqrt{j(j+1)}\approx j+\tfrac{1}{2}$, where $j\gg1$. In
analogy to the quantum case, the probability for measuring the classical spin
vector's $z$-component as $m_{1}$ at $t_{1}$ and $m_{2}$ at $t_{2}$ is%
\begin{equation}
p_{\text{cl},\mathbf{J}}(m_{1},t_{1};m_{2},t_{2})=p_{\text{cl},\mathbf{J}%
}(m_{1},t_{1})\,p_{\text{cl},\mathbf{J}}(m_{2},t_{2})_{m_{1},t_{1}}.
\label{eq pcl}%
\end{equation}
The first probability is given by $p_{\text{cl},\mathbf{J}}(m_{1}%
,t_{1})=\delta(m_{1}-\mathbf{e}_{z}\cdot\mathbf{J}(t_{1}))$, where
$\mathbf{e}_{z}$ is the unit vector in $z$-direction and $\delta$ is Dirac's
delta function. The classical Newtonian time evolution of the spin components
$J_{i}$ ($i=x,y,z$) is given by the Poisson brackets (\ref{eq Poisson}), where
the (classical) Hamilton function is given by (\ref{eq Hamilton}). As
$\mathbf{J}^{2}=j\,(j+1)$ is a constant of motion, only the term
$\omega\,J_{x}$ governs the time evolution. As we already know, the solutions
correspond to a rotation around the $x$-axis. Thus, the spin at $t_{1}$ reads
$\mathbf{J}(t_{1})=\hat{R}_{x}(\omega\,t_{1})\,\mathbf{J}(0)$, with $\hat
{R}_{x}(\gamma)$ the $3\times3$ rotation matrix about the $x$-axis by an angle
$\gamma$. But for the scalar product $\mathbf{e}_{z}\cdot\mathbf{J}(t_{1})$ in
$p_{\text{cl},\mathbf{J}}(m_{1},t_{1})$ it does not matter whether we rotate
$\mathbf{J}(0)$ around $x$ by $\omega\,t_{1}$ or $\mathbf{e}_{z}$ around $x$
by $-\omega\,t_{1}$. Hence, we can write%
\begin{equation}
p_{\text{cl},\mathbf{J}}(m_{1},t_{1})=\delta(m_{1}-\bm{\alpha}\cdot
\mathbf{J}(0))\,,
\end{equation}
where $\bm{\alpha}=\hat{R}_{x}(-\omega\,t_{1})\,\mathbf{e}_{z}$ is the unit
vector with polar angle $\omega\,t_{1}$ and azimuthal angle $-\tfrac{\pi}{2}$.
The remaining probability is%
\begin{equation}
p_{\text{cl},\mathbf{J}}(m_{2},t_{2})_{m_{1},t_{1}}=\delta(m_{2}%
-\bm{\beta }\cdot\mathbf{J}(0))\,,
\end{equation}
i.e.\ it depends only on the initial spin $\mathbf{J}(0)$ and the elapsed time
$t_{2}$ but not on anything that happened at $t_{1}$ because of the
\textit{non-invasiveness of classical measurements}. In other words, there is
no reduction at $t_{1}$ and the condition \textquotedblleft$m_{1}$ was the
outcome at $t_{1}$\textquotedblright\ is unnecessary. Here, $\bm{\beta}=\hat
{R}_{x}(-\omega\,t_{2})\,\mathbf{e}_{z}$ is the unit vector with the polar
angle $\omega\,t_{2}$ and azimuthal angle $-\tfrac{\pi}{2}$. Hence,
eq.\ (\ref{eq pcl}) becomes%
\begin{equation}
p_{\text{cl},\mathbf{J}}(m_{1},t_{1};m_{2},t_{2})=\delta(m_{1}%
-\bm{\alpha }\cdot\mathbf{J})\,\delta(m_{2}-\bm{\beta}\cdot\mathbf{J})
\end{equation}
with $\mathbf{J}\equiv\mathbf{J}(0)$. The Fourier transform reads%
\begin{align}
\tilde{p}_{\text{cl},\mathbf{J}}(\xi,\eta)  &  =\iint\text{e}^{\text{i}(\xi
m_{1}+\eta m_{2})}p_{\text{cl},\mathbf{J}}(m_{1},t_{1};m_{2},t_{2}%
)\,\text{d}m_{1}\text{d}m_{2}\nonumber\\
&  =\text{e}^{\text{i}(\xi\,\bm{\alpha}\cdot\mathbf{J}+\eta\,\bm{\beta}\cdot
\mathbf{J})}\equiv\text{e}^{\text{i}\,\mathbf{k}\cdot\mathbf{J}}%
\end{align}
with $\mathbf{k}\equiv\xi\,\bm{\alpha}+\eta\,\bm{\beta}$. This implies%
\begin{equation}
k^{2}=\mathbf{k}^{2}=\xi^{2}+\eta^{2}+2\,\xi\,\eta\cos\theta\,, \label{eq k2}%
\end{equation}
where $\theta\equiv\omega\,(t_{2}-t_{1})$ is the angle between $\bm{\alpha}$
and $\bm{\beta}$, in total agreement with the low frequency limit
(\ref{eq low frequency}). The classical correlation $\tilde{p}_{\text{cl}%
,\mathbf{J}}(\xi,\eta)$ must be averaged over all possible initial directions
of $\mathbf{J}=\mathbf{J}(\Omega)$. Mimicking the mixed quantum state, the
distribution is isotropic and thus%
\begin{equation}
\tilde{p}_{\text{cl}}(\xi,\eta)=\dfrac{1}{4\pi}\,%
{\displaystyle\iint}
\,\tilde{p}_{\text{cl},\mathbf{J}}(\xi,\eta)\,\text{d}^{2}\Omega=\dfrac
{\sin[(2j+1)\,\tfrac{k}{2}]}{(2j+1)\,\tfrac{k}{2}}\,,
\label{eq classical corr}%
\end{equation}
with d$^{2}\Omega$ the infinitesimal solid angle element, the vector
$\mathbf{k}$ in $\tilde{p}_{\text{cl},\mathbf{J}}(\xi,\eta)$ as the polar
integration axis and $J=j+\frac{1}{2}$. This is the limiting value of the
quantum correlation (\ref{eq quantum corr}). We note that both the quantum and
the classical correlation for measurements of a rotating spin in time,
eqs.\ (\ref{eq quantum corr}) and (\ref{eq classical corr}), have a similar
form as in the case of a generalized singlet state and measurements in
space~\cite{Pere1995}.

\textit{The quantum to classical transition}. Let us now evaluate under which
circumstances the quantum correlation (\ref{eq quantum corr}) becomes
classical (\ref{eq classical corr}), i.e.%
\begin{equation}
\tilde{p}\rightarrow\tilde{p}_{\text{cl}}\,.
\end{equation}
First, we note that the allowed frequencies $\xi$ and $\eta$ are independent
and thus have to be small in the same order, i.e.\ $\xi\sim\eta$, whereas
higher frequencies are cut off. In the following, for the sake of a short and
intuitive notation, we consider all quantities as positive and ignore any
factors of the order of 1 whenever we write $\xi$. Comparing the exact
expression for $\kappa^{2}$, obtained from eq.\ (\ref{eq cos kappa}), with the
low frequency classical limit $k^{2}$, eq.\ (\ref{eq k2}), the leading order
of the error is $\xi^{4}$:%
\begin{equation}
\kappa^{2}=k^{2}+\xi^{4}\,,
\end{equation}
Since $k\sim\xi$ due to eq.\ (\ref{eq k2}), we have%
\begin{equation}
\kappa=\sqrt{k^{2}+\xi^{4}}=k+\xi^{3}\,. \label{eq kappa}%
\end{equation}
Thus, $\kappa\rightarrow k$ is fulfilled if and only if $\xi^{3}\ll k$, or
equivalently, $\xi^{2}\ll1$. Two formal conditions---(i) for the arguments in
numerators and (ii) for the denominators in eqs.\ (\ref{eq quantum corr}) and
(\ref{eq classical corr}), respectively---have to be met to guarantee
$\tilde{p}\rightarrow\tilde{p}_{\text{cl}}$:%
\begin{align}
(2j+1)\,\tfrac{\kappa}{2}  &  \rightarrow(2j+1)\,\tfrac{k}{2}
\label{eq cond 1}\\
(2j+1)\,\sin\tfrac{\kappa}{2}  &  \rightarrow(2j+1)\,\tfrac{k}{2}
\label{eq cond 2}%
\end{align}
Using eq.\ (\ref{eq kappa}), the left-hand side of condition (\ref{eq cond 1})
becomes%
\begin{equation}
(2j+1)\,\tfrac{\kappa}{2}=(2j+1)\,\tfrac{k}{2}+j\,\xi^{3}\,. \label{eq approx}%
\end{equation}
As $j$ is very large in the macroscopic limit, we have neglected the term
$\xi^{3}$ compared to $j\,\xi^{3}$, which is the leading order of the error.
Eq.\ (\ref{eq approx}) is a good approximation for $(2j+1)\,\tfrac{k}{2}$,
i.e.\ the right-hand side of (\ref{eq cond 1}), if and only if the error
$j\,\xi^{3}$ is much smaller than the smallest term in $(2j+1)\,\tfrac{k}{2}$,
which is $\tfrac{k}{2}\sim\xi$. Hence we have to postulate $j\,\xi^{3}\ll\xi$,
or equivalently,%
\begin{equation}
\xi\ll\frac{1}{\sqrt{j}}\,, \label{eq condition xi}%
\end{equation}
and the same condition has to hold for the frequency $\eta$. The evaluation of
condition (\ref{eq cond 2}), using a Taylor expansion of the sine function,
leads to the same condition (\ref{eq condition xi}). This high-frequency
cut-off, ensuring $\tilde{p}\rightarrow\tilde{p}_{\text{cl}}$, implies that
different values of $m$ can be experimentally distinguished only if their
separation $\Delta m$ fulfills%
\begin{equation}
\Delta m\gg\sqrt{j}\,,
\end{equation}
which is the minimum quantum uncertainty for spin coherent states and in
agreement with the previous sections.

\begin{quote}
\textit{If the angular resolution of the instruments is much poorer than the
intrinsic quantum uncertainty, they cannot detect the quantum features of the
spin system, let alone a violation of the Leggett-Garg inequality. The
temporal correlations become classical.}
\end{quote}

\chapter{General conditions for quantum violation of macroscopic realism}

\textbf{Summary:}\bigskip

We first show that a \textit{violation of the Leggett-Garg inequality itself
is possible for arbitrary Hamiltonians} given the ability to distinguish
consecutive eigenstates in sharp quantum measurements. This is understandable
because it is generally accepted that \textquotedblleft microscopically
distinct states\textquotedblright\ do not have objective existence. For
testing macrorealism one needs to apply the Leggett-Garg definition referring
to \textit{macroscopically distinct states}. In our every-day life, to
experience macrorealism it is usually sufficient to employ apparatus
decoherence (where the system is isolated\ and only after a premeasurement,
i.e.\ coupling of system and apparatus, the environment interacts irreversibly
with the apparatus) or the restriction of coarse-grained measurements.

Both mechanisms usually allow to describe the time evolution of any quantum
spin state by a classical time evolution of a statistical mixture. However, we
demonstrate that there exist \textit{\textquotedblleft
non-classical\textquotedblright\ Hamiltonians} for which the time evolution of
this mixture cannot be understood classically. Despite the fact that apparatus
decoherence or coarse-graining allow to describe the state merely by a
classical mixture at every instance of time, a non-classical Hamiltonian
builds up superpositions of macroscopically distinct states and allows to
violate macrorealism. We find the necessary condition for these non-classical
evolutions and illustrate it with the example of an oscillating Schrödinger
cat-like state. System decoherence, i.e.\ the constant monitoring of the
system by an environment, leads to macrorealism but a continuous
spatiotemporal description of non-classical time evolutions in terms of
classical laws of motion remains impossible.

In the last part we argue that non-classical Hamiltonians require interactions
between a large number of particles or are computationally much more complex
than classical Hamiltonians, which might be the reason why they are unlikely
to appear in nature.\bigskip

\noindent This chapter mainly bases on and also uses parts of
References~\cite{Kofl2007a,Kofl2008}:

\begin{itemize}
\item J. Kofler and \v{C}. Brukner\newline\textit{A coarse-grained Schrödinger
cat}\newline In:~\textit{Quantum Communication and Security}, ed.~M.
\.{Z}ukowski, S. Kilin, and J. Kowalik (IOS Press 2007).

\item J. Kofler and \v{C}. Brukner\newline\textit{The conditions for quantum
violation of macroscopic realism}\newline Phys.~Rev.~Lett.~(accepted);
arXiv:0706.0668 [quant-ph].\newpage
\end{itemize}

\section{Violation of the Leggett-Garg inequality for arbitrary Hamiltonians}

In contrast to the Leggett-Garg inequality in the previous chapter,
ineq.\ (\ref{eq Leggett}), with four possible measurement times $t_{i}$, we
will now only use three. Then, any macrorealistic theory predicts a
Leggett-Garg inequality of the Wigner type~\cite{Wign1970}, where
$C_{ij}\equiv\langle A_{i}\,A_{j}\rangle$ denotes the temporal correlation of
a dichotomic quantity $A$ at times $t_{i}$ and $t_{j}$:%
\begin{equation}
K\equiv C_{12}+C_{23}-C_{13}\leq1\,. \label{eq Herbert}%
\end{equation}

\begin{figure}[t]
\begin{center}
\includegraphics{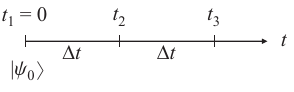}
\end{center}
\par
\vspace{-0.5cm}\caption{At time $t_{1}=0$ the initial state of the quantum
system is $|\psi_{0}\rangle$. Possible later measurement times are
$t_{2}=\Delta t$ and $t_{3}=2\Delta t$, where the system can be asked whether
it is still in its initial state or in the orthogonal subspace.}%
\label{Figure_Leggett_Garg_Wigner}%
\end{figure}

We extend the approach of Peres in Reference~\cite{Pere1995} and look at the
\textquotedblleft survival probability\textquotedblright\ of the system's
initial state at time $t=0$.\footnote{In Reference~\cite{Pere1995}, exercise
12.23 correctly claims that any non-trivial Hamiltonian is incompatible with a
Leggett-Garg inequality. But the given reason is inconclusive (and no proof is
presented). The survival probability, to which is referred to, is written only
in form of an inequality, ineq.~(12.136) in~\cite{Pere1995}. This constraint
is not strong enough and does not exclude time evolutions which would not
violate any Leggett-Garg inequality.} This state be denoted as $|\psi
(0)\rangle\equiv|\psi_{0}\rangle$ (which must not be an energy eigenstate)
and, without measurements, it evolves to%
\begin{equation}
|\psi(t)\rangle=\text{e}^{-\text{i}\hat{H}t}\,|\psi_{0}\rangle
\end{equation}
according to the Schrödinger equation in units where the reduced Planck
constant is $\hbar=1$. Our dichotomic observable is%
\begin{equation}
\hat{A}\equiv2\,|\psi_{0}\rangle\langle\psi_{0}%
|-\leavevmode\hbox{\small1\kern-3.3pt\normalsize1}\,,
\end{equation}
i.e.~we ask whether the system is (still) in the state $|\psi_{0}\rangle$
(outcome $+\equiv+1$) or not (outcome $-\equiv-1$). The temporal correlations
$C_{ij}$, eq.\ (\ref{eq C}), can be written as
\begin{equation}
C_{ij}=p_{i+}\,q_{j+|i+}+p_{i-}\,q_{j-|i-}-p_{i+}\,q_{j-|i+}-p_{i-}%
\,q_{j+|i-}\,, \label{eq Cij}%
\end{equation}
where $p_{i+}$ ($p_{i-}$) is the probability for measuring $+$ ($-$) at
$t_{i}$ and $q_{jl|ik}$ is the probability for measuring $l$ at $t_{j}$ given
that $k$ was measured at $t_{i}$ ($k,l=+,-$). For simplicity we choose
$t_{1}=0$ and equidistant possible measurement times with $\Delta t\equiv
t_{2}-t_{1}=t_{3}-t_{2}$ (Figure~\ref{Figure_Leggett_Garg_Wigner}). Then the
correlation $C_{12}$ is given by $C_{12}=2p(\Delta t)-1$, where%
\begin{equation}
p(t)\equiv|\langle\psi_{0}|\psi(t)\rangle|^{2}%
\end{equation}
is the (survival) probability to find $|\psi_{0}\rangle$ given the state
$|\psi(t)\rangle$. Analogously, we find $C_{13}=2p(2\Delta t)-1$. As
$p_{2+}=1-p_{2-}=p(\Delta t)$, $q_{3+|2+}=1-q_{3-|2+}=p(\Delta t)$, and
$q_{3-|2-}=1-q_{3+|2-}$, the only remaining unknown quantity in $C_{23}$ is
$q_{3+|2-}$. For its computation one needs the reduced state at $t_{2}$, given
that the outcome was $-$, $|\psi_{-}(t_{2})\rangle$. It is obtained by
applying the projector $\leavevmode\hbox{\small1\kern-3.3pt\normalsize1}-|\psi
_{0}\rangle\langle\psi_{0}|$ to the state at $t_{2}$, $|\psi(t_{2})\rangle$,
and normalizing: $|\psi_{-}(t_{2})\rangle=[|\psi(t_{2})\rangle\!-\!\langle
\psi_{0}|\psi(t_{2})\rangle\,|\psi_{0}\rangle]/\!\sqrt{1\!-\!p(\Delta t)}$.
This evolves a time $\Delta t$ to $t_{3}$, resulting in $|\psi_{-}%
(t_{3})\rangle=[|\psi(t_{3})\rangle\!-\!\langle\psi_{0}|\psi(t_{2}%
)\rangle\,|\psi(t_{2})\rangle]/\!\sqrt{1\!-\!p(\Delta t)}$, and $q_{3+|2-}%
=|\langle\psi_{0}|\psi_{-}(t_{3})\rangle|^{2}$ is the probability for the
outcome $+$ in that state. Plugging everything into (\ref{eq Herbert}), one
ends up with the Leggett-Garg inequality%
\begin{equation}
\fbox{$\;\;K=4\,p(\Delta t)\sqrt{p(2\Delta t)}\,\cos\gamma-4\,p(2\Delta
t)+1\leq1\,,\;\;$} \label{eq Herbert P}%
\end{equation}
where $\gamma\equiv2\alpha-\beta$ and $\alpha$ and $\beta$ are the phases in
$\langle\psi_{0}|\psi(t_{2})\rangle=\!\sqrt{p(\Delta t)}\exp($i$\alpha)$ and
$\langle\psi_{0}|\psi(t_{3})\rangle=\!\sqrt{p(2\Delta t)}\exp($i$\beta)$.

Now, independent of the system's dimension, it is sufficient to
consider as initial state a superposition of only two energy eigenstates
$|u_{1}\rangle$ and $|u_{2}\rangle$ with energy eigenvalues $E_{1}$ and
$E_{2}$, respectively: $|\psi_{0}\rangle=(|u_{1}\rangle\!+\!|u_{2}%
\rangle)/\!\sqrt{2}$. Ineq.\ (\ref{eq Herbert P}) becomes%
\begin{equation}
K=2\cos(\Delta E\Delta t)-\cos(2\Delta E\Delta t)\leq1\,,
\label{eq Leggett sup}%
\end{equation}
with $\Delta E\equiv E_{2}-E_{1}$ the energy difference of the two levels.

\begin{quote}
\textit{Any non-trivial Hamiltonian leads to a violation of this inequality.}
\end{quote}

\begin{figure}[t]
\begin{center}
\includegraphics{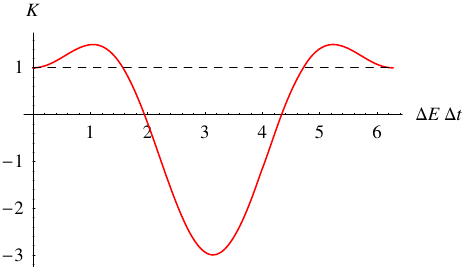}
\end{center}
\par
\vspace{-0.25cm}\caption{Violation of the Wigner-type Leggett-Garg inequality
(\ref{eq Leggett sup}) for arbitrary Hamiltonian evolutions. Every non-trivial
time evolution is in conflict with a classical description as long as one can
project onto individual quantum states of the system.}%
\label{Figure_Violation_H}%
\end{figure}

The left-hand side reaches $K=1.5$ for $\Delta t=\frac{\pi}{3\Delta E}$ and
$\Delta t=\frac{5\pi}{3\Delta E}$ and in $\frac{2\pi\hbar}{\Delta E}$ periods
thereof (Figure~\ref{Figure_Violation_H}).

\section{Macrorealism per se}

If every Hamiltonian leads to a violation of the Leggett-Garg inequality, why
then do we not see this in every-day life? The possible answers \textit{within
quantum theory} are:

\begin{enumerate}
\item It is due to decoherence, i.e.\ the quantum system interacts with an
uncontrollable environment such that it is driven into a statistical
mixture~\cite{Zure1991,Zure2003}.\vspace{-0.2cm}

\item The fact that the resolution of our measurement apparatuses is not
sharp, makes it impossible to project onto individual states and to see the
above demonstrated violation that is always present for
microstates~\cite{Kofl2007b}.
\end{enumerate}

For testing macrorealism---i.e.\ testing the Leggett-Garg inequality under the
restriction of \textit{coarse-grained measurements}---we consider again a
spin-$j$ system (with $j\!\gg\!1$) as a model example. To keep this chapter
self-contained, we briefly repeat that any spin-$j$ state can be written in
the quasi-diagonal form%
\begin{equation}
\hat{\rho}=%
{\displaystyle\iint}
P(\Omega)\,|\Omega\rangle\langle\Omega|\,\text{d}^{2}\Omega
\end{equation}
with d$^{2}\Omega$ the solid angle element and $P$ a normalized and
\textit{not necessarily positive} real function~\cite{Arec1972}. The\textit{
}spin coherent states\textit{ }$|\Omega\rangle\equiv|\vartheta,\varphi\rangle
$, with $\vartheta$ and $\varphi$ the polar and azimuthal angle, are the
eigenstates with maximal eigenvalue of a spin operator pointing into the
direction $\Omega\equiv(\vartheta,\varphi)$~\cite{Radc1971,Atki1971}:
$\hat{\mathbf{J}}_{\Omega}\left\vert \Omega\right\rangle =j\left\vert
\Omega\right\rangle $ in units where $\hbar=1$.

In coarse-grained measurements our resolution is not able to
resolve individual eigenvalues $m$ of a spin component, say the $z$-component
$\hat{J}_{z}$, but bunches together $\Delta m$ \textit{neighboring} outcomes
into \textquotedblleft slots\textquotedblright\ $\bar{m}$, where the
measurement coarseness is much larger than the intrinsic quantum uncertainty
of coherent states, i.e.\ $\Delta m\!\gg\!\sqrt{j}$.

The question arises whether it is problematic to use coarse-grained
(projective) \textit{von Neumann measurements} of the form
\begin{equation}%
{\displaystyle\sum\nolimits_{m\in\{\bar{m}\}}}
\!\left\vert m\right\rangle \!\left\langle m\right\vert ,
\end{equation}
where $\left\vert m\right\rangle $ are the $\hat{J}_{z}$ eigenstates, as
\textquotedblleft classical measurements\textquotedblright\ as we did in the
previous chapter. In contrast to the \textit{positive operator value measure}
(POVM), they have sharp edges and could violate the Leggett-Garg inequality by
distinguishing with certainty between microstates at two sides of a slot
border. E.g., if $\left\vert m\right\rangle $ and $\left\vert
m\!+\!1\right\rangle $ belong to two different slots, macrorealism could be
violated by a simple microscopic time evolution, producing $\cos(\omega
t)\left\vert m\right\rangle +\sin(\omega t)\left\vert m\!+\!1\right\rangle $.
This is the reason why we model our coarse-grained $\hat{J}_{z}$ measurements
as belonging to a (spin coherent state) POVM, where the element corresponding
to the outcome $\bar{m}$ is represented by%
\begin{equation}
\fbox{$\;\;\hat{P}_{\bar{m}}\equiv\dfrac{2j+1}{4\pi}\,%
{\displaystyle\iint\nolimits_{\Omega_{\bar{m}}}}
|\Omega\rangle\langle\Omega|\,$d$^{2}\Omega\,.\;\;$} \label{eq P mbar}%
\end{equation}
Here, $\Omega_{\bar{m}}$ is the angular region of polar angular size
$\Delta\Theta_{\bar{m}}\sim\Delta m/j\gg1/\!\sqrt{j}$ whose projection onto
the $z$-axis corresponds to the slot $\bar{m}$ (Figure~\ref{Figure_Sphere}).
As the $\Omega_{\bar{m}}$ are mutually disjoint and form a partition of the
whole angular region, we have%
\begin{equation}%
{\displaystyle\sum\nolimits_{\bar{m}}}
\hat{P}_{\bar{m}}=\leavevmode\hbox{\small1\kern-3.3pt\normalsize1}\,.
\end{equation}
Due to overcompleteness of the spin coherent states the POVM elements are
overlapping at the slot borders over the angular size $\sim\!1/\!\sqrt{j}$
which is small compared to the angular slot size $\Delta\Theta_{\bar{m}}$.

To find out how $\hat{P}_{\bar{m}}$ looks like in the basis of $\hat{J}_{z}$
eigenstates, we apply $\hat{P}_{\bar{m}}$ to an arbitrary state $\left\vert
k\right\rangle $ and insert the identity operator $%
{\textstyle\sum\nolimits_{k^{\prime}}}
|k^{\prime}\rangle\langle k^{\prime}%
|=\leavevmode\hbox{\small1\kern-3.3pt\normalsize1}$:%
\begin{equation}
\hat{P}_{\bar{m}}\left\vert k\right\rangle =%
{\displaystyle\sum\nolimits_{k^{\prime}}}
\left(  \dfrac{2j+1}{4\pi}\,%
{\displaystyle\iint\nolimits_{\Omega_{\bar{m}}}}
\langle k^{\prime}|\Omega\rangle\langle\Omega|k\rangle\,\text{d}^{2}%
\Omega\right)  |k^{\prime}\rangle\,.
\end{equation}
The scalar products in the above equation are of the form%
\begin{equation}
\langle m|\Omega\rangle=\left(  \!%
\genfrac{}{}{0pt}{1}{2j}{j+m}%
\!\right)  ^{\!1/2}\cos^{j+m}\!\tfrac{\vartheta}{2}\sin^{j-m}\!\tfrac
{\vartheta}{2}\,\text{e}^{-\text{i}m\varphi},
\end{equation}
where $\Omega\equiv(\vartheta,\varphi)$. The azimuthal integration $%
{\textstyle\int\nolimits_{0}^{2\pi}}
\exp[$i$(k\!-\!k^{\prime})\varphi]\,$d$\varphi$ gives $2\pi\delta_{kk^{\prime
}}$. The remaining integrand with $k=k^{\prime}$ is the modulus square
$|\langle k|\Omega\rangle|^{2}$ and independent of $\varphi$, but we can just
substitute the $2\pi$ factor from the performed $\varphi$-integration by a
trivial unperformed $\varphi$-integration to keep the simple writing with an
integration over $\Omega_{\bar{m}}$. Thus,%
\begin{equation}
\hat{P}_{\bar{m}}=%
{\displaystyle\sum\nolimits_{k}}
\left(  \dfrac{2j+1}{4\pi}%
{\displaystyle\iint\nolimits_{\Omega_{\bar{m}}}}
|\langle k|\Omega\rangle|^{2}\,\text{d}^{2}\Omega\right)  |k\rangle\langle k|
\end{equation}
is \textit{diagonal} in the basis of $\hat{J}_{z}$ eigenstates.

The probability for getting the outcome $\bar{m}$ is given by%
\begin{equation}
w_{\bar{m}}=\text{Tr}[\hat{\rho}\hat{P}_{\bar{m}}]=\dfrac{2j+1}{4\pi}\,%
{\displaystyle\iint}
\,\langle\Omega|\hat{\rho}\hat{P}_{\bar{m}}|\Omega\rangle\,\text{d}^{2}%
\Omega\,.
\end{equation}
Inserting $\hat{P}_{\bar{m}}$ from eq.\ (\ref{eq P mbar}), we obtain%
\begin{equation}
w_{\bar{m}}=\left(  \dfrac{2j+1}{4\pi}\right)  ^{\!2}%
{\displaystyle\iint}
\text{d}^{2}\Omega\,%
{\displaystyle\iint\nolimits_{\Omega_{\bar{m}}}}
\text{d}^{2}\Omega^{\prime}\,\langle\Omega^{\prime}|\Omega\rangle\langle
\Omega|\hat{\rho}|\Omega^{\prime}\rangle\,.
\end{equation}
Using that $\tfrac{2j+1}{4\pi}%
{\textstyle\iint}
$d$^{2}\Omega\,|\Omega\rangle\langle\Omega|$ is the identity operator and
renaming $\Omega^{\prime}$ to $\Omega$, we get%
\begin{equation}
\fbox{$\;\;w_{\bar{m}}=%
{\displaystyle\iint\nolimits_{\Omega_{\bar{m}}}}
Q(\Omega)\,$d$^{2}\Omega\,.\;\;$}%
\end{equation}
This is \textit{exactly} the integration of the \textit{positive probability
distribution}%
\begin{equation}
Q(\Omega)\equiv\dfrac{2j+1}{4\pi}\,\langle\Omega|\hat{\rho}|\Omega
\rangle\label{eq Q}%
\end{equation}
associated to the quantum state $\hat{\rho}$ (the well-know $Q$%
-function~\cite{Agar1981}) over the region $\Omega_{\bar{m}}$.

\begin{quote}
\textit{Given an arbitrary quantum state, under coarse-grained measurements
the outcome probabilities at any time can exactly be computed from an ensemble
of classical spins (i.e.\ there exists a hidden variable model). This is
macrorealism per se.}
\end{quote}

\section{Non-invasive measurability}

Upon a coarse-grained measurement with outcome $\bar{m}$, the state $\hat
{\rho}$ is reduced to%
\begin{equation}
\hat{\rho}_{\bar{m}}=\frac{\hat{M}_{\bar{m}}\,\hat{\rho}\,\hat{M}_{\bar{m}}%
}{w_{\bar{m}}}\,,
\end{equation}
where we have chosen a particular (optimal) implementation of the POVM with
the Hermitean Kraus operators%
\begin{equation}
\fbox{$\;\;\hat{M}_{\bar{m}}=\hat{M}_{\bar{m}}^{\dag}=%
{\displaystyle\sum\nolimits_{k}}
\,\sqrt{\dfrac{2j+1}{4\pi}%
{\displaystyle\iint\nolimits_{\Omega_{\bar{m}}}}
|\langle k|\Omega\rangle|^{2}\,\text{d}^{2}\Omega}\,\,|k\rangle\langle
k|\,.\;\;$}%
\end{equation}
It can be easily seen that they satisfy the necessary relation%
\[
\hat{M}_{\bar{m}}^{2}=\hat{P}_{\bar{m}}\,.
\]
We note that, independent of the concrete implementation, the $\hat{P}%
_{\bar{m}}$ (and the Kraus operators) behave almost as projectors for all
states $|\Omega\rangle$ except for those near a slot border:%
\begin{equation}
\hat{P}_{\bar{m}}\,|\Omega\rangle\approx\left\{
\begin{array}
[c]{ll}%
|\Omega\rangle & \text{for }\Omega\text{ inside }\Omega_{\bar{m}}\text{,}\\
\mathbf{0} & \text{for }\Omega\text{ outside }\Omega_{\bar{m}}\text{.}%
\end{array}
\right.
\end{equation}
In a proper classical limit ($\!\sqrt{j}/\Delta m\rightarrow0$) the relative
weight of these border $\Omega$ becomes \textit{vanishingly small}.

We now show that the $Q$-distribution before the measurement is the (weighted)
\textit{mixture} of the $Q$-distributions%
\begin{equation}
Q_{\bar{m}}(\Omega)=\dfrac{2j+1}{4\pi}\,\langle\Omega|\hat{\rho}_{\bar{m}%
}|\Omega\rangle
\end{equation}
of the possible reduced states $\hat{\rho}_{\bar{m}}$:%
\begin{equation}
\fbox{$\;\;Q(\Omega)\approx%
{\displaystyle\sum\nolimits_{\bar{m}}}
w_{\bar{m}}\,Q_{\bar{m}}(\Omega)\,.\;\;$} \label{eq Q1}%
\end{equation}

\begin{quote}
\textit{This demonstrates that a fuzzy measurement can be understood
classically as reducing the previous ignorance about predetermined properties
of the spin system.}
\end{quote}

The approximate sign \textquotedblleft$\approx$\textquotedblright\ reflects
that, depending on the density matrix%
\begin{equation}
\hat{\rho}\equiv%
{\displaystyle\sum\nolimits_{n}}
\,%
{\displaystyle\sum\nolimits_{n^{\prime}}}
c_{nn^{\prime}}|n\rangle\langle n^{\prime}|\,, \label{eq density matrix}%
\end{equation}
this relationship may only approximately hold for the set of those
$\Omega\equiv(\vartheta,\varphi)$ near a slot border. In detail
eq.~(\ref{eq Q1}) reads%
\begin{equation}
\tfrac{2j+1}{4\pi}\,%
{\displaystyle\sum\nolimits_{n}}
\,%
{\displaystyle\sum\nolimits_{n^{\prime}}}
c_{nn^{\prime}}\langle\Omega|n\rangle\langle n^{\prime}|\Omega\rangle
\approx\tfrac{2j+1}{4\pi}\,%
{\displaystyle\sum\nolimits_{n}}
\,%
{\displaystyle\sum\nolimits_{n^{\prime}}}
c_{nn^{\prime}}%
{\displaystyle\sum\nolimits_{\bar{m}}}
\!\sqrt{g_{\bar{m}}(n)\,g_{\bar{m}}(n^{\prime})}\,\langle\Omega|n\rangle
\langle n^{\prime}|\Omega\rangle
\end{equation}
with%
\begin{equation}
g_{\bar{m}}(k)\equiv\dfrac{2j+1}{4\pi}\,%
{\displaystyle\iint\nolimits_{\Omega_{\bar{m}}}}
|\langle k|\Omega\rangle|^{2}\,\text{d}^{2}\Omega
\end{equation}
which is smaller or equal to 1. The two sides are exactly equal if $\hat{\rho
}$ is diagonal as $%
{\textstyle\sum\nolimits_{\bar{m}}}
\,g_{\bar{m}}(n)=1$for all $n$. Deviations only occur if $n$, $n^{\prime}$ and
$j\cos\vartheta$ are all within a distance of order $\!\sqrt{j}$ to each other
and to a slot border. If they are not close to each other, the quantity
$\langle\Omega|n\rangle\langle n^{\prime}|\Omega\rangle$ is exponentially
small and suppression by the factor $%
{\textstyle\sum\nolimits_{\bar{m}}}
\!\sqrt{g_{\bar{m}}(n)\,g_{\bar{m}}(n^{\prime})}$ is not important. If $n$,
$n^{\prime}$ are well within a slot, $%
{\textstyle\sum\nolimits_{\bar{m}}}
\!\sqrt{g_{\bar{m}}(n)\,g_{\bar{m}}(n^{\prime})}$ is almost identical to $1$.

\begin{figure}[t]
\begin{center}
\includegraphics[width=.75\textwidth]{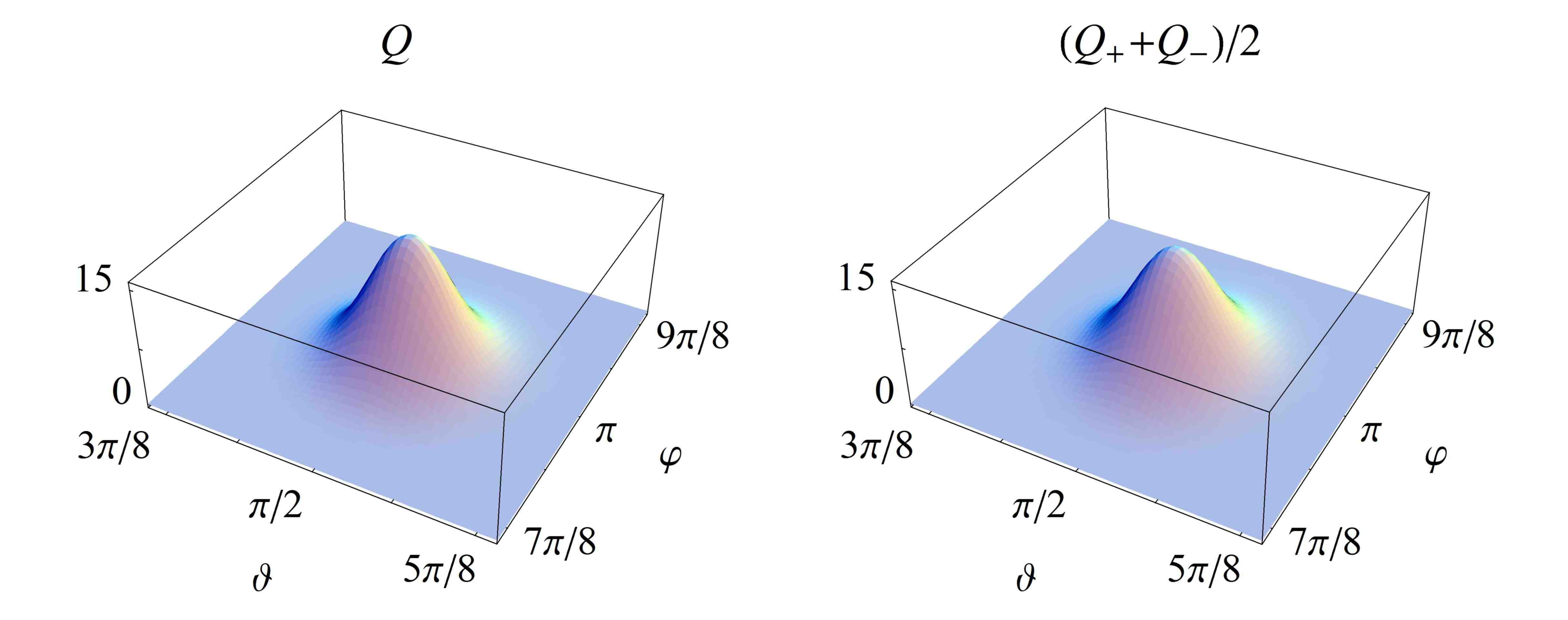}
\end{center}
\par
\vspace{-0.25cm}\caption{Left: The $Q$-distribution of the spin coherent state
$|\frac{\pi}{2},\pi\rangle$ located at the equator. Right: The mixture
$\tfrac{1}{2}(Q_{+}+Q_{-})$ of the $Q$-distributions of the two possible
reduced states in a which-hemisphere measurement. Although this is a
worst-case scenario in the sense that the quantum state is located exactly at
the slot border, the two distributions are very similar. The pictures are
drawn for the spin size $j=100$ but are scale invariant.}%
\label{Figure_Q_on_equator}%
\end{figure}

Consider the worst-case scenario of a \textquotedblleft which-hemisphere
measurement\textquotedblright\ with only two slots $\bar{m}=+$ (northern
hemisphere $\Omega_{+}$) and $\bar{m}=-$ (southern hemisphere $\Omega_{-}$)
and an initial spin coherent state%
\begin{equation}
\hat{\rho}_{\text{eq}}\equiv|\tfrac{\pi}{2},\pi\rangle\langle\tfrac{\pi}%
{2},\pi|
\end{equation}
located exactly at the equator, i.e.\ the slot border. Both outcomes happen
with the same probability $w_{+}=w_{-}=\frac{1}{2}$.
Figure~\ref{Figure_Q_on_equator} illustrates the unmeasured state's
$Q$-distribution, $Q(\Omega)=\tfrac{2j+1}{4\pi}\,\langle\Omega|\hat{\rho
}_{\text{eq}}|\Omega\rangle$, and the mixture $\tfrac{1}{2}\,[Q_{+}%
(\Omega)+Q_{-}(\Omega)]$ of the $Q$-distributions%
\begin{equation}
Q_{\pm}(\Omega)=\dfrac{2j+1}{4\pi}\,\dfrac{\langle\Omega|\hat{M}_{\pm}%
\,\hat{\rho}_{\text{eq}}\,\hat{M}_{\pm}|\Omega\rangle}{w_{\pm}}%
\end{equation}
of the reduced states, i.e.\ the left and right-hand side of eq.\ (\ref{eq Q1}%
), respectively. Independent of the spin size $j$ the $Q$-distributions with
and without measurement have a very large overlap,%
\begin{equation}%
{\displaystyle\iint}
\sqrt{\tfrac{1}{2}\,[Q_{+}(\Omega)\!+\!Q_{-}(\Omega)]\,Q(\Omega)}%
\,\text{d}^{2}\Omega\approx0.997\,,
\end{equation}
indicating the good quality of eq.~(\ref{eq Q1}). Here, the overlap between
two normalized probability distributions $f$ and $g$ is defined by $%
{\textstyle\iint}
\sqrt{f(\Omega)\,g(\Omega)}\,$d$^{2}\Omega\in\lbrack0,1]$.

\begin{figure}[t]
\begin{center}
\includegraphics{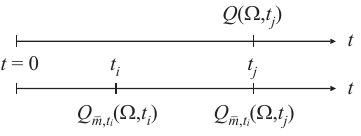}
\end{center}
\par
\vspace{-0.33cm}\caption{Non-invasive measurability is fulfilled if the
$Q$-distribution at some time $t_{j}$, without any measurement before, is the
same as the weighted mixture of the $Q$-distributions stemming from the
possible reduced states from a measurement at some earlier time $t_{i}$.}%
\label{Figure_Non-invasiveness}%
\end{figure}

However, for non-invasiveness we need more than eq.\ (\ref{eq Q1}). Consider
the initial distribution of classical spins, $Q(\Omega,t_{0})$, corresponding
to an initial quantum state $\hat{\rho}(t_{0})$. We first compute the
$Q$-distribution of the state $\hat{\rho}(t_{j})$ for an undisturbed evolution
without measurement until some time $t_{j}$,%
\begin{equation}
Q(\Omega,t_{j})=\dfrac{2j+1}{4\pi}\,\langle\Omega|\hat{\rho}(t_{j}%
)|\Omega\rangle\,.
\end{equation}
This has to be compared with the \textit{mixture} of all possible reduced
distributions upon measurement at a time $t_{i}$ ($t_{0}\leq t_{i}<t_{j}$)
with outcomes $\bar{m}$ which evolved to $t_{j}$, denoted as%
\begin{equation}
Q_{\bar{m},t_{i}}(\Omega,t_{j})=\dfrac{2j+1}{4\pi}\,\dfrac{\langle\Omega
|\hat{U}_{t_{j}-t_{i}}\hat{M}_{\bar{m}}\,\hat{\rho}(t_{i})\,\hat{M}_{\bar{m}%
}\hat{U}_{t_{j}-t_{i}}^{\dag}|\Omega\rangle}{w_{\bar{m},t_{i}}}\,,
\end{equation}
with $w_{\bar{m},t_{i}}\equiv\;$Tr$[\hat{\rho}(t_{i})\hat{P}_{\bar{m}}]$ the
probability for outcome $\bar{m}$ at time $t_{i}$ and $\hat{U}_{t}\equiv
\exp(-$i$\hat{H}t)$ the time evolution operator (see
Figure~\ref{Figure_Non-invasiveness}). The system evolves macrorealistically
if these two quantities coincide for all $t_{i}$ and $t_{j}$, i.e.\ if%
\begin{equation}
\fbox{$\;\;Q(\Omega,t_{j})\approx%
{\displaystyle\sum\nolimits_{\bar{m}}}
w_{\bar{m},t_{i}}\,Q_{\bar{m},t_{i}}(\Omega,t_{j})\,.\;\;$} \label{eq Q cond}%
\end{equation}

\begin{quote}
\textit{This is the condition for non-invasive measurability (together with
induction).}
\end{quote}

In a dichotomic scenario, the outcomes $+$ and $-$ correspond to finding the
spin system in one out of two slots $\bar{m}=\pm$. This is represented by a
measurement of two complementary regions $\Omega_{+}$ and $\Omega_{-}$ (for
instance the northern and southern hemisphere in a \textquotedblleft which
hemisphere measurement\textquotedblright). Then, e.g., the probability for
measuring $-$ at $t_{3}$ if $+$ was measured at $t_{1}$ is given by%
\begin{equation}
q_{3-|1+}=%
{\displaystyle\iint\nolimits_{\Omega_{-}}}
Q_{+,t_{1}}(\Omega,t_{3})\,\text{d}^{2}\Omega
\end{equation}
with $Q_{+,t_{1}}(\Omega,t_{3})$ the $Q$-distribution of the state which was
reduced at $t_{1}$ with outcome $+$ and evolved to $t_{3}$. If
condition~(\ref{eq Q cond}) is satisfied, it implies that the probabilities
can be decomposed into \textquotedblleft classical paths\textquotedblright.
This means that, e.g., $q_{3-|1+}$ is just the sum of the two possible paths
via $+$ and $-$ at $t_{2}$:%
\begin{equation}
q_{3-|1+}=q_{2+|1+}\,q_{3-|2+,1+}+q_{2-|1+}\,q_{3-|2-,1+}\,,
\end{equation}
where $q_{3-|2+,1+}$ ($q_{3-|2-,1+}$) denotes the probability to measure $-$
at $t_{3}$ given that $+$ was measured at $t_{1}$ and $+$ ($-$) at $t_{2}$.
Thus, eq.~(\ref{eq Q cond}) allows to derive Leggett-Garg inequalities such as
ineq.~(\ref{eq Herbert}).

\section{The sufficient condition for macrorealism}

We can now establish the sufficient condition for macrorealism that holds even
for isolated systems, namely%
\begin{equation}
\fbox{$\;\;\hat{P}_{\bar{m}}\,\hat{U}_{t}\,|\Omega\rangle\approx\left\{
\begin{array}
[c]{ll}%
\hat{U}_{t}\,|\Omega\rangle & \text{for one }\bar{m},\\
\mathbf{0} & \text{for all the others,}%
\end{array}
\right.  $} \label{eq U cond}%
\end{equation}
for all $t$ and $\Omega$, allowing deviations at slot borders.

\begin{quote}
\textit{This means that the time evolution does not produce superpositions of
macroscopically distinct states.}
\end{quote}

Vice versa, if $\hat{U}_{t}\,|\Omega\rangle$ produced states of the form
$\alpha\,|\Omega^{\prime}\rangle+\beta\,|\Omega^{\prime\prime}\rangle$ where
neither $|\alpha|$ nor $|\beta|$ is close to zero and with $|\Omega^{\prime
}\rangle$ and $|\Omega^{\prime\prime}\rangle$ belonging to macroscopically
different outcomes (different slots), eq.~(\ref{eq U cond}) would not be
fulfilled. Eq.~(\ref{eq U cond}) implies that $\hat{P}_{\bar{m}}$, and hence
$\hat{M}_{\bar{m}}$, quasi behave as projectors and that%
\begin{equation}
\langle\Omega|\hat{U}_{t_{j}-t_{i}}\,\hat{\rho}(t_{i})\,\hat{U}_{t_{j}-t_{i}%
}^{\dag}|\Omega\rangle\approx%
{\displaystyle\sum\nolimits_{\bar{m}}}
\langle\Omega|\hat{U}_{t_{j}-t_{i}}\hat{M}_{\bar{m}}\,\hat{\rho}(t_{i}%
)\,\hat{M}_{\bar{m}}\hat{U}_{t_{j}-t_{i}}^{\dag}|\Omega\rangle
\end{equation}
This directly leads to eq.~(\ref{eq Q cond}). Thus, eq.~(\ref{eq U cond})
$\rightarrow$ eq.~(\ref{eq Q cond}) $\rightarrow$ macrorealism.

\section{An oscillating Schrödinger cat}

We denote those Hamiltonians for which eq.~(\ref{eq U cond}) is satisfied
under coarse-grained measurements as \textit{classical}. An example is the
rotation, say around $x$, $\hat{H}_{\text{rot}}=\omega\hat{J}_{x}$, with
$\hat{J}_{x}$ the spin $x$-component and $\omega$ the angular precession
frequency, which satisfies eq.~(\ref{eq U cond}) and moreover allows a
Newtonian description of the time evolution as shown in the previous chapter.
But there is no \textit{a priori} reason why all Hamiltonians should satisfy
eq.~(\ref{eq U cond}).

\begin{quote}
\textit{Can one find non-classical Hamiltonians violating macrorealism despite
coarse-grained measurements?}
\end{quote}

The necessary condition for such a situation is that the Hamiltonian builds up
coherences between states belonging to different slots. One explicit (extreme)
example is%
\begin{equation}
\fbox{$\;\;\hat{H}=\;$i$\,\omega\left(  \left\vert -j\right\rangle
\!\left\langle +j\right\vert -\left\vert +j\right\rangle \!\left\langle
-j\right\vert \right)  ,\;\;$} \label{eq Schroe}%
\end{equation}
which, given the special initial state $|\Psi(0)\rangle=\left\vert
+j\right\rangle $, produces a time-dependent Schrödinger cat-like
superposition of two \textit{distant} (orthogonal) spin-$j$ coherent states
$\left\vert +j\right\rangle $ (`north') and $\left\vert -j\right\rangle $
(`south'):%
\begin{equation}
|\Psi(t)\rangle=\cos(\omega t)\left\vert +j\right\rangle +\sin(\omega
t)\left\vert -j\right\rangle . \label{eq psi}%
\end{equation}
Under fuzzy measurements or apparatus decoherence after a
premeasurement~\cite{Zure2003}, the state~(\ref{eq psi}) appears like a
statistical mixture at every instance of time:%
\begin{equation}
\hat{\rho}_{\text{mix}}(t)=\cos^{2}(\omega t)\left\vert +j\right\rangle
\!\left\langle +j\right\vert +\sin^{2}(\omega t)\left\vert -j\right\rangle
\!\left\langle -j\right\vert . \label{eq mix}%
\end{equation}
While the two states $\hat{\rho}_{\text{sup}}(t)\equiv|\Psi(t)\rangle
\langle\Psi(t)|$ and $\hat{\rho}_{\text{mix}}(t)$ have dramatically different
$P$-functions and can be distinguished by sharp measurements, they are de
facto equivalent on the coarse-grained level. The $Q$-distributions,
$Q_{\text{sup}}$ for $\hat{\rho}_{\text{sup}}(t)$ and $Q_{\text{mix}}$ for
$\hat{\rho}_{\text{mix}}(t)$, are given by eq.~(\ref{eq Q}). The coherence
terms stemming from $\hat{\rho}_{\text{sup}}(t)$ are of the form
$\langle\Omega\left\vert +j\right\rangle \!\left\langle -j\right\vert
\Omega\rangle$ and vanish exponentially fast with the spin length $j$ for all
$\Omega$. For $j\gg1$ the $Q$-distributions are practically identical,
i.e.\ $Q\equiv Q_{\text{mix}}\approx Q_{\text{sup}}$:%
\begin{equation}
Q(\Omega,t)=\dfrac{2j+1}{4\pi}\left[  \cos^{2}(\omega t)\cos^{4j}%
(\tfrac{\Theta_{1}}{2})+\sin^{2}(\omega t)\cos^{4j}(\tfrac{\Theta_{2}}%
{2})\right]  ,
\end{equation}
where $\Theta_{1}=\vartheta$ ($\Theta_{2}=\pi-\vartheta$) is the angle between
$\Omega\equiv(\vartheta,\varphi)$ and $+z$ ($-z$).

In general the $P$-function reads~\cite{Agar1981,Agar1993}%
\begin{equation}
P(\Omega,t)=\sum_{k=0}^{2j}\,\sum_{q=-k}^{k}\,\rho_{kq}\,Y_{kq}(\Omega
,t)\,\dfrac{(-1)^{k-q}\,\sqrt{(2j-k)!\,(2j+k+1)!}}{\sqrt{4\pi}\,(2j)!}\,.
\end{equation}
Here, $Y_{kq}$ are the spherical harmonics and%
\begin{equation}
\rho_{kq}(t)=\sqrt{2k+1}\,\,%
{\displaystyle\sum\nolimits_{m}}
(-1)^{j-m}\,c_{m,m-q}(t)\left(
\genfrac{}{}{0pt}{1}{j}{-m+q}%
\genfrac{}{}{0pt}{1}{k}{-q}%
\genfrac{}{}{0pt}{1}{j}{m}%
\right)  \!,
\end{equation}
where the last bracket denotes the Wigner 3$j$ symbol. The $c_{nn^{\prime}%
}(t)$ are the coefficients in the representation (\ref{eq density matrix}) of
the density matrix one is interested in.\begin{figure}[t]
\begin{center}
\includegraphics[width=.75\textwidth]{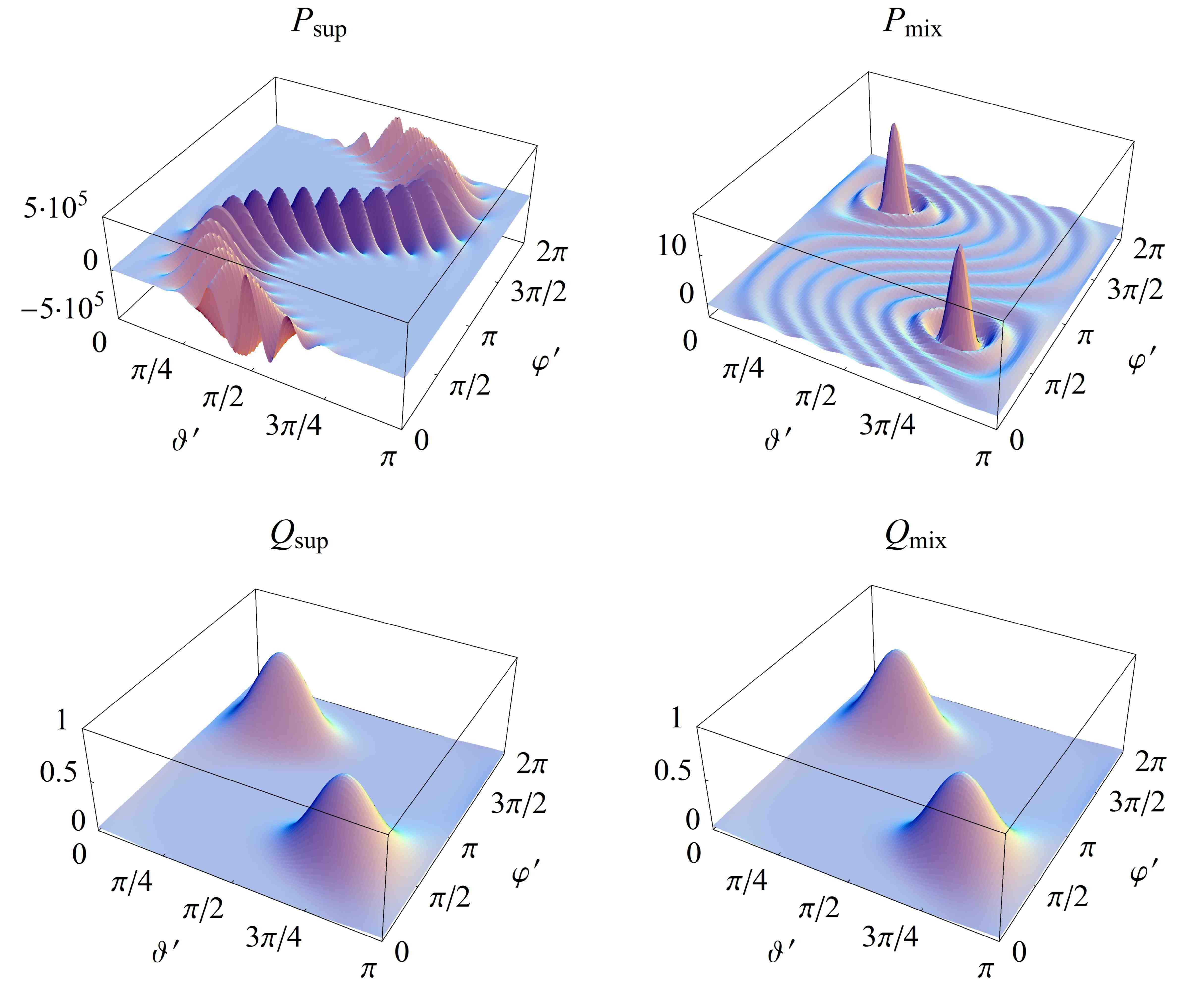}
\end{center}
\par
\vspace{-0.25cm}\caption{Top left: The $P$-function $P_{\text{sup}}$ at time
$t=\frac{\pi}{4\omega}$ of the equal-weight superposition (\ref{eq psi}) of
two opposite spin coherent states $\left\vert +j\right\rangle $ and
$\left\vert -j\right\rangle $ for spin length $j=10$, plotted in a rotated
coordinate system in which $\left\vert +j\right\rangle =|\frac{\pi}{4}%
,\frac{3\pi}{2}\rangle$. It is wildly oscillating with very large positive and
negative regions. Top right: The $P$-function $P_{\text{mix}}$ of the
corresponding statistical mixture (\ref{eq mix}). Bottom: In every-day life
the angular measurement resolution is much weaker than $1/\!\sqrt{j}$ (which
is equivalent to $\Delta m\!\gg\!\sqrt{j}$\ in a $\hat{J}_{z}$ measurement).
Then we cannot distinguish anymore between the superposition state and the
classical mixture, as both effectively lead to the same (positive)
$Q$-distribution $Q_{\text{sup}}\approx Q_{\text{mix}}$. Nevertheless, the
time evolution of such a mixture can violate macrorealism even under classical
coarse-grained measurements.}%
\label{Figure_P_and_Q}%
\end{figure}

The $P$ and $Q$-functions of $\hat{\rho}_{\text{sup}}$ and $\hat{\rho
}_{\text{mix}}$ at $t=\frac{\pi}{4\omega}$ are shown in
Figure~\ref{Figure_P_and_Q} for a certain choice of parameters. The
$P$-function of the superposition is pathologically oscillating. The
$Q$-distributions show just two peaks, corresponding to a classical mixture in
which half of the spins are pointing into the north direction and the other
into the south. Going to larger and larger values of $j$, i.e. from
Schrödinger kittens to cats, makes it more and more difficult to observe the
quantum nature of superposition states like~(\ref{eq psi}). The angular
resolution which is necessary to distinguish a superposition from the
corresponding classical mixture is of the order of $1/\!\sqrt{j}$.

\begin{figure}[t]
\begin{center}
\includegraphics[width=.8\textwidth]{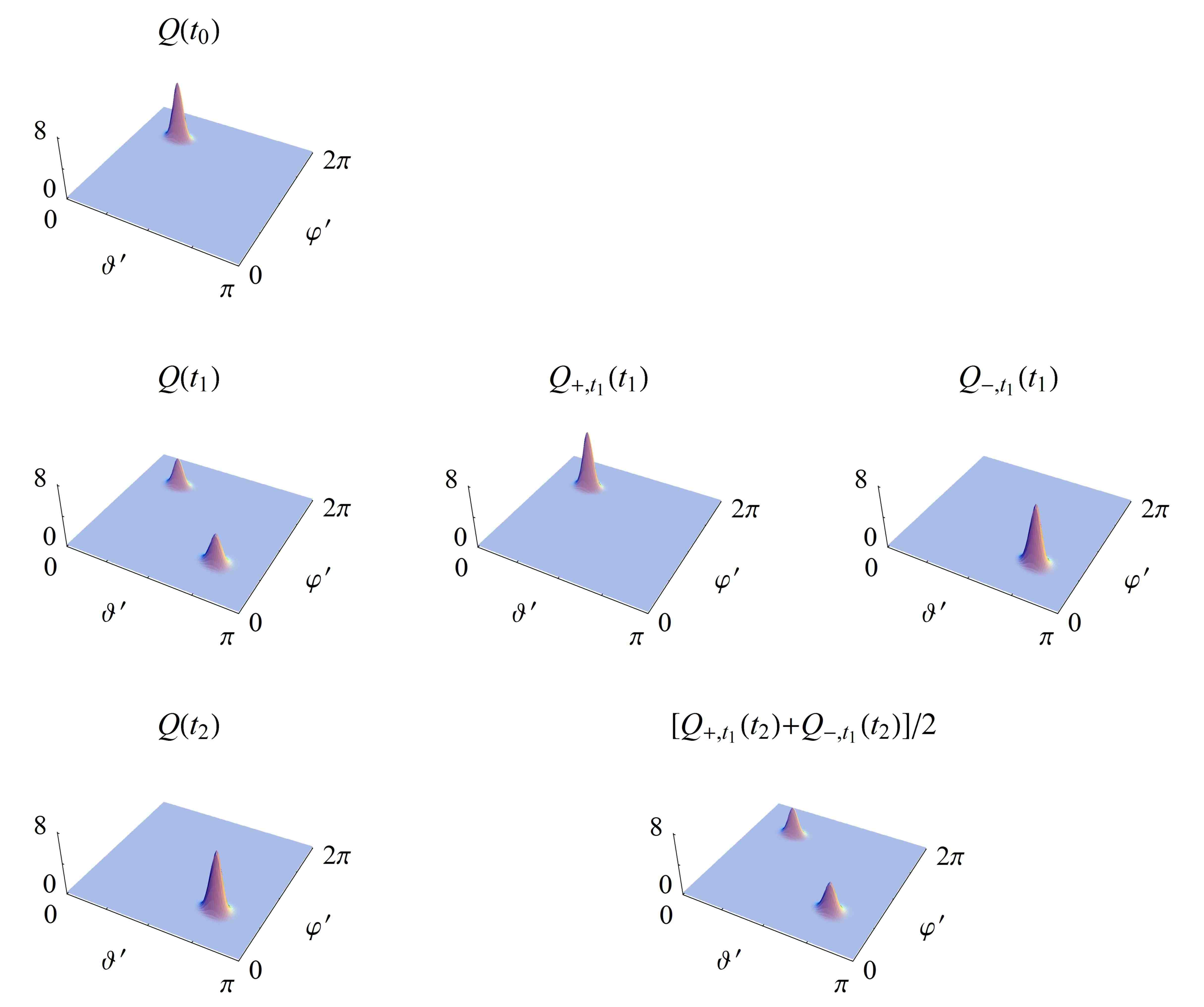}
\end{center}
\par
\vspace{-0.25cm}\caption{The time evolution of the $Q$-distribution for the
non-classical (oscillating Schrödinger cat) Hamiltonian (\ref{eq Schroe}),
shown for a spin $j=100$. At time $t_{0}=0$ the quantum state is $\left\vert
+j\right\rangle $ (`north') with the $Q$-distribution $Q(t_{0})$. At the later
time $t_{1}=\frac{\pi}{4\omega}$ the spin is in an equal-weight superposition
of $\left\vert +j\right\rangle $ and $\left\vert -j\right\rangle $ and the
$Q$-distribution $Q(t_{1})$ shows two peaks at north and south. If no
measurement takes place, the system reaches the state $\left\vert
-j\right\rangle $ (`south') at $t_{2}=\frac{\pi}{2\omega}$, represented by
$Q(t_{2})$. If, on the other hand, one performs a measurement at $t_{1}$, the
state is reduced to either $\left\vert +j\right\rangle $ or $\left\vert
-j\right\rangle $, with $Q$-distributions $Q_{+,t_{1}}(t_{1})$ and
$Q_{-,t_{1}}(t_{1})$, respectively. Either of these two states will evolve
into an equal-weight superposition until $t_{2}$. The weighted mixture at that
time, $[Q_{\pm,t_{1}}(t_{2})\!+\!Q_{\pm,t_{1}}(t_{2})]/2$ is different from
the undisturbed evolution. Eqs.~(\ref{eq U cond}) and (\ref{eq Q cond}) are
not fulfilled. Although at every instance of time the system can be described
by a classical mixture, the time evolution of this mixture allows to violate
the Leggett-Garg inequality and is in conflict with macrorealism.}%
\label{Figure_Q_in_time}%
\end{figure}

However, under the Hamiltonian (\ref{eq Schroe}) even simple dichotomic
which-hemisphere measurements are sufficient to violate macrorealism. The
temporal correlation function (for times $t_{i}$ and $t_{j}$) reads%
\begin{equation}
C_{ij}\approx\cos[\omega(t_{j}\!-\!t_{i})]
\end{equation}
with an exponentially small correction due to the tiny chance that, e.g.,
$\left\vert +j\right\rangle $ can be found in the southern hemisphere. The
system effectively behaves as a spin-$\tfrac{1}{2}$ particle which violates
the Leggett-Garg inequality. It can be easily seen that without any
measurement the initial state $\left\vert +j\right\rangle $ evolves to the
state $\left\vert -j\right\rangle $ at $t_{2}=\frac{\pi}{2\omega}$. If, on the
other hand, there is a measurement at $t_{1}=\frac{\pi}{4\omega}$ in between,
where we have an equal weight superposition of $\left\vert +j\right\rangle $
and $\left\vert -j\right\rangle $, the state is projected to either
$\left\vert +j\right\rangle $ or $\left\vert -j\right\rangle $. Either of
these reduced states will evolve to an equal weight superposition of
$\left\vert +j\right\rangle $ and $\left\vert -j\right\rangle $ at $t_{2}$.
Thus, eq.~(\ref{eq Q cond}) is not fulfilled for fuzzy measurements and the
evolution (\ref{eq Schroe}), as shown in Figure~\ref{Figure_Q_in_time}.

\begin{quote}
\textit{Despite the fact that apparatus decoherence or coarse-graining allow
to describe the state effectively by a classical mixture at every instance of
time, the non-classical Hamiltonian indeed builds up superpositions of
macroscopically distinct states and allows to violate macrorealism.}
\end{quote}

To get macrorealism one would have to coarse-grain always those states which
are \textit{connected by the Hamiltonian} (in time) and not necessarily in
real space. In the present case it is (at least) the outcomes $+j$ and $-j$
that have to be coarse-grained into one and the same slot, which is of course
highly counter-intuitive. Such a coarse-graining would lead to a different
kind of macrorealistic physics than the classical laws we know, bringing
systems through space and time continuously.

\section{Continuous monitoring by an environment}

Until now we have considered isolated systems. Under coarse-grained
measurements or, mathematically equivalent, apparatus decoherence (where the
system is isolated and only after a premeasurement of the system an
uncontrollable environment couples to the apparatus), classical Hamiltonians
lead to macrorealism whereas non-classical Hamiltonians allow to violate it.
The question arises:

\begin{quote}
\textit{What happens if the Hamiltonian is non-classical and the system is
continuously monitored by an environment}?
\end{quote}

\begin{figure}[t]
\begin{center}
\includegraphics{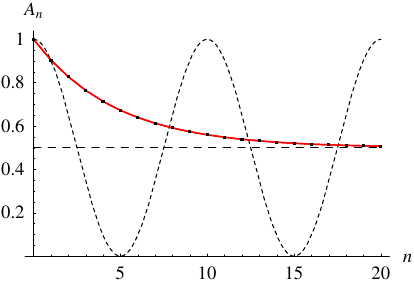}
\end{center}
\par
\vspace{-0.25cm}\caption{The survival probability $A_{n}$ (black squares in
red curve) in the decohered state $\hat{\rho}(n\Delta t)$, eq.~(\ref{eq rho n}%
), as a function of the number of steps, $n$. The free evolution time interval
between instances of rapid decoherence is chosen to be $\Delta t=\frac{\pi
}{10\omega}$, where $\omega$ is the angular frequency in the non-classical
Hamiltonian (\ref{eq Schroe}). The state is driven into an equal weight
mixture, eq.~(\ref{eq rho equal}), with $A_{\infty}=\frac{1}{2}$ (approaching
the dashed line asymptotically). For comparison, we also draw the function
$\cos^{2}(\omega t)=\cos^{2}(\omega n\Delta t)$ which is the probability to
find the state along north, i.e.\ $\left\vert +j\right\rangle $, at time $t$
if no environmental decoherence takes place (dashed curve).}%
\label{Figure_Decoherence}%
\end{figure}

Given the non-classical Hamiltonian (\ref{eq Schroe}) with the time evolution
operator%
\begin{align}
\hat{U}_{t}  &  =+\cos(\omega t)\left(  \left\vert +j\right\rangle
\!\left\langle +j\right\vert \!+\!\left\vert -j\right\rangle \!\left\langle
-j\right\vert \right) \nonumber\\
&  \;\;\;\;+\sin(\omega t)\left(  \left\vert -j\right\rangle \!\left\langle
+j\right\vert \!-\!\left\vert +j\right\rangle \!\left\langle -j\right\vert
\right)  +%
{\textstyle\sum\nolimits_{m=-j+1}^{j-1}}
\left\vert m\right\rangle \!\left\langle m\right\vert ,
\end{align}
let us approximate the effects of system decoherence by the following
simplified model: The initial state along north,%
\begin{equation}
\hat{\rho}(0)=\left\vert +j\right\rangle \!\left\langle +j\right\vert ,
\end{equation}
freely evolves without decoherence a short time $\Delta t$ to%
\begin{align}
\hat{\rho}(\Delta t)  &  =\hat{U}_{\Delta t}\,\hat{\rho}(0)\,\hat{U}_{\Delta
t}^{\dag}\nonumber\\
&  =\cos^{2}(\omega\Delta t)\left\vert +j\right\rangle \!\left\langle
+j\right\vert +\sin^{2}(\omega\Delta t)\left\vert -j\right\rangle
\!\left\langle -j\right\vert +\text{coh.\thinspace terms\thinspace,}
\label{eq rho Delta t}%
\end{align}
where the coherence terms are of the form $\left\vert +j\right\rangle
\!\left\langle -j\right\vert $ and $\left\vert -j\right\rangle \!\left\langle
+j\right\vert $. Now we assume that the macroscopic spin system decoheres very
rapidly (in the standard pointer basis of $\left\vert +j\right\rangle $ and
$\left\vert -j\right\rangle $), for instance due to the fact that a single
qubit from the environment couples to it in a c-not manner~\cite{Zure2003},
becomes inaccessible immediately afterwards, and does not interact (recohere)
with it anymore. If it is impossible to make (joint) measurements on the
environmental qubit (and our spin system), the partial trace over the qubit of
the total density matrix has to be performed, which kills the coherence terms
in eq.~(\ref{eq rho Delta t}), leading to the decohered state of the system:%
\begin{equation}
\hat{\rho}(\Delta t)=\cos^{2}(\omega\Delta t)\left\vert +j\right\rangle
\!\left\langle +j\right\vert +\sin^{2}(\omega\Delta t)\left\vert
-j\right\rangle \!\left\langle -j\right\vert .
\end{equation}
Assuming again free time evolution for a duration of $\Delta t$, this
decohered state will evolve to%
\begin{align}
\hat{\rho}(2\Delta t)  &  =\hat{U}_{\Delta t}\,\hat{\rho}(\Delta t)\,\hat
{U}_{\Delta t}^{\dag}\\
&  =[\cos^{4}(\omega\Delta t)\!+\!\sin^{4}(\omega\Delta t)]\left\vert
+j\right\rangle \!\left\langle +j\right\vert +2\cos^{2}(\omega\Delta
t)\sin^{2}(\omega\Delta t)\left\vert -j\right\rangle \!\left\langle
-j\right\vert +\text{coh.\thinspace terms\thinspace.}\nonumber
\end{align}
Repeating the alternating sequence of rapid decoherence and free time
evolution, we obtain the general expression for the (decohered) state at time
$n\Delta t$:%
\begin{equation}
\hat{\rho}(n\Delta t)=A_{n}\left\vert +j\right\rangle \!\left\langle
+j\right\vert +(1\!-\!A_{n})\left\vert -j\right\rangle \!\left\langle
-j\right\vert . \label{eq rho n}%
\end{equation}
The survival probability to find the state along north, $A_{n}$, can be
retrieved from the recurrence relation%
\begin{align}
A_{0}  &  =1\,,\\
A_{n}  &  =a\,A_{n-1}+(1\!-\!a)\,(1\!-\!A_{n-1})
\end{align}
with integer $n$ and%
\begin{equation}
a\equiv\cos^{2}(\omega\Delta t)\,.
\end{equation}
If $\Delta t$ is not too small (to avoid a quantum Zeno-like freezing of the
initial state~\cite{Misr1977}) but smaller than the dynamical timescale of the
Hamiltonian, $\tfrac{1}{\omega}$, the probability $A_{n}$ decays to
$A_{\infty}=\tfrac{1}{2}$ in a way which can be very well approximated by
\begin{equation}
A(t)=\tfrac{1}{2}\,(1-\text{e}^{-\nu t}) \label{eq exp decay}%
\end{equation}
with $\tfrac{1}{\nu}$ the characteristic decay time. For $t\gg\tfrac{1}{\nu}$
the state becomes an equal weight statistical mixture%
\begin{equation}
\hat{\rho}(\infty)=\frac{\left\vert +j\right\rangle \!\left\langle
+j\right\vert +\left\vert -j\right\rangle \!\left\langle -j\right\vert }{2}
\label{eq rho equal}%
\end{equation}
as illustrated in Figure~\ref{Figure_Decoherence}.

The conclusion of this simple and crude model of decoherence---namely that the
state is driven into a mixture where half of the spins point to north and half
of the spins point to south---is expected to remain valid under more realistic
circumstances where one does not separate into free evolution and rapid
decoherence. As long as the environmental microscopic degrees of freedom only
couple to the macroscopic spin system but do not disturb its diagonal
elements, the system does not leave the subspace spanned by $\left\vert
+j\right\rangle $ and $\left\vert -j\right\rangle $, and never populates any
of the other states $\left\vert m\right\rangle $.

Importantly, despite the non-classical Hamiltonian, the exponential decay of
$A_{n}$ due to decoherence does not allow to violate the Leggett-Garg
inequality any longer. If no (coarse-grained) measurement takes place, the
spin's $Q$-distribution at time $t_{j}$---i.e.\ the left-hand side of
eq.~(\ref{eq Q cond})---is given by%
\begin{equation}
Q(t_{j})=A(t_{j})\,Q_{\text{north}}+[1\!-\!A(t_{j})]\,Q_{\text{south}},
\end{equation}
where $Q_{\text{north}}$ ($Q_{\text{south}}$) is the $Q$-distribution of a
spin pointing to the north (south). If a measurement takes place at the
intermediate time $t_{i}$ $(0<t_{i}<t_{j})$,\footnote{Note:\ After finding the
spin along \textit{south} at $t_{i}$, which happens with probability
$1\!-\!A(t_{i})$, $A(t_{j}-t_{i})$ is the (survival) probability to find the
spin again along south at time $t_{j}$.} the weighted mixture of the reduced
and evolved $Q$-distributions---i.e.\ the right-hand side of
eq.~(\ref{eq Q cond})---takes on exactly the same form for all choices of
$t_{i}$ and $t_{j}$, given the exponential decay $A(t)=\tfrac{1}{2}%
(1-$e$^{-\nu t})$, eq.~(\ref{eq exp decay}), is used.\footnote{It is
interesting that any other form of the survival probability other than
exponential decay violates the non-invasiveness condition (\ref{eq Q cond}).}
Hence, the system's time evolution fulfills the condition (\ref{eq Q cond})
for non-invasive measurability, and consequently macrorealism is satisfied.

\begin{quote}
\textit{However, decoherence cannot account for a continuous spatiotemporal
description of the spin system in terms of classical laws of motion.}
\end{quote}

To see this, it is enough to use coarse-grained measurements corresponding to
only three different angular regions, one covering the northern part, one the
equatorial region, and one the southern part. The initial spin along north can
be found pointing to the south at some later time, \textit{although it did not
go through the equatorial region}. No classical Hamilton function can achieve
such discontinuous \textquotedblleft jumps\textquotedblright\ of a spin
vector. For claiming that classicality emerges from quantum physics, it is
simply not enough to show that the density matrix is driven into a diagonal
form due to tracing out the environmental degrees of freedom. One must also
demonstrate that the time evolution of this mixed state can be described in
terms of classical laws of motion. Classical physics is within the class of
macrorealistic theories but it is more restrictive than macrorealism itself.

The key point here is that in classical physics we have Newton's or Hamilton's
\textit{differential equations for observable quantities} such as spin
directions. Under all circumstances these equations evolve the observables
continuously through real space. In quantum mechanics, however, the situation
is very different. The Schrödinger equation evolves the state vector
continuously through Hilbert space but one cannot give a spatiotemporal
description of the system's observables independent of
observation.\begin{figure}[t]
\begin{center}
\includegraphics{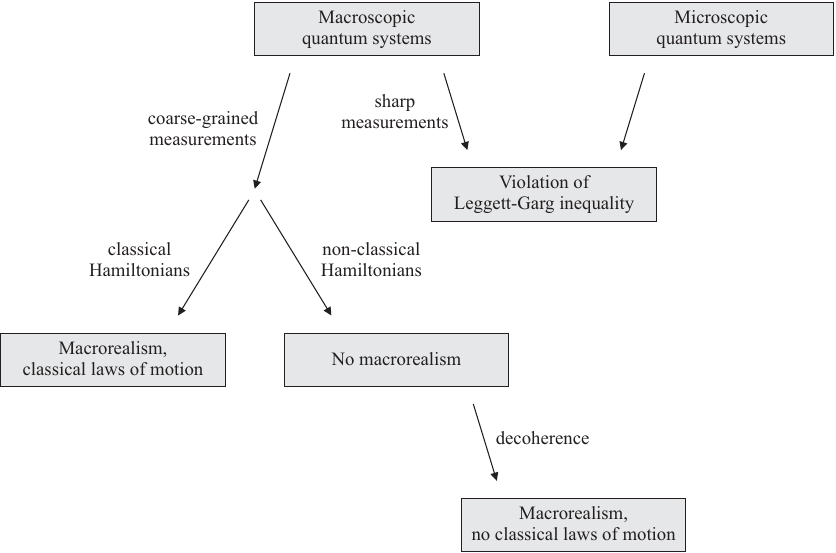}
\end{center}
\par
\vspace{-0.cm}\caption{Microscopic systems as well as macroscopic systems
under sharp measurements allow to violate the Leggett-Garg inequality. Since
no macroscopic \textquotedblleft classical\textquotedblright\ observables are
involved, one cannot speak about a violation of macrorealism in these cases.
Under coarse-grained measurements of a macroscopic system and classical
Hamiltonians not only macrorealism is valid but also classical laws of motion
emerge. Non-classical Hamiltonians allow to violate macrorealism even under
coarse-grained measurements. Decoherence then establishes macrorealism but
cannot account for a description of the system's time evolution in terms of
classical laws of motion.}%
\label{Figure_Branching_diagram}%
\end{figure}

Figure~\ref{Figure_Branching_diagram} gives an overview of the conclusions we
have reached until this point concerning the Leggett-Garg inequality,
macrorealism, and classical laws of motion with respect to sharp and
coarse-grained measurements as well as classical and non-classical Hamiltonians.

\section{Non-classical Hamiltonians are complex}

Finally, we suggest a possible reason why non-classical evolutions might be
unlikely to be realized by nature: Such evolutions (i) either require
Hamiltonians with many-particle interactions or (ii) a specific sequence of a
large number of computational steps if only few-particle interactions are
used. Then, they are of \textit{high computational complexity}. In the first
case the number of interacting particles, and in the second the number of
computational steps has to scale linearly with the size of the Schrödinger cat
state. Both cases intuitively seem to be of very low probability to happen spontaneously.

\begin{figure}[t]
\begin{center}
\includegraphics{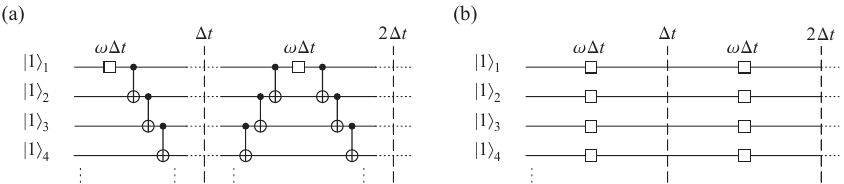}
\end{center}
\par
\vspace{-0.25cm}\caption{(a) In order to simulate the time evolution
(\ref{eq unitary2}) of a qubit chain one has to rotate the first qubit by a
small angle $\omega\Delta t$ and sequentially make c-nots. For the next time
interval $\Delta t$ one has to undo the c-nots, rotate the first qubit again
and make all the c-nots again. With this procedure one gets a sequence of
states which approximate (\ref{eq unitary2}). (b) In contrast, the simulation
of an interval $\Delta t$ of a spin rotation of the whole chain can be
achieved in a single global transformation on all qubits simultaneously.}%
\label{Figure_Simulation_circuit}%
\end{figure}

Consider our spin-$j$ as a macroscopic ensemble of $N=2j$ spin-$\tfrac{1}{2}$
particles (i.e.\ qubits) such as, e.g., any magnetic material is constituted
by many individual microscopic spins. For violating macrorealism it is
necessary to build up superpositions of two macroscopically distinct coherent
states. For large $j$ their angular separation $\Delta\theta$ can be very
small and only has to obey the coarse-graining condition $\Delta\theta
\!\gg1/\!\sqrt{j}$. This guarantees quasi-orthogonality as their (modulus
square) overlap is $\cos^{4j}(\Delta\theta/2)\sim\;$e$^{-j\,\Delta\theta^{2}}%
$. Without loss of generality we consider again the particular
Hamiltonian~(\ref{eq Schroe}). If $|0\rangle$ and $|1\rangle$ denote the
individual qubit states `up' and `down' along $z$, then $|11...1\rangle$ and
$|00...0\rangle$ form the total spin coherent states $\left\vert
+j\right\rangle $ and $\left\vert -j\right\rangle $, respectively. The
Hamiltonian represents $N$-particle interactions of the form
\begin{equation}
\hat{H}=\tfrac{\text{i\thinspace}\omega}{2}\,(\hat{\sigma}_{-}^{\otimes
N}\!-\!\hat{\sigma}_{+}^{\otimes N})\,,
\end{equation}
where $\hat{\sigma}_{\pm}\equiv\hat{\sigma}_{x}\pm\,$i$\,\hat{\sigma}_{y}$
with $\hat{\sigma}_{x}$ and $\hat{\sigma}_{y}$ the Pauli operators. As an
alternative one can simulate the evolution governed by this many-body
interaction by means of a series of (in nature typically appearing) few-qubit
interactions (gates), using the methods of quantum computation
science~\cite{Niel2000}. The task is to simulate%
\begin{equation}
|11...1\rangle\;\rightarrow\;\cos(\omega t)\,|11...1\rangle+\sin(\omega
t)\,|00...0\rangle\,. \label{eq unitary2}%
\end{equation}
Assuming next-neighbor qubit interactions, we start from the state
$|11...1\rangle$ and rotate the first qubit `1' by a small angle $\omega\Delta
t$: $|1\rangle_{1}\rightarrow\cos(\omega\Delta t)\,|1\rangle_{1}+\sin
(\omega\Delta t)\,|0\rangle_{1}$. Then we perform a controlled-not (c-not)
gate between this qubit `1' and its neighbor `2' such that $|x\rangle
_{1}|y\rangle_{2}\rightarrow|x\rangle_{1}|x\!\oplus\!y\rangle_{2}$
($x,y=0,1$). Afterwards c-nots between qubits are performed such that all
other qubits are reached. The whole procedure is depicted in
Figure~\ref{Figure_Simulation_circuit}(a). This procedure brings us to the
state at time $\Delta t$: $|11...1\rangle\rightarrow\cos(\omega\Delta
t)\,|11...1\rangle+\sin(\omega\Delta t)\,|00...0\rangle$. To simulate the next
time interval $\Delta t$, we have to undo all the c-nots, rotate the first
qubit again by $\omega\Delta t$, and make all the c-nots again, leading to the
correct state at time $2\Delta t$. With this procedure we get a sequence of
states, simulating the evolution (\ref{eq unitary2}). One needs $O(N)$
computational steps per interval $\Delta t$. This is known to be optimal in
the case where only neighboring qubits can interact~\cite{Brav2006}. Relaxing
this condition and permitting two-qubit interactions between all possible
qubits, allows to decrease the number of sequential steps but does not change
the total number $O(N)$ of necessary gates per interval.

Note for comparison, however, that the rotation (say around $x$)%
\begin{equation}
\hat{H}_{\text{rot}}=\tfrac{\omega}{2}\,%
{\displaystyle\sum\nolimits_{k=1}^{N}}
\hat{\sigma}_{x}^{(k)}\,,
\end{equation}
with $k$ labeling the qubits, does not require multi-particle interactions.
Moreover, the simulation of an interval $\Delta t$ of a spin rotation of the
whole chain, i.e.%
\begin{equation}
|111...\rangle\rightarrow\lbrack\cos(\omega\Delta t)\,|1\rangle+\sin
(\omega\Delta t)\,|0\rangle]^{\otimes N},
\end{equation}
can be achieved in a \textit{single global transformation} on all qubits
simultaneously, as shown in Figure~\ref{Figure_Simulation_circuit}(b).

\begin{quote}
\textit{While both evolutions are rotations in Hilbert space (and require only
polynomial resources), the simulation of the non-classical cosine-law between
states that are distant in real space is---for macroscopically large }%
$N$\textit{---computationally much more complex than the classical rotation in
real space.}\footnote{One should, however, mention the possibility that an
\textit{external} field may produce an effectively simple non-classical
Hamiltonian for \textit{N} qubits where the field interacts with the
collective modes $\left\vert +j\right\rangle =|11...1\rangle$ and $\left\vert
-j\right\rangle =|00...0\rangle$.}
\end{quote}

\section{Information and randomness}

In a perfect von Neumann measurement of a spin component with sharp resolution
an individual state $|m\rangle$ out of the $2j+1\approx2j$ possible ones
carries%
\begin{equation}
I_{\text{sharp}}\approx\log_{2}(2j)=1+\log_{2}j
\end{equation}
bits of information. Coarse-grained measurements correspond to the fact that
we cannot resolve individual eigenvalues $m$ but only whole bunches of size
$\Delta m\gg\!\sqrt{j}$. The finding that an outcome lies in a certain slot of
size $\Delta m=c\,\sqrt{j}$ (with $c\gg1$) carries only%
\begin{equation}
I_{\text{c.-g.}}\approx\log_{2}(\tfrac{2j}{c\sqrt{j}})=1-\log_{2}c+\tfrac
{1}{2}\log_{2}j
\end{equation}
bits of information. For large $j$, i.e.\ $j\gg c\gg1$, the information gain
in a sharp quantum measurement is approximately $\log_{2}j$ bits, whereas in
the classical case it is (at most) only half of that, namely $\tfrac{1}{2}%
\log_{2}j$ bits\ \cite{Kofl2007a}:%
\begin{equation}
I_{\text{c.-g.}}\approx\tfrac{1}{2}\,I_{\text{sharp}}\,.
\end{equation}

Finally, we note that---given coarse-grained measurements---it is objectively
random which of the two states, `north' or `south', one will find in a spin
measurement in the Schrödinger cat state~(\ref{eq psi}). Classical physics
emerges out of the quantum world but the randomness in the classical mixture
is still irreducible. Which possibility becomes factual is objectively random
and does not have a causal reason, according to the Copenhagen interpretation.

In the last chapter we will address the question of quantum randomness in more
detail and link it with mathematical undecidability.

\chapter{Entanglement between macroscopic observables}

\textbf{Summary:}\bigskip

We investigate entanglement between collective operators of two blocks of
oscillators in an infinite linear harmonic chain. These operators are defined
as averages over local operators (individual oscillators) in the blocks. On
the one hand, this approach of \textquotedblleft physical
blocks\textquotedblright\ meets realistic experimental conditions, where
measurement apparatuses do not interact with single oscillators but rather
with a whole bunch of them, i.e.\ where in contrast to usually studied
\textquotedblleft mathematical blocks\textquotedblright\ not every possible
measurement on them is allowed. On the other, this formalism naturally allows
the generalization to blocks which may consist of several non-contiguous
regions. We quantify entanglement between the collective operators by a
measure based on the Peres-Horodecki criterion and show how it can be
extracted and transferred to two qubits. Entanglement between two blocks is
found even in the case where none of the oscillators from one block is
entangled with an oscillator from the other, showing genuine bipartite
entanglement between collective operators. Allowing the blocks to consist of a
periodic sequence of subblocks, we verify that entanglement scales at most
with the total boundary region. We also apply the approach of collective
operators to scalar quantum field theory.

What can we learn about entanglement between individual particles in
macroscopic samples by observing only the collective properties of the
ensembles? Using only a few experimentally feasible collective properties, we
establish an entanglement measure between two samples of spin-$\tfrac{1}{2}$
particles (as representatives of two-dimensional quantum systems). This is a
tight lower bound for the average entanglement between all pairs of spins in
general and is equal to the average entanglement for a certain class of
systems. We compute the entanglement measures for explicit examples and show
how to generalize the method to more than two samples and multi-partite
entanglement. On the fundamental side, our method demonstrates that there is
no reason in principle why purely quantum correlations could not have an
effect on the global properties of objects. On the practical side, it enables
us to characterize the structure of entanglement in large spin systems by
performing only a few feasible measurements of their collective properties,
independently of the symmetry and mixedness of the state.

Since this analysis uses sharp measurements, it is not in disagreement with
our quantum-to-classical approach resting upon coarse-grained
measurements.\bigskip

\noindent This chapter mainly bases on and also uses parts of
References~\cite{Kofl2006b,Kofl2006c}:

\begin{itemize}
\item J. Kofler, V. Vedral, M. S. Kim, and \v{C}. Brukner\newline%
\textit{Entanglement between collective operators in a linear harmonic
chain}\newline Phys.~Rev.~A \textbf{73}, 052107 (2006).

\item J. Kofler and \v{C}. Brukner\newline\textit{Entanglement distribution
revealed by macroscopic observations}\newline Phys.~Rev.~A \textbf{74},
050304(R) (2006).\newpage
\end{itemize}

\section{The linear harmonic chain}

Quantum entanglement is a physical phenomenon in which the quantum states of
two or more systems can only be described with reference to each other, even
though the individual systems may be spatially separated. This leads to
correlations between observables of the systems that cannot be understood on
the basis of classical (local realistic) theories~\cite{Bell1964}. Its
importance today exceeds the realm of the foundations of quantum physics and
entanglement has become an important physical resource, like energy, that
allows performing communication and computation tasks with efficiency which is
not achievable classically~\cite{Niel2000}. Moving to higher-dimensional
entangled systems or entangling more systems with each other, will eventually
push the realm of quantum physics well into the macroscopic world. It will
therefore be important to investigate under which conditions entanglement
within or between \textquotedblleft macroscopic\textquotedblright\ objects,
each consisting of a sample containing a large number of the constituents, can arise.

Recently, it was shown that macroscopic entanglement can arise
\textquotedblleft naturally\textquotedblright\ between constituents of various
complex physical systems. Examples of such systems are chains of interacting
spin systems~\cite{Niel2000,Arne2001}, harmonic
oscillators~\cite{Aude2002,Sera2005} and quantum fields~\cite{Rezn2003}.
Entanglement can have an effect on the macroscopic properties of these
systems~\cite{Ghos2003,Wies2005,Bruk2006} and can be in principle extractable
from them for quantum information
processing~\cite{Rezn2003,Pate2004,Retz2005,deCh2006}.

With the aim of better understanding macroscopical entanglement we will
investigate entanglement between \textit{collective operators}. A simple and
natural system is the ground state of a linear chain of harmonic oscillators
furnished with harmonic nearest-neighbor interaction. The mathematical
entanglement properties of this system were extensively investigated
in~\cite{Aude2002,Bote2004,Sera2005,Pate2005}. Entanglement was computed in
the form of logarithmic negativity for general bisections of the chain and for
contiguous blocks of oscillators that do not comprise the whole chain. It was
shown that the log-negativity typically decreases exponentially with the
separation of the groups and that the larger the groups, the larger the
maximal separation for which the log-negativity is non-zero~\cite{Aude2002}.
It also was proven that an area law holds for harmonic lattice systems,
stating that the amount of entanglement between two complementary regions
scales with their boundary~\cite{Cram2005,Wolf2008}.

In a real experimental situation, however, we are typically not able to
determine the complete mathematical amount of entanglement (as measured, e.g.,
by log-negativity) which is non-zero even if two blocks share only one
arbitrarily weak entangled pair of oscillators. Our measurement apparatuses
normally cannot resolve single oscillators, but rather interact with a whole
bunch of them in one way, potentially even in \textit{non-contiguous regions},
thus measuring certain \textit{global properties}. Here we will study
entanglement between \textquotedblleft physical blocks\textquotedblright\ of
harmonic oscillators---existing only if there is entanglement between the
\textit{collective operators} defined on the entire blocks---as a function of
their size, relative distance and the coupling strength. Our aim is to
quantify (experimentally accessible) entanglement between global properties of
two groups of harmonic oscillators. Surprisingly, we will see that such
collective entanglement can be demonstrated even in the case where none of the
oscillators from one block is entangled with an oscillator from the other
block (i.e.\ it cannot be understood as a cumulative effect of entanglement
between pairs of oscillators), which is in agreement with
Reference~\cite{Aude2002}. This shows the existence of bipartite entanglement
between collective operators.

Because of the area law~\cite{Cram2005,Wolf2008}\ the amount of entanglement
is relatively small in the first instance. We suggest a way to overcome this
problem by allowing the collective blocks to consist of a \textit{periodic
sequence of subblocks}. Then the total boundary region between them is
increased and we verify that indeed a larger amount of entanglement is found
for periodic blocks, where the entanglement scales at most with the
\textit{total} boundary region. We give an analytical approximation of this
amount of entanglement and motivate how it can in principle be extracted from
the chain~\cite{Rezn2003,Pate2004,Retz2005,deCh2006}.

Methodologically, we will quantify the entanglement between collective
operators of two blocks of harmonic oscillators by using a measure for
continuous variable systems based on the Peres-Horodecki
criterion~\cite{Pere1996,Horo1997,Simo2000,Kim2002}. The collective operators
will be defined as sums over local operators for all single oscillators
belonging to the block. The infinite harmonic chain is assumed to be in the
ground state and since the blocks do not comprise the whole chain, they are in
a mixed state.

\begin{figure}[t]
\begin{center}
\includegraphics{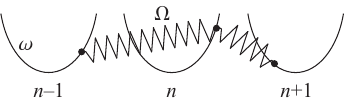}
\end{center}
\par
\vspace{-0.33cm}\caption{In a linear harmonic chain the oscillators are
situated in a harmonic potential with frequency $\omega$. Each oscillator,
labeled with the index $n$, is coupled with its nearest neighbors, $n-1$ and
$n+1$, by a harmonic potential with the coupling frequency $\Omega$.}%
\label{Figure_Chain}%
\end{figure}

We investigate a linear harmonic chain, where each of the $N$ oscillators is
situated in a harmonic potential with frequency $\omega$ and each oscillator
is coupled with its neighbors by a harmonic potential with the coupling
frequency $\Omega$ (Figure~\ref{Figure_Chain}). The oscillators have mass $m$
and their positions and momenta are denoted as $\overline{q}_{i}$ and
$\overline{p}_{i}$, respectively. Assuming periodic boundary conditions
($\overline{q}_{N+1}\equiv\overline{q}_{1}$), the Hamilton function
reads~\cite{Schw2003}%
\begin{equation}
H=%
{\displaystyle\sum\limits_{j=1}^{N}}
\left(  \frac{\overline{p}_{j}^{2}}{2\,m}+\frac{m\,\omega^{2}\,\overline
{q}_{j}^{2}}{2}+\frac{m\,\Omega^{2}\,(\overline{q}_{j}-\overline{q}_{j-1}%
)^{2}}{2}\right)  \!.
\end{equation}
We canonically introduce dimensionless variables:\ $q_{j}\equiv
C\,\overline{q}_{j}$ and $p_{j}\equiv\overline{p}_{j}/C$, where $C$ is given
by $C\equiv\sqrt{m\omega(1+2\,\Omega^{2}/\omega^{2})^{1/2}}$ \cite{Bote2004}.
By this means the Hamilton function becomes%
\begin{equation}
\fbox{$\;\;H=\dfrac{E_{0}}{2}\,%
{\displaystyle\sum\limits_{j=1}^{N}}
\,(p_{j}^{2}+q_{j}^{2}-\alpha\,q_{j}\,q_{j+1})\,,\;\;$}%
\end{equation}
with the abbreviations $\alpha\equiv2\,\Omega^{2}/(2\,\Omega^{2}+\omega^{2})$
and $E_{0}\equiv\sqrt{2\,\Omega^{2}+\omega^{2}}$. The (single) coupling
constant is restricted to values $0<\alpha<1$, where $\alpha\rightarrow0$ in
the weak coupling limit ($\Omega/\omega\rightarrow0$) and $\alpha\rightarrow1$
in the strong coupling limit ($\Omega/\omega\rightarrow\infty$).

\begin{figure}[t]
\begin{center}
\includegraphics{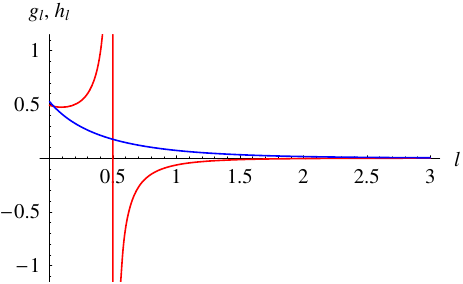}
\end{center}
\par
\vspace{-0.33cm}\caption{The two-point vacuum correlation functions
(\ref{eq gl}) and (\ref{eq hl}) for the linear harmonic chain with coupling
$\alpha=0.5$. The blue curve shows $h_{|i-j|}\equiv\left\langle 0\right\vert
\hat{q}_{i}\,\hat{q}_{j}\left\vert 0\right\rangle $, and the red curve shows
$h_{|i-j|}\equiv\left\langle 0\right\vert \hat{p}_{i}\,\hat{p}_{j}\left\vert
0\right\rangle $ with $l\equiv|i-j|$. Only integer values for $l$ are
meaningful.}%
\label{Figure_Chain_two-point}%
\end{figure}

In the language of second quantization the positions and momenta are converted
into operators ($q_{j}\rightarrow\hat{q}_{j}$, $p_{j}\rightarrow\hat{p}_{j}$)
and are expanded into modes of their annihilation and creation operators,
$\hat{a}$ and $\hat{a}^{\dagger}$, respectively:%
\begin{align}
\hat{q}_{j}  &  =\frac{1}{\sqrt{N}}\,%
{\displaystyle\sum\limits_{k=0}^{N-1}}
\,\frac{1}{\sqrt{2\,\nu(\theta_{k})}}\left[  \hat{a}(\theta_{k})\,\text{e}%
^{\text{i}\theta_{k}j}+\text{H.c.}\right]  \!,\label{eq qj}\\
\hat{p}_{j}  &  =\frac{-\text{i}}{\sqrt{N}}\,%
{\displaystyle\sum\limits_{k=0}^{N-1}}
\,\sqrt{\frac{\nu(\theta_{k})}{2}}\left[  \hat{a}(\theta_{k})\,\text{e}%
^{\text{i}\theta_{k}j}-\text{H.c.}\right]  \!. \label{eq pj}%
\end{align}
Here $\theta_{k}\equiv2\,\pi\,k/N$ (with $k=0,1,...,N-1$) is the dimensionless
pseudo-momentum and
\begin{equation}
\nu(\theta_{k})\equiv\sqrt{1-\alpha\cos\theta_{k}}%
\end{equation}
is the dispersion relation. The annihilation and creation operators fulfil the
well known commutation relation $\left[  \hat{a}(\theta_{k}),\hat{a}^{\dagger
}(\theta_{k^{\prime}})\right]  =\delta_{kk^{\prime}}$, since $[\hat{q}%
_{i},\hat{p}_{j}]=\;$i$\,\delta_{ij}$ has to be guaranteed. The ground state
(vacuum), denoted as $\left\vert 0\right\rangle $, is defined by $\hat
{a}(\theta_{k})\left\vert 0\right\rangle =0$ holding for all $\theta_{k}$. The
two-point vacuum correlation functions are%
\begin{align}
g_{|i-j|}  &  \equiv\left\langle 0\right\vert \hat{q}_{i}\,\hat{q}%
_{j}\left\vert 0\right\rangle \equiv\left\langle \!\right.  \hat{q}_{i}%
\,\hat{q}_{j}\left.  \!\right\rangle =(2N)^{-1}\,%
{\displaystyle\sum\nolimits_{k=0}^{N-1}}
\,\nu^{-1}(\theta_{k})\cos(l\,\theta_{k}),\label{eq gl}\\
h_{|i-j|}  &  \equiv\left\langle 0\right\vert \hat{p}_{i}\,\hat{p}%
_{j}\left\vert 0\right\rangle \equiv\left\langle \!\right.  \hat{p}_{i}%
\,\hat{p}_{j}\left.  \!\right\rangle =(2N)^{-1}\,%
{\displaystyle\sum\nolimits_{k=0}^{N-1}}
\,\nu(\theta_{k})\cos(l\,\theta_{k}), \label{eq hl}%
\end{align}
where $l\equiv|i-j|$. In the limit of an infinite chain ($N\rightarrow\infty
$)---which we will study below---and for $l<N/2$ they can be expressed in
terms of the hypergeometric function $_{2}F_{1}$~\cite{Bote2004}%
:\ $g_{l}=[z^{l}/(2\mu)]\tbinom{l-1/2}{l}\,_{2}F_{1}(1/2,l+1/2,l+1,z^{2})$,
$h_{l}=(\mu z^{l}/2)\tbinom{l-3/2}{l}\,_{2}F_{1}(-1/2,l-1/2,l+1,z^{2})$, where
$z\equiv(1-\sqrt{1-\alpha^{2}})/\alpha$ and $\mu\equiv1/\sqrt{1+z^{2}}$
(Figure~\ref{Figure_Chain_two-point}).

\subsection{Defining collective operators}

In the following, we are interested in entanglement between two
\textquotedblleft physical blocks\textquotedblright\ of oscillators, where the
blocks are represented by a specific form of \textit{collective operators}
which are normalized sums of individual operators. By means of such a
formalism we seek to fulfil experimental conditions and constraints, since
\textit{finite experimental resolution implies naturally the measurement of,
e.g., the average momentum of a bunch of oscillators rather than the momentum
of only one}. On the other hand, this formalism can easily take account of
blocks that\textit{ consist of non-contiguous regions}, leading to interesting
results which will be shown below. We want to point out that this convention
of the term \textit{block} is not the same as it is normally used in the
literature. In contrast to the latter, for which one allows any possible
measurement, our simulation of realizable experiments already lacks some
information due to the averaging.

\begin{figure}[t]
\begin{center}
\includegraphics{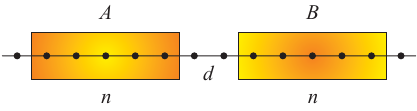}
\end{center}
\par
\vspace{-0.33cm}\caption{Two blocks $A$ and $B$ of a harmonic chain. Each
block consists of $n$ oscillators and the blocks are separated by $d$
oscillators.}%
\label{Figure_Chain_blocks}%
\end{figure}

Let us now consider two non-overlapping blocks of oscillators, $A$ and $B$,
within the closed harmonic chain in its ground state, where each block
contains $n$ oscillators. The blocks are separated by $d\geq0$ oscillators
(Figure~\ref{Figure_Chain_blocks}). We assume $n,d\ll N$ and $N\rightarrow
\infty$ for the numerical calculations of the two-point correlation
functions.

By a Fourier transform we map the $n$ oscillators of each block onto $n$
(\textquotedblleft orthogonal\textquotedblright) \textit{frequency-dependent}
collective operators%
\begin{align}
\hat{Q}_{A}^{(k)}  &  \equiv\dfrac{1}{\sqrt{n}}\;%
{\displaystyle\sum\limits_{j\in A}}
\;\hat{q}_{j}\;\text{e}^{\tfrac{2\pi\text{i}jk}{n}},\\
\hat{P}_{A}^{(k)}  &  \equiv\dfrac{1}{\sqrt{n}}\;%
{\displaystyle\sum\limits_{j\in A}}
\;\hat{p}_{j}\;\text{e}^{-\tfrac{2\pi\text{i}jk}{n}},
\end{align}
with the frequencies $k=0,...,n-1$, and analogously for block $B$. The
commutator of the collective position and momentum operators is%
\begin{equation}
\lbrack\hat{Q}_{A}^{(k)},\hat{P}_{A}^{(k^{\prime})}]=\text{i}\,\delta
_{kk^{\prime}}\,. \label{eq commutator QP}%
\end{equation}
This means that collective operators for different frequencies $k\neq
k^{\prime}$ commute. For different blocks the commutator vanishes:\ $[\hat
{Q}_{A}^{(k)},\hat{P}_{B}^{(k^{\prime})}]=0$.

If the individual positions and momenta of all oscillators are written into a
vector%
\begin{equation}
\hat{\mathbf{x}}\equiv(\hat{q}_{1},\hat{p}_{1},\hat{q}_{2},\hat{p}%
_{2},...,\hat{q}_{N},\hat{p}_{N})^{\text{T}},
\end{equation}
with T denoting the transpose, then there holds the commutation relation%
\begin{equation}
\lbrack\hat{x}_{i},\hat{x}_{j}]=\text{i}\,\Omega_{ij}%
\end{equation}
with $\mathbf{\Omega}$ the $n$-fold direct sum of $2\!\times\!2$ symplectic
matrices:%
\begin{equation}
\mathbf{\Omega}\equiv%
{\displaystyle\bigoplus\limits_{j=1}^{n}}
\left(  \!%
\begin{array}
[c]{cc}%
0 & 1\\
-1 & 0
\end{array}
\!\right)  \!.
\end{equation}
A matrix $\mathbf{S}$ transforms $\hat{\mathbf{x}}$ into a vector of
collective (and uninvolved individual) oscillators:%
\begin{equation}
\hat{\mathbf{X}}\equiv\mathbf{S}\,\hat{\mathbf{x}}=(\{\hat{Q}_{A}^{(k)}%
,\hat{P}_{A}^{(k)}\}_{k},\{\hat{Q}_{B}^{(k)},\hat{P}_{B}^{(k)}\}_{k},\{\hat
{q}_{j},\hat{p}_{j}\}_{j})^{\text{T}}.
\end{equation}
Here $\{\hat{Q}_{A}^{(k)},\hat{P}_{A}^{(k)}\}_{k}=(\hat{Q}_{A}^{(0)},\hat
{P}_{A}^{(0)},...,\hat{Q}_{A}^{(n-1)},\hat{P}_{A}^{(n-1)})$ denotes all
collective oscillators of block $A$ and analogously for block $B$, whereas
$\{\hat{q}_{j},\hat{p}_{j}\}_{j}$ denotes the $2\,(N-2n)$ position and
momentum entries of those $N-2n$ oscillators which are not part of one of the
two blocks. The matrix $\mathbf{S}$ corresponds to a Gaussian
operation~\cite{Eise2003}. It has determinant det$\,\mathbf{S}=1$ and
preserves the symplectic structure%
\begin{equation}
\mathbf{\Omega}=\mathbf{S}^{\text{T}}\,\mathbf{\Omega}\,\mathbf{S}\,,
\label{eq sympl struct}%
\end{equation}
and hence%
\begin{equation}
\lbrack\hat{X}_{i},\hat{X}_{j}]=\text{i}\,\Omega_{ij} \label{eq comm sympl}%
\end{equation}
for all $i,j$, in particular verifying eq.~(\ref{eq commutator QP}). This
means that the Gaussianness of the ground state of the harmonic chain
(i.e.\ the fact that the state is completely characterized by its first and
second moments) is preserved by the (Fourier) transformation to the
frequency-dependent collective operators.

\subsection{Quantifying entanglement between collective operators}

In reality, we are typically not capable of single particle resolution
measurements but only of measuring the collective operators with one
frequency, namely $k=0$, i.e.\ the \textquotedblleft average\textquotedblright%
\ over the individual oscillators. Note that in general the correlations of
higher-frequency collective operators, e.g., $\left\langle \!\right.  (\hat
{Q}_{A}^{(k)})^{2}\left.  \!\right\rangle $ or $\left\langle \!\right.
\hat{Q}_{A}^{(k)}\hat{Q}_{B}^{(k)}\left.  \!\right\rangle $ with $k\neq0$, are
not real numbers. Therefore, as a natural choice, we denote as the collective
operators%
\begin{equation}
\fbox{$%
\begin{array}
[c]{c}%
\hat{Q}_{A}\equiv\hat{Q}_{A}^{(0)}=\dfrac{1}{\sqrt{n}}\;%
{\displaystyle\sum\limits_{j\in A}}
\;\hat{q}_{j}\,,\\
\;\hat{P}_{A}\equiv\hat{P}_{A}^{(0)}=\dfrac{1}{\sqrt{n}}\;%
{\displaystyle\sum\limits_{j\in A}}
\;\hat{p}_{j}\,,\rule{0pt}{16pt}\!
\end{array}
$} \label{eq Def Q}%
\end{equation}
and analogously for block $B$. It seems to be a very natural situation that
the experimenter only has access to these collective properties and we are
interested in the amount of (physical) entanglement one can extract from the
system if only the collective observables $\hat{Q}_{A,B}$ and $\hat{P}_{A,B}$
are measured.

Reference \cite{Simo2000} derives a separability criterion which is based on
the Peres-Horodecki criterion~\cite{Pere1996,Horo1997} and the fact that---in
the continuous variables case---the partial transposition allows a geometric
interpretation as mirror reflection in phase space. Following largely the
notation in the original paper, we introduce the vector%
\begin{equation}
\hat{\mathbf{\xi}}\equiv(\hat{Q}_{A},\hat{P}_{A},\hat{Q}_{B},\hat{P}_{B})
\end{equation}
of collective operators. The commutation relations have the compact form
$[\hat{\xi}_{\alpha},\hat{\xi}_{\beta}]=\;$i$\,K_{\alpha\beta}$ with
$\mathbf{K}\equiv%
{\textstyle\bigoplus\nolimits_{j=1}^{2}}
\!\left(
\genfrac{}{}{0pt}{1}{0}{-1}%
\genfrac{}{}{0pt}{1}{1}{0}%
\right)  $. The separability criterion bases on the covariance matrix (of
first and second moments)%
\begin{equation}
V_{\alpha\beta}\equiv\dfrac{1}{2}\left\langle \!\right.  \Delta\hat{\xi
}_{\alpha}\Delta\hat{\xi}_{\beta}+\Delta\hat{\xi}_{\beta}\Delta\hat{\xi
}_{\alpha}\left.  \!\right\rangle ,
\end{equation}
where $\Delta\hat{\xi}_{\alpha}\equiv\hat{\xi}_{\alpha}-\langle\hat{\xi
}_{\alpha}\rangle$ with $\langle\hat{\xi}_{\alpha}\rangle=0$ in our case (the
ground state is a Gaussian with its mean at the origin of phase space).

The covariance matrix $\mathbf{V}$ is real (which would not be the case for
higher-frequency collective operators) and symmetric:\ $\left\langle
\!\right.  \hat{Q}_{A}\hat{Q}_{B}\left.  \!\right\rangle =\left\langle
\!\right.  \hat{Q}_{B}\hat{Q}_{A}\left.  \!\right\rangle $ and $\left\langle
\!\right.  \hat{P}_{A}\hat{P}_{B}\left.  \!\right\rangle =\left\langle
\!\right.  \hat{P}_{B}\hat{P}_{A}\left.  \!\right\rangle $, coming from the
fact that the two-point correlation functions (\ref{eq gl}) and (\ref{eq hl})
only depend on the absolute value of the position index difference. On the
other hand, using eqs.~(\ref{eq qj}) and (\ref{eq pj}), we verify that
$\left\langle \!\right.  \hat{q}_{i}\,\hat{p}_{j}\left.  \!\right\rangle
=\;$i$\,(2\,N)^{-1}\,%
{\textstyle\sum\nolimits_{k=0}^{N-1}}
\exp[$i$\,\theta_{k}(i-j)]$ and $\left\langle \!\right.  \hat{p}_{j}\,\hat
{q}_{i}\left.  \!\right\rangle =-$i$\,(2\,N)^{-1}\,%
{\textstyle\sum\nolimits_{k=0}^{N-1}}
\exp[$i$\,\theta_{k}(j-i)]$. For $i\neq j$ both summations vanish ($\theta
_{k}\equiv2\,\pi\,k/N$ and $i,j$ integer) and for $i=j$ they are the same but
with opposite sign. Thus, in all cases $\left\langle \!\right.  \hat{q}%
_{i}\,\hat{p}_{j}\left.  \!\right\rangle =-\left\langle \!\right.  \hat{p}%
_{j}\,\hat{q}_{i}\left.  \!\right\rangle $. These symmetries also hold for the
collective operators and hence we obtain%
\begin{equation}
\mathbf{V}=\left(  \!%
\begin{array}
[c]{cccc}%
G & 0 & G_{AB} & 0\\
0 & H & 0 & H_{AB}\\
G_{AB} & 0 & G & 0\\
0 & H_{AB} & 0 & H
\end{array}
\!\right)  \!. \label{eq V}%
\end{equation}
The matrix elements are%
\begin{align}
G  &  \equiv\left\langle \!\right.  \hat{Q}_{A}^{2}\left.  \!\right\rangle
=\left\langle \!\right.  \hat{Q}_{B}^{2}\left.  \!\right\rangle =\frac{1}{n}\,%
{\displaystyle\sum\limits_{j\in A}}
\,%
{\displaystyle\sum\limits_{i\in A}}
\,g_{|j-i|}\,,\\
H  &  \equiv\left\langle \!\right.  \hat{P}_{A}^{2}\left.  \!\right\rangle
=\left\langle \!\right.  \hat{P}_{B}^{2}\left.  \!\right\rangle =\frac{1}{n}\,%
{\displaystyle\sum\limits_{j\in A}}
\,%
{\displaystyle\sum\limits_{i\in A}}
\,h_{|j-i|}\,,\\
G_{AB}  &  \equiv\left\langle \!\right.  \hat{Q}_{A}\hat{Q}_{B}\left.
\!\right\rangle =\frac{1}{n}\,%
{\displaystyle\sum\limits_{j\in A}}
\,%
{\displaystyle\sum\limits_{i\in B}}
\,g_{|j-i|}\,,\\
H_{AB}  &  \equiv\left\langle \!\right.  \hat{P}_{A}\hat{P}_{B}\left.
\!\right\rangle =\frac{1}{n}\,%
{\displaystyle\sum\limits_{j\in A}}
\,%
{\displaystyle\sum\limits_{i\in B}}
\,h_{|j-i|}\,.
\end{align}
To quantify entanglement between two collective blocks we use the degree of
entanglement $\varepsilon$, given by the absolute sum of the negative
eigenvalues of the partially transposed density operator:\ $\varepsilon
\equiv\;$Tr$|\hat{\rho}^{\text{T}_{B}}|-1$, i.e.\ by measuring how much the
mirror reflected state $\hat{\rho}^{\text{T}_{B}}$, where the momenta in block
$B$ are reversed, fails to be positive definite. This measure (proportional to
the negativity) is based on the Peres-Horodecki
criterion~\cite{Pere1996,Horo1997} and was shown to be an entanglement
monotone~\cite{Lee2000,Vida2002}. For covariance matrices of the form
(\ref{eq V}) it reads~\cite{Kim2002}%
\begin{equation}
\fbox{$\;\;\varepsilon=\max\left(  0,\dfrac{(\delta_{1}\delta_{2})_{0}}%
{\delta_{1}\delta_{2}}-1\right)  \;\;$} \label{eq epsilon}%
\end{equation}
with%
\begin{align*}
\delta_{1}  &  \equiv G-|G_{AB}|\,,\\
\delta_{2}  &  \equiv H-|H_{AB}|\,.
\end{align*}
In general, the numerator is defined by the square of the Heisenberg
uncertainty relation%
\begin{equation}
(\delta_{1}\delta_{2})_{0}\equiv\,|\!\left\langle \!\right.  [\hat{Q}%
_{A,B},\hat{P}_{A,B}]\left.  \!\right\rangle \!|^{2}=\frac{1}{4}\,,
\end{equation}
where the last equal sign holds due to eq.~(\ref{eq commutator QP}). We note
that $\varepsilon$ is a \textit{degree} of entanglement (in the sense of
necessity and sufficiency) only for Gaussian states which are completely
characterized by their first and second moments, as for example the ground
state of the harmonic chain we are studying. However, we left out the
higher-frequency collective operators (and all the oscillators which are not
part of the blocks) and therefore, the entanglement $\varepsilon$ has to be
understood as the Gaussian part of the amount of entanglement which exists
between (and can be extracted from) the two blocks when only the collective
properties $\hat{Q}_{A,B}$ and $\hat{P}_{A,B}$, as defined in
eq.~(\ref{eq Def Q}), are accessible.

There also exists an entanglement witness in form of a separability criterion
based on variances, where
\begin{equation}
\Delta\equiv\left\langle \!\right.  (\hat{Q}_{A}-\hat{Q}_{B})^{2}\left.
\!\right\rangle +\left\langle \!\right.  (\hat{P}_{A}+\hat{P}_{B})^{2}\left.
\!\right\rangle =2\,(G-G_{AB}+H+H_{AB})<2
\end{equation}
is a sufficient condition for the state to be entangled~\cite{Duan2000}. We
note that the above negativity measure (\ref{eq epsilon}) is \textquotedblleft
stronger\textquotedblright\ than this witness in the whole parameter space
($\alpha,n$). In particular, there are cases where $\varepsilon>0$ although
$\Delta\geq2$. This is in agreement with the finding that the variance
criterion is weaker than a generalized negativity criterion~\cite{Shch2005}.

We further note that the amount of entanglement (\ref{eq epsilon}) is
invariant under a change of potential redefinitions of the collective
operators, e.g., $\hat{Q}_{A}\equiv\,%
{\textstyle\sum\nolimits_{j\in A}}
\,\hat{q}_{j}$ or $\hat{Q}_{A}\equiv(1/n)\,%
{\textstyle\sum\nolimits_{j\in A}}
\,\hat{q}_{j}$, as then the modified scaling in the correlations ($G$,
$G_{AB}$, $H$, and $H_{AB}$) is exactly compensated by the modified scaling of
the Heisenberg uncertainty in the numerator.\begin{figure}[t]
\begin{center}
\includegraphics{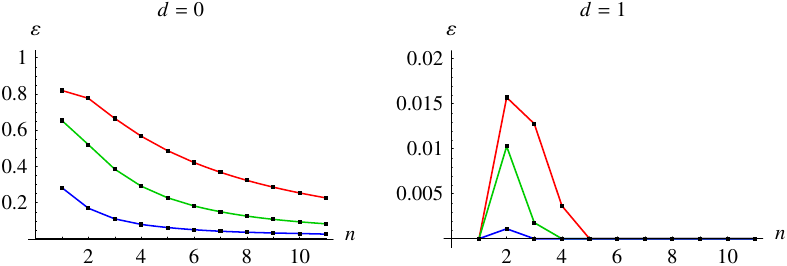}
\end{center}
\par
\vspace{-0.33cm}\caption{Degree of collective entanglement $\varepsilon$ for
two blocks of oscillators as a function of their size $n$. Left: The blocks
are neighboring ($d=0$) and entanglement exists for all $n$ and coupling
strengths $\alpha$. Plotted are $\alpha=0.99$ (red), $\alpha=0.9$ (green) and
$\alpha=0.5$ (blue). Right: The same for two blocks which are separated by one
oscillator ($d=1$). The two blocks are unentangled for $n=1$ but can be
entangled, if one increases the block size ($n>1$), although none of the
individual pairs between the blocks is entangled.}%
\label{Figure_Chain_entanglement}%
\end{figure}

Figure~\ref{Figure_Chain_entanglement} shows the results for $d=0$ and $d=1$.
In the first case---if the blocks are neighboring---there exists entanglement
for all possible coupling strengths $\alpha$ and block sizes $n$. In the
latter case---if there is one oscillator between the blocks---due to the
strongly decaying correlation functions $g$ an $h$ there is no entanglement
between two single oscillators ($n=1$), but there exists entanglement for
larger blocks (up to $n=4$, depending on $\alpha$). The statement that
entanglement can emerge by going to larger blocks was also found
in~\cite{Aude2002}. But there the blocks were abstract objects, containing all
the information of their constituents. In the case of collective operators,
however, increasing the block size (averaging over more oscillators) is also
connected with a loss of information. In spite of this loss and the mixedness
of the state, two blocks can be entangled, although none of the individual
pairs between the blocks is entangled---indicating true bipartite entanglement
between collective operators (multipartite entanglement between individual
oscillators). For $d\geq2$, however, no entanglement can be found anymore.

\begin{figure}[t]
\begin{center}
\includegraphics{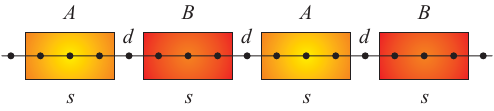}
\end{center}
\par
\vspace{-0.33cm}\caption{Two periodic blocks of a harmonic chain $A$ and $B$.
Each block can consist of $m$ subblocks with $s$ oscillators each, separated
by $d$ oscillators. In the picture $d=1$, $m=2$, $s=3$ and the number of
oscillators per block is $n=ms=6$.}%
\label{Figure_Chain_blocks_periodic}%
\end{figure}

These results are in agreement with the general statement that entanglement
between a region and its complement scales with the size of the
boundary~\cite{Cram2005,Wolf2008}. In the present case of two blocks in a
one-dimensional chain (Figure~\ref{Figure_Chain_blocks}) the boundary is
constant and as the blocks are made larger, the entanglement decreases since
it is distributed over more and more oscillators. We therefore propose to
increase the number of boundaries by considering two non-overlapping blocks,
where we allow a \textit{periodic continuation} of the situation above,
i.e.\ a sequence of $m\geq1$ subblocks, separated by $d$ oscillators and each
consisting of $s\geq1$ oscillators, where $ms=n$
(Figure~\ref{Figure_Chain_blocks_periodic}).

\begin{figure}[t]
\begin{center}
\includegraphics{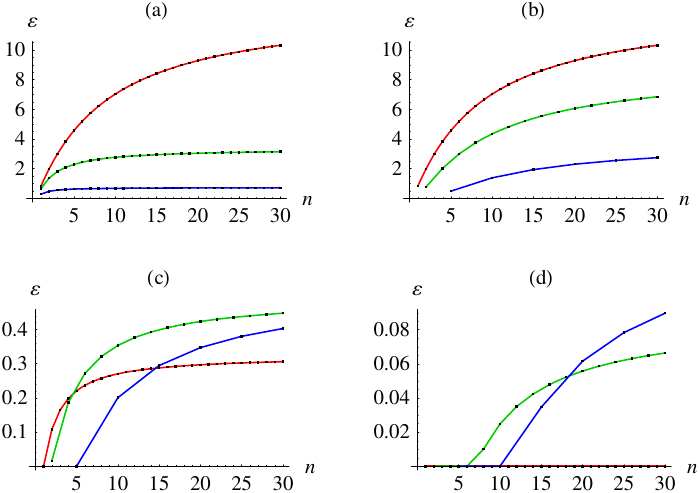}
\end{center}
\par
\vspace{-0.33cm}\caption{Degree of collective entanglement $\varepsilon$ for
two periodic blocks of oscillators as a function of their total size $n$. (a)
Neighboring one-particle subblocks ($d=0$, $s=1$). Entanglement monotonically
increases with $n$ and becomes larger as the coupling strength $\alpha$
increases. Plotted are $\alpha=0.99$ (red), $\alpha=0.9$ (green) and
$\alpha=0.5$ (blue). (b) The coupling is fixed to $\alpha=0.99$ for this and
the subsequent graphs. There is no separation, $d=0$. Plotted are the cases
$s=1,2,5$ in red, green, blue, respectively. For fixed $n$ the entanglement is
more or less proportional to the number of boundaries, i.e.\ inversely
proportional to the subblock size $s$. (c) and (d) correspond to the cases
$d=1$ and $d=2$, respectively. The dependence on the size of the subblocks is
more complicated as there is a trade-off between having a large number of
boundaries (i.e.\ small $s$) and the fact that one should have large subblocks
as individual separated oscillators are not entangled.}%
\label{Figure_Chain_entanglement_periodic}%
\end{figure}

The degree of entanglement between two periodic blocks of non-separated
($d=0$) one-particle subblocks ($s=1$) is larger for stronger coupling
constant $\alpha$ and grows with the overall number of oscillators $n$
(Figure~\ref{Figure_Chain_entanglement_periodic}a). For given $\alpha$ and $n$
and no separation between the subblocks ($d=0$) the entanglement is larger for
the case of small subblocks, as then there are many of them, causing a large
total boundary (Figure~\ref{Figure_Chain_entanglement_periodic}b).
Entanglement can be even found for larger separation ($d=1,2$) with a more
complicated dependence on the size $s$ of the subblocks. There is a trade-off
between having a large number of boundaries and the fact that one should have
large subblocks as individual separated oscillators are not entangled
(Figure~\ref{Figure_Chain_entanglement_periodic}c,d). For $d\geq3$ no
entanglement can be found anymore. (In a realistic experimental situation,
where the separation $d$ is not sharply defined, e.g., where there are
weighted contributions for $d=0,1,...,d_{\text{max}}$, entanglement can
persist even for $d_{\text{max}}\geq3$, depending on the weighting
factors.)

For the sake of completeness we give a rough approximation of the entanglement
between two periodic blocks. Let us assume that the subblocks are directly
neighbored, $d=0$. Furthermore, we consider couplings $\alpha$ such that we
may neglect higher than next neighbor correlations ($\alpha\lesssim0.5$),
i.e.\ we only take into account $g_{0}$, $g_{1}$, $h_{0}$ and $h_{1}$. The
correlations read%
\begin{align}
G  &  =\dfrac{1}{n}\,%
{\displaystyle\sum\limits_{j\in A}}
\,%
{\displaystyle\sum\limits_{i\in A}}
\,g_{|j-i|}\approx g_{0}+\dfrac{2m\,(s-1)}{n}\,g_{1}\,,\\
G_{AB}  &  =\dfrac{1}{n}\,%
{\displaystyle\sum\limits_{j\in A}}
\,%
{\displaystyle\sum\limits_{i\in B}}
\,g_{|j-i|}\approx\dfrac{2m-1}{n}\,g_{1}\,,
\end{align}
and analogously for $H$ and $H_{AB}$. The first equation reflects that there
are $n$ self-correlations and $m\,(s-1)$ nearest neighbor pairs (which are
counted twice) within one block, i.e.\ $s-1$ pairs per subblock. The second
equation represents the fact that there are $2m-1$ boundaries where blocks $A$
and $B$ meet. Using $s=n/m$, the entanglement (\ref{eq epsilon}) becomes (note
that $g_{1}>0$ and $h_{1}<0$)%
\begin{equation}
\varepsilon\approx\dfrac{1}{4\,[g_{0}+(2-\tfrac{4\,m-1}{n})\,g_{1}%
]\,[h_{0}+(2-\tfrac{1}{n})\,h_{1}]}-1\,.
\end{equation}
For given $n$ this approximation obviously increases with the total number of
boundaries, $m$. It can be considered as an estimate for a situation like in
Figure~\ref{Figure_Chain_entanglement_periodic}b, if a smaller coupling is
used such that the neglect of higher correlations becomes justified.

We close this section by annotating that the entanglement (\ref{eq epsilon})
between collective blocks of oscillators---being the Gaussian part---can in
principle (for sufficient control of the block separation $d$) be transferred
to two remote qubits via a Jaynes-Cummings type
interaction~\cite{Rezn2003,Pate2004,Retz2005}. For the interaction with
periodic blocks \textquotedblleft gratings\textquotedblright\ have to be
employed in the experimental setup. The interaction Hamiltonian is of the form%
\begin{equation}
\hat{H}_{\text{int}}\sim(\text{e}^{-\text{i}\omega_{1}t}\,\hat{\sigma}_{1}%
^{+}+\text{e}^{+\text{i}\omega_{1}t}\,\,\hat{\sigma}_{1}^{+})\,\hat{Q}%
_{A}+(\text{e}^{-\text{i}\omega_{2}t}\,\hat{\sigma}_{2}^{+}+\text{e}%
^{+\text{i}\omega_{2}t}\,\hat{\sigma}_{2}^{+})\,\hat{Q}_{B}\,,
\end{equation}
where $\omega_{i}$ is the Rabi frequency and $\hat{\sigma}_{i}^{+}%
=(\hat{\sigma}_{i}^{-})^{\dagger}=\left\vert e\right\rangle \!_{i}%
\,_{i}\!\left\langle g\right\vert $ is the bosonic operator (with $\left\vert
g\right\rangle \!_{i}$ and $\left\vert e\right\rangle \!_{i}$ the ground and
the excited state) of the $i$-th qubit ($i=1,2$).

\subsection{Collective operators for scalar quantum fields}

The continuum limit of the linear harmonic chain is the (1+1)-dimensional
Klein-Gordon field $\phi(x,t)$ with the canonical momentum field
$\pi(x,t)=\dot{\phi}(x,t)$. It satisfies the Klein-Gordon equation (in natural
units $\hbar=c=1$) with mass $m$%
\begin{equation}
\ddot{\phi}-\nabla^{2}\phi+m^{2}\,\phi=0\,.
\end{equation}
With the canonical quantization procedure $\phi$ and $\pi$ become operators
satisfying the non-trivial commutation relation $[\hat{\phi}(x,t),\hat{\pi
}(x^{\prime},t)]=\;$i$\,\delta(x-x^{\prime})$. The field operator can be
expanded into a Fourier integral over elementary plane wave
solutions~\cite{Bjo2003}%
\begin{align}
\hat{\phi}(x,t)  &  =%
{\displaystyle\int}
\,\frac{\text{d}k}{\sqrt{4\pi\omega_{k}}}\left[  \hat{a}(k)\,\text{e}%
^{\text{i}kx-\text{i}\omega_{k}t}+\text{H.c.}\right]  \!,\label{eq field}\\
\hat{\pi}(x,t)  &  =-\text{i}\,%
{\displaystyle\int}
\,\frac{\text{d}k\,\omega_{k}}{\sqrt{4\pi}}\left[  \hat{a}(k)\,\text{e}%
^{\text{i}kx-\text{i}\omega_{k}t}-\text{H.c.}\right]  \!,
\end{align}
where $k$ is the wave number and $\omega_{k}=+\sqrt{k^{2}+m^{2}}$ is the
dispersion relation. The annihilation and creation operators fulfil $\left[
\hat{a}(k),\hat{a}^{\dagger}(k^{\prime})\right]  =\delta(k-k^{\prime})$. We
write the field operator as a sum of two contributions $\hat{\phi}=\hat{\phi
}^{(+)}+\hat{\phi}^{(-)}$, where $\hat{\phi}^{(+)}$ ($\hat{\phi}^{(-)}$) is
the contribution with positive (negative) frequency. Thus, $\hat{\phi}^{(+)}$
corresponds to the term with the annihilation operator in eq.~(\ref{eq field}%
). The vacuum correlation function is given by the (equal-time) commutator of
the positive and the negative frequency part:%
\begin{equation}
\left\langle 0\right\vert \hat{\phi}(x,t)\,\hat{\phi}(y,t)\left\vert
0\right\rangle =[\hat{\phi}^{(+)}(x,t),\hat{\phi}^{(-)}(y,t)]\,.
\end{equation}
It is a peculiarity of the idealization of quantum field theory that for $x=y$
this propagator diverges in the ground state:%
\begin{equation}
\left\langle 0\right\vert \hat{\phi}^{2}(x,t)\left\vert 0\right\rangle
\rightarrow\infty\,. \label{eq divergence}%
\end{equation}
The same is true for $\left\langle 0\right\vert \hat{\pi}^{2}(x,t)\left\vert
0\right\rangle $ and hence we cannot easily build an entanglement measure like
for the harmonic chain, since the analogs of the two-point correlation
functions $g_{0}$ and $h_{0}$, eqs.~(\ref{eq gl}) and (\ref{eq hl}), are
divergent now. Automatically, we are motivated to study the more physical
situation and consider extended space-time regions, which means that we should
integrate the field (and conjugate momentum) over some spatial area. We define
the collective field operators%
\begin{equation}
\fbox{$%
\begin{array}
[c]{c}%
\hat{\Phi}_{L}(x_{0},t)\equiv\dfrac{1}{\sqrt{L}}\,%
{\displaystyle\int\nolimits_{-L/2}^{L/2}}
\,\hat{\phi}(x+x_{0},t)\,\text{d}x\,,\\
\hat{\Pi}_{L}(x_{0},t)\equiv\dfrac{1}{\sqrt{L}}\,%
{\displaystyle\int\nolimits_{-L/2}^{L/2}}
\,\hat{\pi}(x+x_{0},t)\,\text{d}x\,.\,\rule{0pt}{18pt}\!
\end{array}
$} \label{eq Phi}%
\end{equation}
Therefore, $\hat{\Phi}_{L}(x_{0},t)$ and $\hat{\Pi}_{L}(x_{0},t)$ are
equal-time operators which are spatially averaged over a length $L$, centered
at position $x_{0}$. The commutator is%
\begin{equation}
\lbrack\hat{\Phi}_{L}(x_{0},t),\hat{\Pi}_{L}(x_{0},t)]=\frac{1}{L}\,%
{\displaystyle\int\nolimits_{-L/2}^{L/2}}
\,%
{\displaystyle\int\nolimits_{-L/2}^{L/2}}
\,\text{i}\,\delta(x-y)\,\text{d}x\,\text{d}y=\text{i}\,,
\label{eq commutator PhiPi}%
\end{equation}
which is in complete analogy to eq.~(\ref{eq commutator QP}). If $\hat{\Phi
}_{L}$ and $\hat{\Pi}_{L}$ correspond to separated regions without overlap,
i.e.\ $|x_{0}-y_{0}|>L$, then of course $[\hat{\Phi}_{L}(x_{0},t),\hat{\Pi
}_{L}(y_{0},t)]=0$. The spatial integration in eq.~(\ref{eq Phi}) can be
carried out analytically:%
\begin{align}
\hat{\Phi}_{L}(x_{0},t)  &  =\frac{1}{\sqrt{\pi L}}\,%
{\displaystyle\int\nolimits_{-\infty}^{\infty}}
\,\frac{\text{d}k}{k\sqrt{\omega_{k}}}\,\sin(\tfrac{k\,L}{2})\left[  \hat
{a}(k)\,\text{e}^{\text{i}kx_{0}-\text{i}\omega_{k}t}+\text{H.c.}\right]
\!,\label{eq field Phi}\\
\hat{\Pi}_{L}(x_{0},t)  &  =\frac{-\text{i}}{\sqrt{\pi L}}\,%
{\displaystyle\int\nolimits_{-\infty}^{\infty}}
\,\frac{\text{d}k\,\sqrt{\omega_{k}}}{k}\,\sin(\tfrac{k\,L}{2})\left[  \hat
{a}(k)\,\text{e}^{\text{i}kx_{0}-\text{i}\omega_{k}t}-\text{H.c.}\right]  \!.
\label{eq field Pi}%
\end{align}
The final step is to calculate the propagators of the field and the conjugate
momentum.\ We find%
\begin{align}
D_{\hat{\Phi},L}(r)  &  \equiv\left\langle 0\right\vert \hat{\Phi}_{L}%
(x_{0},t)\,\hat{\Phi}_{L}(y_{0},t)\left\vert 0\right\rangle \nonumber\\
&  =\frac{1}{\pi L}\,%
{\displaystyle\int\nolimits_{-\infty}^{\infty}}
\,\frac{\text{d}k}{k^{2}\sqrt{k^{2}+m^{2}}}\,\sin^{2}(\tfrac{k\,L}{2}%
)\cos(k\,r)\,,\label{eq DPhi}\\
D_{\hat{\Pi},L}(r)  &  \equiv\left\langle 0\right\vert \hat{\Pi}_{L}%
(x_{0},t)\,\hat{\Pi}_{L}(y_{0},t)\left\vert 0\right\rangle \nonumber\\
&  =\frac{1}{\pi L}\,%
{\displaystyle\int\nolimits_{-\infty}^{\infty}}
\,\frac{\text{d}k\,\sqrt{k^{2}+m^{2}}}{k^{2}}\,\sin^{2}(\tfrac{k\,L}{2}%
)\cos(k\,r)\,, \label{eq DPi}%
\end{align}
with $r\equiv|x_{0}-y_{0}|$ the distance between the centers of the two
regions, reflecting the spatial symmetry. Thus $D_{\hat{\Phi},L}(0)$ and
$D_{\hat{\Pi},L}(0)$ are the analogs of $\left\langle \!\right.  \hat{Q}%
_{A,B}^{2}\left.  \!\right\rangle $ and $\left\langle \!\right.  \hat{P}%
_{A,B}^{2}\left.  \!\right\rangle $ (\textit{intra}-block correlations within
the same block), respectively, whereas $D_{\hat{\Phi},L}(r>L)$ and
$D_{\hat{\Pi},L}(r>L)$ correspond to $\left\langle \!\right.  \hat{Q}_{A}%
\hat{Q}_{B}\left.  \!\right\rangle $ and $\left\langle \!\right.  \hat{P}%
_{A}\hat{P}_{B}\left.  \!\right\rangle $ (\textit{inter}-block correlations
between separated blocks).

The expressions (\ref{eq DPhi}) and (\ref{eq DPi}) are finite, especially for
$r=0$. Mathematically, the integration over a finite spatial region $L$
corresponds to a cutoff, which removes the divergence we faced in
eq.~(\ref{eq divergence}). However, the expressions are ill defined for
$L\rightarrow0$.

Applying the entanglement measure (\ref{eq epsilon}) with $G=D_{\hat{\Phi}%
,L}(0)$, $H=D_{\hat{\Pi},L}(0)$, $G_{AB}=D_{\hat{\Phi},L}(r)$, and
$H_{AB}=D_{\hat{\Pi},L}(r)$ does not indicate entanglement for any choice of
$L$ and $r>L$. The same is true for the generalized case of blocks consisting
of periodic subregions of space, showing an inherent difference between the
harmonic chain and its continuum limit. This might be due to the fact, that
any spatial integration immediately corresponds to an infinitely large block
in the discrete harmonic chain and that the information loss (compared to the
mathematical indeed existing exponentially small entanglement~\cite{Rezn2003})
due to the collective operators already is too large. Nonetheless, defining
collective operators like in eq.~(\ref{eq Phi}) and use of the measure
(\ref{eq epsilon}) may reveal entanglement between spatially separated regions
for other quantum field states.

\section{Spin ensembles}

Observation of quantum entanglement between increasingly larger objects is one
of the most promising avenues of experimental quantum physics. Eventually, all
these developments might lead to a full understanding of the simultaneous
coexistence of a macroscopic classical world and an underlying quantum realm.
Macroscopic samples typically contain $N\sim10^{20}$ particles. Because the
system's Hilbert space grows exponentially with the number of constituent
particles, a complete microscopic picture of entanglement in large systems
seems to be in general intractable. The question arises: What can we learn
about entanglement between constituent particles of macroscopic samples, if
only limited experimentally accessible knowledge about the samples is available?

There is a strong motivation in addressing this question because of recent
experimental progress in creating and manipulating entangled states of
increasing complexity, such as spin-squeezed states of two atomic
ensembles~\cite{Juls2001}. In such experiments one typically measures only
expectation values of \textit{collective operators} of two separated samples.
It is known that the two samples of spins can be characterized as either
entangled or separable by measuring collective spin operators~\cite{Sore2001}.
Furthermore, such measurements are shown to be sufficient to determine
entanglement measures of Gaussian states~\cite{Sher2005} and of a pair of
particles that is extracted from a totally symmetric spin state (invariant
under exchange of particles)~\cite{Wang2002}. It appears that collective
operators cannot be used to fully characterize entanglement in composite
systems without strong requirements on the symmetry of the state.

Here we present a general and practical method for entanglement detection
between two samples of spins. It solely employs \textit{collective spin
properties of the samples} and works irrespectively of the number of spin
particles constituting the samples and with no assumption about the symmetry
or mixedness of the state. The method is based on an entanglement measure
which is a \textit{tight lower bound for average entanglement between all
pairs of spins }belonging to the two samples. This measure is \textit{equal}
to the average entanglement for a certain class of systems which need not be
totally symmetric. We generalize the method to obtain the entanglement measure
between $M$ separated spin samples based on collective measurements. The
results apply for any entanglement monotone that is a convex measure on the
set of density matrices (e.g., concurrence~\cite{Hill1997},
negativity~\cite{Zycz1998,Vida2002}, three-way tangle~\cite{Coff2000}).

\subsection{Bipartite entanglement}

\begin{figure}[t]
\begin{center}
\includegraphics{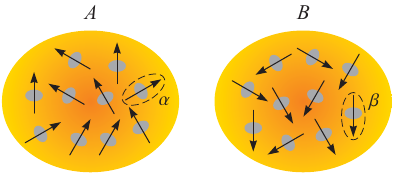}
\end{center}
\par
\vspace{-0.25cm}\caption{Two contiguous and non-overlapping spin subsystems
$A$ and $B$, each of which contains a large number $n$ of spins. What can we
learn about entanglement of a pair $(\alpha,\beta)$ of spins chosen at random
where $\alpha\in A$ and $\beta\in B$, if individual spins are experimentally
not accessible but only the collective properties of the samples $A$ and $B$?
What can we learn about entanglement between $A$ and $B$ from such collective
measurements?}%
\label{Figure_Spin_subsystems}%
\end{figure}

Consider an ensemble of spin-$\frac{1}{2}$ particles which is separated into
two ensembles $A$ and $B$ (Figure~\ref{Figure_Spin_subsystems}). Each of these
ensembles contains a large number of spins which we denote as $n$. Because of
the large dimensions,%
\begin{equation}
d=2^{n}\,,
\end{equation}
of the samples' Hilbert spaces, the structure of entanglement between the two
samples is considerably more complex than between two single spins. While
there are experimentally viable methods for detecting entanglement, they still
require a large number of parameters to be determined (proportional to $d^{2}%
$~\cite{Horo2002}).

The problem simplifies in situations in which each of the ensembles of $n$
spins can be treated as one large total spin of length $\frac{n}{2}$. This
means that, within the ensembles, the individual spin-$\frac{1}{2}$ particles
form symmetrized states (Dicke states). Though this reduces the dimension of
the Hilbert space of $A$ ($B$) to%
\begin{equation}
d^{\prime}=n+1\,,
\end{equation}
entanglement determination is still demanding for large $n$ both
experimentally and theoretically: Analytical solutions exist only for pure
states in general and for mixed states only for small $n$~\cite{Schl2005}.

In this section, we give a method to detect entanglement between large spin
samples by measuring only a \textit{small number of collective spin
properties} (sample spin components and their correlations), which is
independent of the sample size $n$. The collective spin operators are%
\begin{equation}
\fbox{$%
\begin{array}
[c]{c}%
\hat{S}_{i}^{A}\equiv\dfrac{\hbar}{2}\,\,%
{\displaystyle\sum\limits_{\alpha\in A}}
\,\hat{\sigma}_{i}^{(\alpha)}\,,\\
\hat{S}_{i}^{B}\equiv\dfrac{\hbar}{2}\,\,%
{\displaystyle\sum\limits_{\beta\in B}}
\,\hat{\sigma}_{i}^{(\beta)}\,.\rule{0pt}{16pt}\!
\end{array}
$} \label{eq SiA}%
\end{equation}
The index $i$ denotes the spatial component of the spins: $i\in\{1\!\equiv
\!x,2\!\equiv\!y,3\!\equiv\!z\}$. The Pauli matrix of the spin at site
$\alpha\in A$ is given by $\hat{\sigma}_{i}^{(\alpha)}$, and analogously for a
spin $\beta$ from subsystem $B$. Note that the collective operators satisfy
the usual commutation relations%
\begin{equation}
\lbrack\hat{S}_{i}^{A},\hat{S}_{j}^{A}]=\text{i}\,\hbar\,\varepsilon
_{ijk}\,\hat{S}_{k}^{A}\,,
\end{equation}
since $[\hat{\sigma}_{i}^{(\alpha)},\hat{\sigma}_{j}^{(\alpha^{\prime})}%
]=2\,$i$\,\delta_{\alpha\alpha^{\prime}}\,\varepsilon_{ijk}\,\hat{\sigma}%
_{k}^{(\alpha)}$.

The spin expectation values and correlations are%
\begin{align}
S_{i}^{A}  &  \equiv\left\langle \!\right.  \hat{S}_{i}^{A}\left.
\!\right\rangle =\frac{\hbar}{2}\,\,%
{\displaystyle\sum\limits_{\alpha\in A}}
\,g_{i}(\alpha)\,,\label{eq S}\\
T_{ij}^{AB}\,  &  \equiv\left\langle \!\right.  \hat{S}_{i}^{A}\hat{S}_{j}%
^{B}\left.  \!\right\rangle =\frac{\hbar^{2}}{4}\,\,%
{\displaystyle\sum\limits_{\alpha\in A}}
\,%
{\displaystyle\sum\limits_{\beta\in B}}
\,h_{ij}(\alpha,\beta)\,, \label{eq T}%
\end{align}
and analogously for $S_{i}^{B}\equiv\frac{\hbar}{2}\,{\sum\nolimits_{\beta\in
B}}\,g_{i}(\beta)$, where $S_{i}^{A},S_{i}^{B}\!\in\![-\frac{n\hbar}{2}%
,\frac{n\hbar}{2}]$, $T_{ij}^{AB}\!\in\![-\frac{n^{2}\hbar^{2}}{4},\frac
{n^{2}\hbar^{2}}{4}]$. These are only 15 numbers. Here $g_{i}(\alpha)$,
$g_{i}(\beta)$ and $h_{ij}(\alpha,\beta)$ are the (dimensionless) expectation
values and pair correlations of two single spins ($\alpha,\beta$) \textit{to
which it is assumed there is no experimental access}:
\begin{align}
g_{i}(\alpha)  &  \equiv\left\langle \!\right.  \hat{\sigma}_{i}^{(\alpha
)}\left.  \!\right\rangle _{\hat{\rho}_{\alpha\beta}},\\
h_{ij}(\alpha,\beta)  &  \equiv\left\langle \!\right.  \hat{\sigma}%
_{i}^{(\alpha)}\hat{\sigma}_{j}^{(\beta)}\left.  \!\right\rangle _{\hat{\rho
}_{\alpha\beta}}.
\end{align}
They are obtained from the actual $4\!\times\!4$ density matrix of the two
spins $\alpha$ and $\beta$:%
\begin{equation}
\hat{\rho}_{\alpha\beta}\equiv\frac{1}{4}\left[  \rule{0pt}{16pt}\!\right.
\leavevmode\hbox{\small1\kern-3.3pt\normalsize1}^{\!(\alpha)}\otimes
\leavevmode\hbox{\small1\kern-3.3pt\normalsize1}^{\!(\beta)}+%
{\displaystyle\sum\limits_{k=1}^{3}}
\,g_{k}(\alpha)\,\hat{\sigma}_{k}^{(\alpha)}\otimes
\leavevmode\hbox{\small1\kern-3.3pt\normalsize1}^{\!(\beta)}+%
{\displaystyle\sum\limits_{l=1}^{3}}
\,\leavevmode\hbox{\small1\kern-3.3pt\normalsize1}^{\!(\alpha)}\otimes
g_{l}(\beta)\,\hat{\sigma}_{l}^{(\beta)}+%
{\displaystyle\sum\limits_{k,l=1}^{3}}
h_{kl}(\alpha,\beta)\,\hat{\sigma}_{k}^{(\alpha)}\otimes\hat{\sigma}%
_{l}^{(\beta)}\left.  \rule{0pt}{16pt}\!\right]  \label{eq rhoalphabeta}%
\end{equation}
with $\leavevmode\hbox{\small1\kern-3.3pt\normalsize1}^{\!(\alpha)}$
($\leavevmode\hbox{\small1\kern-3.3pt\normalsize1}^{\!(\beta)}$) the
$2\!\times\!2$ identity matrix in the Hilbert space of spin $\alpha$ ($\beta$).

Out of the experimentally accessible quantities (\ref{eq S}) and (\ref{eq T})
we will construct a $4\!\times\!4$ density matrix of \textit{two virtual
qubits} which describes the collective properties of the two spin ensembles.
The \textit{a priori} justification for this method is:

\begin{enumerate}
\item A general treatment of the problem between two large samples of spins is
intractable because of the high dimensionality.\vspace{-0.2cm}

\item We have a fully developed theory of entanglement for two-qubit systems.
Therefore, this approach is a natural way to say something about the
entanglement between two spin systems if only collective observables are measured.
\end{enumerate}

We first introduce the normalized (dimensionless) average subsystem
expectation values (magnetization per particle) and correlations:
\begin{align}
s_{i}^{a}  &  \equiv\frac{1}{n}\,\,%
{\displaystyle\sum\limits_{\alpha\in A}}
\,g_{i}(\alpha)=\frac{2}{n\hbar}\,S_{i}^{A}\,,\label{eq si}\\
t_{ij}^{ab}  &  \equiv\frac{1}{n^{2}}\,\,%
{\displaystyle\sum\limits_{\alpha\in A}}
\,%
{\displaystyle\sum\limits_{\beta\in B}}
\,h_{ij}(\alpha,\beta)=\frac{4}{n^{2}\hbar^{2}}\,T_{ij}^{AB}\,, \label{eq tij}%
\end{align}
where $s_{i}^{a},t_{ij}^{ab}\in\lbrack-1,1]$. These are the coefficients of
the virtual density matrix:
\begin{equation}
\fbox{$\;\;\hat{\rho}_{ab}\equiv\dfrac{1}{4}\left[  \rule{0pt}{16pt}\!\right.
\leavevmode\hbox{\small1\kern-3.3pt\normalsize1}^{\!a}\otimes
\leavevmode\hbox{\small1\kern-3.3pt\normalsize1}^{\!b}+%
{\displaystyle\sum\limits_{k=1}^{3}}
\,s_{k}^{a}\,\hat{\sigma}_{k}^{a}\otimes
\leavevmode\hbox{\small1\kern-3.3pt\normalsize1}^{\!b}+%
{\displaystyle\sum\limits_{l=1}^{3}}
\,\leavevmode\hbox{\small1\kern-3.3pt\normalsize1}^{\!a}\otimes s_{l}%
^{b}\,\hat{\sigma}_{l}^{b}+%
{\displaystyle\sum\limits_{k=1}^{3}}
\,%
{\displaystyle\sum\limits_{l=1}^{3}}
\,t_{kl}^{ab}\,\hat{\sigma}_{k}^{a}\otimes\hat{\sigma}_{l}^{b}\left.
\rule{0pt}{16pt}\!\right]  \;\;$} \label{eq rho12}%
\end{equation}
with $a$ denoting the first and $b$ the second virtual collective qubit,
associated with subsystems $A$ and $B$, respectively. Here,
$\leavevmode\hbox{\small1\kern-3.3pt\normalsize1}^{\!a}$,
$\leavevmode\hbox{\small1\kern-3.3pt\normalsize1}^{\!b}$, $\hat{\sigma}%
_{k}^{a}$ and $\hat{\sigma}_{l}^{b}$ are $2\!\times\!2$ identity and Pauli
matrices for the collective qubits $a$ and $b$.

The question is whether the density matrix (\ref{eq rho12}) is positive
semi-definite, i.e.\ whether it is a physical state of two qubits. The answer
is affirmative and the proof follows from the consideration of an equal-weight
statistical mixture of \textit{one} (virtual) qubit pair which can be in any
of the $n^{2}$ states $\hat{\rho}_{\alpha\beta}$. The density matrix of this
mixture is the mixture of density matrices of all possible pairs
($\alpha,\beta$):%
\begin{equation}
\hat{\rho}_{\text{mix}}=\frac{1}{n^{2}}\,\,{\sum\nolimits_{\alpha,\beta}%
}\,\hat{\rho}_{\alpha\beta}\,.
\end{equation}
It can easily be seen that%
\begin{equation}
\hat{\rho}_{\text{mix}}=\hat{\rho}_{ab}%
\end{equation}
for both are uniquely determined by the same expectations and correlations:
\begin{align}
\langle\hat{\sigma}_{i}^{(\alpha)}\rangle_{\hat{\rho}_{\text{mix}}}  &
=\langle\hat{\sigma}_{i}^{a}\rangle_{\hat{\rho}_{ab}}=s_{i}^{a}=\frac
{2}{n\hbar}\,S_{i}^{A}\,,\\
\langle\hat{\sigma}_{i}^{(\alpha)}\hat{\sigma}_{j}^{(\beta)}\rangle_{\hat
{\rho}_{\text{mix}}}  &  =\langle\hat{\sigma}_{i}^{a}\hat{\sigma}_{j}%
^{b}\rangle_{\hat{\rho}_{ab}}=t_{ij}^{ab}=\frac{4}{n^{2}\hbar^{2}}%
\,T_{ij}^{AB}\,.
\end{align}
Thus, $\hat{\rho}_{ab}$ is indeed a density matrix. Note that without the
normalizations as given in (\ref{eq si}) and (\ref{eq tij}), the method would
not work.

Encapsulated in the following two propositions, we relate the entanglement
properties of the virtual qubits to those of the spin samples.

\textit{Proposition 1}. \textit{For any entanglement measure }$E$\textit{ that
is convex on the set of density matrices the entanglement of the virtual
density matrix }$E_{ab}\equiv E(\hat{\rho}_{ab})$\textit{ is a lower bound for
the average entanglement between all pairs }$\bar{E}_{\alpha\beta}\equiv
\tfrac{1}{n^{2}}\,\sum\nolimits_{\alpha,\beta}\,E(\hat{\rho}_{\alpha\beta}%
)$\textit{.}

Proof: This is an immediate consequence of the convexity of $E$:%
\begin{equation}
\fbox{$\;\;E_{ab}\equiv E(\tfrac{1}{n^{2}}{%
{\textstyle\sum\nolimits_{\alpha,\beta}}
}\hat{\rho}_{\alpha\beta})\leq\dfrac{1}{n^{2}}\,{%
{\displaystyle\sum\nolimits_{\alpha,\beta}}
}E(\hat{\rho}_{\alpha\beta})\equiv\bar{E}_{\alpha\beta}\,.\;\;$}%
\end{equation}

Remarks: First, the result holds for entanglement measures that are convex. In
certain cases this is directly implied by the definition of the entanglement
measure for mixed states, which involves a convex roof $E(\hat{\rho}%
)\equiv\text{min}_{p_{i},\psi_{i}}\sum_{i}p_{i}\,E(|\psi_{i}\rangle\langle
\psi_{i}|)$, where the minimization is taken over those probabilities $p_{i}$
and pure states $|\psi_{i}\rangle$ that realize the density matrix $\hat{\rho
}=\sum_{i}p_{i}|\psi_{i}\rangle\langle\psi_{i}|$ and $E(|\psi_{i}%
\rangle\langle\psi_{i}|)$ is the entanglement measure of the pure state
$|\psi_{i}\rangle$. Second, the proposition implies that if $E_{ab}>0$ then at
least for one pair ($\alpha,\beta$) we must have $E(\hat{\rho}_{\alpha\beta
})>0$:%
\begin{equation}
\hat{\rho}_{ab}\;\text{entangled}\;\Rightarrow\;\exists(\alpha,\beta
)\text{:}\;\hat{\rho}_{\alpha\beta}\;\text{entangled\thinspace}.
\end{equation}
Thus, a non-zero value of $E_{ab}$ is a sufficient condition for entanglement
between the two samples $A$ and $B$. Third, the maximal pairwise
concurrence~\cite{Hill1997} for symmetric states is found to be $2/n$ and is
achieved for the W-state~\cite{Koas2000}. It is conjectured that this remains
valid also when the symmetry constraint is removed. This suggests $E_{ab}%
\leq\bar{E}_{\alpha\beta}\leq2/n$, if concurrence is used as an entanglement
measure. The existence of this upper bound can be seen as a consequence of the
monogamy of entanglement.

We refer to $E_{ab}$ as \textit{pairwise collective entanglement} as it is
determined solely by the expectation and correlation values of the collective
spin observables. The question arises: Under what conditions is $E_{ab}$
\textit{equal} to the average entanglement $\bar{E}_{\alpha\beta}$?
Identifying systems for which the equality holds, would allow feasible
experimental determination of the entanglement distribution in large samples
by observation of their macroscopic properties only. It can easily be seen
that, if the state is symmetric under exchange of particles \textit{within}
each of the samples, one has $E_{ab}=\bar{E}_{\alpha\beta}=E(\hat{\rho
}_{\alpha\beta})$ for every pair of particles $(\alpha,\beta)$. In what
follows, we identify an important class of systems for which $E_{ab}%
\!=\!\bar{E}_{\alpha\beta}$ though the corresponding states need \textit{not}
be symmetric under exchange of particles.

\textit{Proposition 2}. Consider a system (i) with $g_{z}(\alpha)=g_{z}^{A}$
for all $\alpha\in A$ and $g_{z}(\beta)=g_{z}^{B}$ for all $\beta\in B$
(translational invariance within the subsystems), (ii) with $h_{xx}%
(\alpha,\beta)=\varepsilon\,h_{yy}(\alpha,\beta)$ with $\varepsilon
=\;$const$\;=+1$ or $-1$ (all pairs are in absolute value equally correlated
in the $x$ and $y$-direction, where this correlation may be different in size
for different pairs), (iii) with constant sign of the $z$-correlations,
i.e.\ sgn$[h_{zz}(\alpha,\beta)]=-\varepsilon$ for all $(\alpha,\beta)$, and
(iv) where all the remaining expectation values and correlations ($g_{x}$,
$g_{y}$, $h_{ij}$ with $i\neq j$) are zero. \textit{Non-vanishing average
entanglement }$\bar{E}_{\alpha\beta}$\textit{ as measured by the negativity is
equal to the pairwise collective entanglement }$E_{ab}$\textit{, if and only
if the correlation functions }$h_{xx}(\alpha,\beta)$ and $h_{yy}(\alpha
,\beta)$\textit{ are each constant for all pairs and all pairs have a
non-positive eigenvalue of their partial transposed density matrix.}

Proof: The negativity~\cite{Zycz1998,Vida2002} of a density matrix $\hat{\rho
}$ is defined as%
\begin{equation}
E(\hat{\rho})\equiv\frac{\text{Tr}|\hat{\rho}^{\text{pT}}|-1}{2}\,,
\end{equation}
where $\text{Tr}|\hat{\rho}^{\text{pT}}|$ stands for the trace norm of the
partially transposed density matrix $\hat{\rho}^{\text{pT}}$. Hence the
negativity is equal to the modulus of the sum of the negative eigenvalues of
$\rho^{\text{pT}}$.

It is important to stress that proposition 2 holds also for states that do not
need to be totally symmetric, i.e.\ the $h_{zz}(\alpha,\beta)$ may be
different for different pairs of particles. In general, under the above
symmetry, the state of the virtual qubit pair is of the form%
\begin{equation}
\hat{\rho}_{ab}={\small
\begin{pmatrix}
u_{+++} & 0 & 0 & v_{-}\\
0 & u_{+--} & v_{+} & 0\\
0 & v_{+} & u_{-+-} & 0\\
v_{-} & 0 & 0 & u_{--+}%
\end{pmatrix}
\!,}%
\end{equation}
where $u_{\pm\pm\pm}\equiv\frac{1}{4}\,(1\pm s_{z}^{A}\pm s_{z}^{B}\pm
t_{zz}^{ab})$, $v_{\pm}\equiv\frac{1}{4}\,t_{xx}^{ab}\,(1\pm\varepsilon)$. The
state of an arbitrary pair $(\alpha,\beta)$ of particles has a similar
structure. For example, the state of an arbitrary pair of particles extracted
from a spin chain with $xxz$ Heisenberg interaction has such a form.

Depending on the sign $\varepsilon$ of the $zz$-correlations, only one
eigenvalue of $\hat{\rho}_{\alpha\beta}^{\text{pT}}$ and $\hat{\rho}%
_{ab}^{\text{pT}}$, respectively, can be negative:%
\begin{align}
\mu_{\alpha\beta}  &  =\dfrac{1}{4}\left[  1-\sqrt{(g_{z}^{A}+\varepsilon
\,g_{z}^{B})^{2}+4\,h_{xx}^{2}(\alpha,\beta)}+\varepsilon\,h_{zz}(\alpha
,\beta)\right]  \!,\\
\nu_{ab}  &  =\dfrac{1}{4}\left[  1-\sqrt{(s_{z}^{a}+\varepsilon\,s_{z}%
^{b})^{2}+4\,(t_{xx}^{ab})^{2}}+\varepsilon\,t_{zz}^{ab}\right]  \!.
\label{eq evn1}%
\end{align}
The corresponding negativities are given by
\begin{align}
E(\hat{\rho}_{\alpha\beta})  &  =|\!\min(0,\mu_{\alpha\beta})|\,,\\
E_{ab}  &  =|\!\min(0,\nu_{ab})|\,.
\end{align}
One can express $\nu_{ab}$ as given by%
\begin{equation}
\nu_{ab}=\bar{\mu}+\Delta\,,
\end{equation}
where%
\begin{equation}
\bar{\mu}\equiv\frac{1}{n^{2}}\,{\sum\nolimits_{\alpha,\beta}}\,\mu
_{\alpha\beta}%
\end{equation}
and%
\begin{equation}
\Delta\equiv\frac{1}{4n^{2}}\,{\sum\nolimits_{\alpha,\beta}}\,\sqrt{(g_{z}%
^{A}+\varepsilon\,g_{z}^{B})^{2}+4\,h_{xx}^{2}(\alpha,\beta)}-\dfrac{1}%
{4}\,\sqrt{(s_{z}^{a}+\varepsilon\,s_{z}^{b})^{2}+4\,(t_{xx}^{ab})^{2}}\,.
\end{equation}
The quantity $\Delta$ is the difference between the entanglement measures
$\bar{E}_{\alpha\beta}$ and $E_{ab}$, i.e.\ $E_{ab}=\bar{E}_{\alpha\beta
}-\Delta$, for the case that $\nu_{ab}\leq0$ and $\bar{E}_{\alpha\beta}%
\equiv\tfrac{1}{n^{2}}\,{\sum\nolimits_{\alpha,\beta}}\,|\!\min(0,\mu
_{\alpha\beta})|=|\!\min(0,\bar{\mu})|$. This is true, if and only if
$\mu_{\alpha\beta}\leq0$ for all ($\alpha,\beta$), i.e.\ all pairs are either
entangled or have eigenvalue zero. According to proposition 1, $\Delta$ is
non-negative, i.e.%
\begin{equation}
\sqrt{c^{2}+4\,(t_{xx}^{ab})^{2}}\leq\dfrac{1}{n^{2}}\,%
{\displaystyle\sum\nolimits_{\alpha,\beta}}
\sqrt{c^{2}+4\,h_{xx}^{2}(\alpha,\beta)}\,. \label{eq red1}%
\end{equation}
Here we abbreviated $c\equiv g_{z}^{A}+\varepsilon\,g_{z}^{B}=s_{z}%
^{a}+\varepsilon\,s_{z}^{b}$, where the latter equal sign is due to
(\ref{eq si}). Inequality (\ref{eq red1}) becomes an equality, i.e.\ $\Delta
=0$, if and only if $h_{xx}(\alpha,\beta)$ is the same for all pairs
$(\alpha,\beta)$ such that $t_{xx}^{ab}=h_{xx}$. Therefore, the pairwise
collective entanglement $E_{ab}$ equals the average entanglement $\bar
{E}_{\alpha\beta}$, if and only if for all individual pairs $\mu_{\alpha\beta
}\leq0$ and $h_{xx}(\alpha,\beta)=\varepsilon\,h_{yy}(\alpha,\beta)=\;$const
for all pairs. $\square$

We illustrate the method with some explicit examples:

\subsubsection*{\textit{1. Dicke states}}

We consider the Dicke state (generalized W-state)%
\begin{equation}
\left\vert N;k\right\rangle \equiv\left(
\genfrac{}{}{0pt}{1}{N}{k}%
\right)  ^{\!-1/2}\,\hat{P}_{S}\,|\underbrace{0...0}_{N-k}\underbrace
{1...1}_{k}\rangle\label{eq W state}%
\end{equation}
with $N\geq2$ spins, $k$ excitations $|1\rangle$ and $N-k$ non-excited spins
$|0\rangle$, where $0\leq k\leq N$. $\hat{P}_{S}$ is the symmetrization
operator. Here, we identify the collective spin operators $\hat{S}_{i}$,
eq.~(\ref{eq SiA}), which act on $2^{n}$ dimensional Hilbert spaces, with spin
operators $\hat{S}_{i}$ acting only on the $2s+1=n+1$ dimensions of the
(symmetric) Dicke states.

Within the system we consider two subsystems $A$ and $B$ each of size $n$.
Because of the total symmetry of the state one has $E_{ab}=\bar{E}%
_{\alpha\beta}=E(\hat{\rho}_{\alpha\beta})$ for any size $n$. Only for the
cases where just a single spin ($k=0$) or all spins are excited ($k=N$), there
is no entanglement between two arbitrary pairs or arbitrary sized blocks,
respectively~\cite{Vedr2004}.

\begin{figure}[t]
\begin{center}
\includegraphics{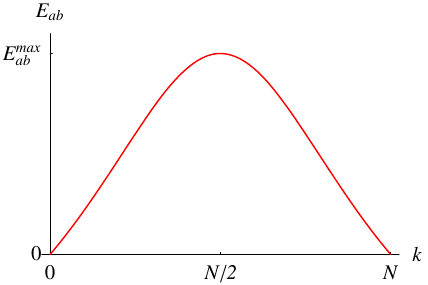}
\end{center}
\par
\vspace{-0.25cm}\caption{Entanglement $E_{ab}$ between two arbitrary sized
spin subensembles $A$ and $B$, where the system is in a Dicke state
(\ref{eq W state}) of $N$ particles with $k$ excitations. The entanglement is
computed as the negativity of two virtual qubits $a$ and $b$, representing the
subensembles. It is non-zero except for the cases where not a single spin or
where all spins are excited. The maximum entanglement $E_{ab}^{\text{max}%
}=\frac{1}{2}\,\frac{1}{N-1}$ is reached for $k=N/2$.}%
\label{Figure_Spin_W_state}%
\end{figure}

The reduced two-qubit density matrix has the form%
\begin{equation}
\hat{\rho}_{\alpha\beta}(N;k)=d\,|00\rangle\!\left\langle 00\right\vert
+e\,|11\rangle\!\left\langle 11\right\vert +2f\,|\psi^{+}\rangle\!\left\langle
\psi^{+}\right\vert ,
\end{equation}
where $|\psi^{+}\rangle\equiv\frac{1}{\sqrt{2}}(|01\rangle+|10\rangle)$ and%
\begin{equation}
d=\dfrac{(N-k)\,(N-k-1)}{N\,(N-1)}\,,\quad e=\dfrac{k\,(k-1)}{N\,(N-1)}%
\,,\quad f=\dfrac{k\,(N-k)}{N\,(N-1)}\,,
\end{equation}
with $d+e+2f=1$. The reduced density matrix $\hat{\rho}_{\alpha\beta}$ is the
same for all pairs of spins ($\alpha,\beta$), independent of their position
and distance from each other. We need to calculate the expectation values of
the spin operator and correlations between the two sites $\alpha$ and $\beta$.
They are independent of $\alpha$ and $\beta$:%
\begin{equation}
g_{z}=e-d\,,\quad h_{xx}=h_{yy}=2f\,,\quad h_{zz}=d+e-2f\,.
\end{equation}
All the others ($g_{x},g_{y},h_{ij}$ with $i\neq j)$ are zero. This is a
Heisenberg $xxz$ type situation, for which all our proofs above hold. The
corresponding normalized collective values of two $n$-particle blocks $A$ and
$B$, (\ref{eq si}) and (\ref{eq tij}), are%
\begin{align}
s_{i}  &  =\frac{1}{n}\,S_{i}^{A,B}=\frac{1}{n}\,\,%
{\displaystyle\sum\limits_{\alpha\in A}^{n}}
\,g_{i}=g_{i}\,,\\
t_{ij}  &  =\frac{1}{n^{2}}\,T_{ij}^{AB}=\frac{1}{n^{2}}\,\,%
{\displaystyle\sum\limits_{\alpha\in A}^{n}}
\,\,%
{\displaystyle\sum\limits_{\beta\in B}^{n}}
\,h_{ij}=h_{ij}\,,
\end{align}
i.e.\ the same as the actual values for two arbitrary individual spins. This
means that for the W-state the averages are the values themselves, since there
is no dependence at all on the position of the spins and therefore on the
distance between them. Therefore, the collective block entanglement (for all
block sizes $n\leq\frac{N}{2}$) equals the entanglement between two
(arbitrary) spins:%
\begin{equation}
E_{ab}=E(\hat{\rho}_{\alpha\beta})\,.
\end{equation}
The corresponding (possibly negative) eigenvalues of the partial transposed
matrices are%
\begin{equation}
\nu_{ab}=\mu_{\alpha\beta}=\tfrac{1}{2}\,(d+e)-\tfrac{1}{2}\,\sqrt
{(e-d)^{2}+4f^{2}}\,.
\end{equation}
The entanglement%
\begin{equation}
E_{ab}=|\nu_{ab}|
\end{equation}
is non-zero for all excitations $1\leq k\leq N-1$. Only for the cases where
not a single spin or where all spins are excited, $E_{ab}=0$ and there is no
entanglement between two arbitrary pairs or arbitrary sized blocks,
respectively. The global maximum of the entanglement is reached for
$k=\frac{N}{2}$, which is the case where half of the spins are excited ($N$
should be even here) and its value is%
\begin{equation}
E_{ab}^{\text{max}}\equiv E_{ab}(k\!=\!\tfrac{N}{2})=\frac{1}{2}\,\frac
{1}{N-1}\,,
\end{equation}
for all block sizes $n$. It vanishes in the limit of an infinite chain
$N\rightarrow\infty$. Figure~\ref{Figure_Spin_W_state} shows the block
entanglement $E_{ab}$ as a function of the number of excitations
$k$.

\subsubsection*{\textit{2. Generalized singlet states}}

The two subsystems $A$ and $B$, each forming a spin $s=\frac{n}{2}$, are in a
generalized singlet state:%
\begin{equation}
\left\vert \psi\right\rangle =\frac{1}{\sqrt{2s+1}}\,\,%
{\displaystyle\sum\limits_{m=-s}^{s}}
(-1)^{s-m}\left\vert m\right\rangle _{A}\left\vert -m\right\rangle _{B},
\label{eq general singlet}%
\end{equation}
where $\left\vert m\right\rangle =\left\vert 2s;s+m\right\rangle $\ denotes
the eigenstates of the spin operator's $z$-component. The collective two-qubit
coefficients are $t_{ii}^{ab}=-\frac{n+2}{3n}$ and the $s_{i}^{a,b}$ and
$t_{ij}^{ab}$ with $i\neq j$ are all zero. The eigenvalue (\ref{eq evn1}) of
$\hat{\rho}_{ab}^{\text{pT}}$ becomes $\nu_{ab}=-\frac{1}{2n}$, and the
collective entanglement (negativity) is
\begin{equation}
E_{ab}=\frac{1}{2n}=\frac{1}{4s}\,.
\end{equation}
It is non-zero for all sizes $n$ of the subsystems and vanishes only in the
limit $n\rightarrow\infty$. This agrees with the statements in
References~\cite{Merm1980,Pere1995} that the generalized singlet state
(\ref{eq general singlet}) can violate a Bell inequality no matter how large
the spin is.

\subsubsection*{\textit{3. Generalized singlet state with an admixture of
non-symmetric correlations}}

\begin{figure}[t]
\begin{center}
\includegraphics{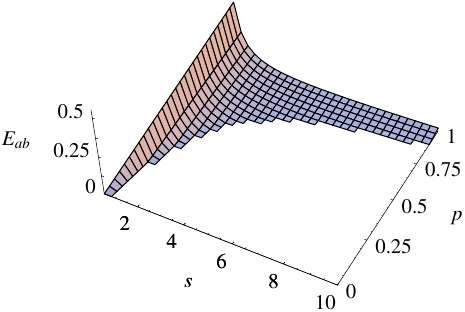}
\end{center}
\par
\vspace{-0.25cm}\caption{Entanglement $E_{ab}$ between two collective spins in
the generalized singlet state with an admixture of non-symmetric noise
(\ref{eq singlet plus z-corr}) as a function of spin length $s$ and proportion
$p$ of the singlet state in the mixture. The entanglement is non-zero for
sufficiently large $p$ and decreases inversely proportionally to $s$.}%
\label{Figure_Spin_singlet_state}%
\end{figure}

Consider the state%
\begin{equation}
\left\vert \psi\right\rangle _{p}=p\left\vert \psi\right\rangle \!\left\langle
\psi\right\vert +(1-p)\,\,%
{\displaystyle\bigotimes\limits_{\alpha=1}^{n}}
\,\,\dfrac{1}{2}\left(  \left\vert 01\right\rangle _{\alpha,\beta=\alpha
}\!\left\langle 10\right\vert +\left\vert 10\right\rangle _{\alpha
,\beta=\alpha}\!\left\langle 01\right\vert \right)
\label{eq singlet plus z-corr}%
\end{equation}
with $p\in\lbrack0,1]$. This is a mixture of the generalized singlet state
(\ref{eq general singlet}) and $n$ perfectly $z$-correlated pairs
($\alpha,\beta\!=\!\alpha$). This state is \textit{not} symmetric under
particle exchange in the $zz$-correlations. The expectation values
$s_{i}^{a,b}$ and correlations $t_{ij}^{ab}$ with $i\neq j$ remain zero. The
correlations $t_{xx}=t_{yy}$ are reduced by a factor $p$ compared to those of
the state (\ref{eq general singlet}). The correlations in $z$-direction,
however, are modified and read $t_{zz}^{ab}=-p\,\frac{n-1}{3n}-\frac{1}{n}$.
Therefore, there is a critical number of particles $n_{c}\equiv\left\lceil
\!\right.  \frac{1+p}{1-p}\left.  \!\right\rceil $, with $\!\left\lceil
\!\right.  .\left.  \!\right\rceil $ the ceiling function, beyond which there
is no collective pairwise entanglement. Only for $n<n_{c}$ we have positive%
\begin{equation}
E_{ab}=\frac{1+p-n\,(1-p)}{4n}\,.
\end{equation}
Note that (\ref{eq singlet plus z-corr}) is in accordance with proposition 2
and thus $E_{ab}=\bar{E}_{\alpha\beta}$.
Figure~\ref{Figure_Spin_singlet_state}\ shows $E_{ab}$ as a function of the
spin length $s=\frac{n}{2}$ and the mixing parameter $p$. $E_{ab}$ is non-zero
in regions where $p>\frac{2s-1}{2s+1}$ and decreases inversely proportionally
to $s$.

\subsection{Multi-partite entanglement}

\begin{figure}[t]
\begin{center}
\includegraphics{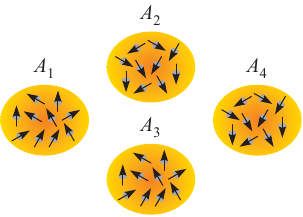}
\end{center}
\par
\vspace{-0.25cm}\caption{Schematic of a multi-partite scenario. There are $M$
separated spin subensembles $A_{1},...,A_{M}$, each containing a large number
of individual spins.}%
\label{Figure_Spin_multipartite}%
\end{figure}

Our method can be generalized to define multi-partite entanglement of $M$
collective spins belonging to $M$ separated samples $A_{1},...,A_{M}$, each
containing a large number of spins $n$ (Figure~\ref{Figure_Spin_multipartite}%
). The collective spin of subsystem $A_{p}$ is%
\begin{equation}
\hat{S}_{i_{p}}^{A_{p}}\equiv\frac{\hbar}{2}\,\,%
{\displaystyle\sum\limits_{\alpha_{p}\in A_{p}}}
\hat{\sigma}_{i_{p}}^{(\alpha_{p})}.
\end{equation}
Here $p=1,...,M$ and $i_{p}\in\{0,x\!\equiv\!1,y\!\equiv
\!2,z\!\equiv\!3\}$ and $\hat{\sigma}_{0}^{(\alpha_{p})}%
=\leavevmode\hbox{\small1\kern-3.3pt\normalsize1}$\ denotes the $2\!\times\!2$
identity matrix. We suppose again that the experimenter measures the
collective spin components (where all $i_{p}$ but one are zero) in each sample
and their (higher-order) correlations (where more than one $i_{p}$ is unequal
to zero) $T_{i_{1}...i_{M}}^{A_{1}...A_{M}}\,\equiv\left\langle \!\right.
\hat{S}_{i_{1}}^{A_{1}}...\hat{S}_{i_{M}}^{A_{M}}\left.  \!\right\rangle $:%
\begin{equation}
T_{i_{1}...i_{M}}^{A_{1}...A_{M}}\,=\frac{\hbar^{M}}{2^{M}}\,\,%
{\displaystyle\sum\limits_{\alpha_{1}\in A_{1}}}
...%
{\displaystyle\sum\limits_{a_{M}\in A_{M}}}
h_{i_{1}...i_{M}}(\alpha_{1},...,\alpha_{M})\,,
\end{equation}
where%
\begin{equation}
h_{i_{1}...i_{M}}(\alpha_{1},...,\alpha_{M})\equiv\left\langle \!\right.
\hat{\sigma}_{i_{1}}^{(\alpha_{1})}...\hat{\sigma}_{i_{M}}^{(\alpha_{M}%
)}\left.  \!\right\rangle _{\hat{\rho}_{\alpha_{1}...\alpha_{M}}}%
\end{equation}
denotes the actual spin expectation values and higher-order correlations in
the physical system. These are $4^{M}-1$ numbers. The virtual correlations are
denoted as%
\begin{equation}
t_{i_{1}...i_{M}}^{a_{1}...a_{M}}\equiv\frac{2^{M}}{(n\hbar)^{M}}%
\,T_{i_{1}...i_{M}}^{A_{1}...A_{M}}\,. \label{eq ti1iM}%
\end{equation}
Note that $\hat{S}_{0}^{A_{p}}%
=n\,\leavevmode\hbox{\small1\kern-3.3pt\normalsize1}$ is $n$ times the
$2\!\times\!2$ identity matrix. A normalized $p$-particle correlation
$t_{i_{1}...i_{M}}^{a_{1}...a_{M}}$ with $p\leq M$ indeed scales with $n^{-p}$
as $p$ ($M\!-\!p$) is the number of subscripts unequal (equal) to zero. The
virtual $2^{M}\!\times\!2^{M}$ collective $M$-qubit density matrix is given by%
\begin{equation}
\fbox{$\;\;\hat{\rho}_{a_{1}...a_{M}}\equiv\dfrac{1}{2^{M}}\,\,%
{\displaystyle\sum\limits_{i_{1}=0}^{3}}
...%
{\displaystyle\sum\limits_{i_{M}=0}^{3}}
\,t_{i_{1}...i_{M}}^{a_{1}...a_{M}}\,\hat{\sigma}_{i_{1}}^{a_{1}}%
\otimes...\otimes\hat{\sigma}_{i_{M}}^{a_{M}}\,,\;\;$}%
\end{equation}
with $\hat{\sigma}_{i_{p}}^{a_{p}}$ denoting the $2\!\times\!2$ Pauli or
identity matrix of the $p$-th virtual collective qubit, associated with the
$n$-particle subsystem $A_{p}$.

We show that the virtual matrix $\hat{\rho}_{a_{1}...a_{M}}$ is a density
matrix. Analogously to the two-subsystems case, we consider an equal-weight
statistical mixture of \textit{one} $m$-tuple of qubits ($\alpha
_{1},...,\alpha_{M}$) which can be in any of the $n^{M}$ (mixed) states
$\hat{\rho}_{\alpha_{1}...\alpha_{M}}$. The density matrix of this mixture is
the mixture of density matrices of all possible $m$-tuples:%
\[
\hat{\rho}_{\text{mix}}=\frac{1}{n^{M}}\,{\sum\nolimits_{\alpha_{1}%
,...,\alpha_{m}}}\,\hat{\rho}_{\alpha_{1}...\alpha_{M}}\,.
\]
We have%
\[
\hat{\rho}_{\text{mix}}=\hat{\rho}_{a_{1}...a_{M}}%
\]
as both give the same expectations and correlations:
\[
\langle\hat{\sigma}_{i_{1}}^{(\alpha_{1})}...\hat{\sigma}_{i_{M}}^{(\alpha
_{M})}\rangle_{\hat{\rho}_{\text{mix}}}=\langle\hat{\sigma}_{i_{1}}^{a_{1}%
}...\hat{\sigma}_{i_{M}}^{a_{M}}\rangle_{\hat{\rho}_{a_{1}...a_{M}}}%
=t_{i_{1}...i_{M}}^{a_{1}...a_{M}}%
\]

Hence, analogously to proposition 1, any convex multi-partite entanglement
measure (e.g., $M$-way tangle) which is applied to the collective matrix,
$E(\hat{\rho}_{a_{1}...a_{M}})$, gives a \textit{lower bound for the average
multi-partite entanglement} $\bar{E}_{\alpha_{1}...\alpha_{M}}$. For states
which are symmetric within the samples equality holds: $E(\hat{\rho}%
_{a_{1}...a_{M}})=\bar{E}_{\alpha_{1}...\alpha_{M}}$. For totally symmetric
states this is also equal to the multi-particle entanglement of $M$ extracted
particles from the system. Since being a legitimate entanglement measure, the
collective multi-partite entanglement $E(\hat{\rho}_{a_{1}...a_{M}})$ obeys
the usual constraints for entanglement sharing such as the
Coffman-Kundu-Wootters inequality \cite{Coff2000}.

Importantly, in the examples considered above our entanglement measure scales
at most with $1/n$ and vanishes in the limit of infinitely large subsystem
sizes $n$. This is a generic property that follows from the commutation
relation for \textit{normalized} spins in this limit. Taking $\hat{s}%
_{i}\equiv\frac{1}{n}\,\hat{S}_{i}=\frac{\hbar}{2n}\sum_{\alpha}\hat{\sigma
}_{i}^{(\alpha)}$ one obtains
\begin{equation}
\lim_{n\rightarrow\infty}[\hat{s}_{x},\hat{s}_{y}]=\lim_{n\rightarrow\infty
}\text{i}\,\frac{\hbar}{2n}\,\hat{s}_{z}=0\,.
\end{equation}
This is sometimes interpreted as suggesting that averaged collective
observables, like the magnetization per particle, represent \textquotedblleft
macroscopic\textquotedblright\ or classical-like, properties of samples. Note,
however, that for any $n$ there are $n^{2}$ pairs between the subsystems so
that the number of pairs multiplied by the pairwise collective entanglement
can scale with $n$, showing the existence of entanglement for arbitrarily
large $n$.

\chapter{Mathematical undecidability and quantum randomness}

\textbf{Summary:}\bigskip

The mathematics of the early twentieth century was concerned with the question
whether a complete and consistent set of axioms for all of mathematics is
conceivable~\cite{Hilb1935}. In 1931 Gödel showed that this is fundamentally
impossible~\cite{Goed1931}. In every consistent axiomatic system that is
capable of expressing elementary arithmetic there are propositions which can
neither be proved nor disproved within the system, i.e.\ they are
\textquotedblleft undecidable\textquotedblright. Via the halting problem
Turing brought Gödel's mathematical proof into the world of physical
machines~\cite{Turi1936}. Chaitin went even one step further and argued that
mathematical undecidability is not bound to self-referential statements but
arises whenever a proposition to be proved and the axioms contain together
more information than the set of axioms itself~\cite{Chai1982}.

Here we propose a new link between mathematical undecidability and quantum
physics. We demonstrate that the states of elementary quantum systems are
capable of encoding mathematical axioms. Quantum mechanics imposes an upper
limit on how much information can be encoded in a quantum
state~\cite{Hole1973,Zeil1999}, thus limiting the information content of the
set of axioms. We show that quantum measurements are capable of revealing
whether a given proposition is decidable or not within this set. This allows
for an experimental test of mathematical undecidability by realizing in the
laboratory the actual quantum states and operations required. We demonstrate
experimentally both the encoding of axioms, using polarization states of
photons, and that the decidability of propositions can be checked by
performing suitable quantum measurements. We theoretically find and
experimentally confirm that whenever a mathematical proposition is undecidable
within the system of axioms encoded in the state, the measurement associated
with the proposition gives random outcomes. Our results support the view that
quantum randomness is irreducible~\cite{Calu2005} and a manifestation of
mathematical undecidability.

Despite its overwhelming success, quantum physics is still heavily debated for
its interpretation has remained to be unclear. One of the reasons is that, in
contrast to all other theories, it is lacking clear and unambiguous
foundational principles. The link with pure mathematics might provide such a
principle and can be seen as a novel approach towards a reconstruction of
quantum theory~\cite{Weiz2002}.\bigskip

\noindent This chapter mainly bases on and also uses parts of
Reference~\cite{Pate2008}:

\begin{itemize}
\item T. Paterek, R. Prevedel, J. Kofler, P. Klimek, M. Aspelmeyer, A.
Zeilinger, and \v{C}. Brukner\newline\textit{Mathematical undecidability and
quantum randomness}\newline Submitted (2008).\newpage
\end{itemize}

\section{Logical complementarity and mathematical undecidability}

We begin our argumentation following the idea of the information-theoretical
formulation of Gödel's theorem~\cite{Chai1982}: Given a set of axioms that
contains a certain amount of information, it is impossible to deduce the truth
value of a proposition which, together with the axioms, contains more
information than the set of axioms itself. To give an example, consider
Boolean functions of a single binary argument:%
\begin{equation}
x\in\{0,1\}\;\rightarrow\;y=f(x)\in\{0,1\}
\end{equation}
\begin{figure}[b]
\begin{center}
\includegraphics[width=.275\textwidth]{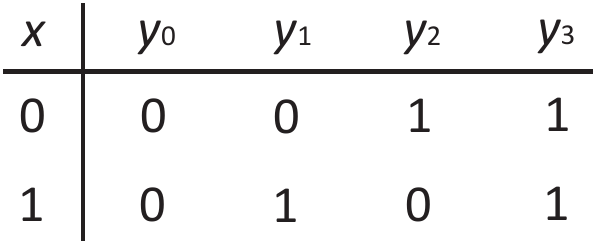}
\end{center}
\par
\vspace{-0.25cm}\caption{The four Boolean functions $y=f(x)$ of a binary
argument, i.e.\ $f(x)=0,1$ with $x=0,1$. The different functions are labeled
by $y_{k}$ with $k=0,1,2,3$.}%
\label{Figure_Undecidability_functions}%
\end{figure}There are four such functions, $y_{k}$ ($k=0,1,2,3$), shown in
Figure~\ref{Figure_Undecidability_functions}. We shall discuss the following
(binary) propositions about their properties:%
\begin{align*}
\text{(A)}\quad\text{\textquotedblleft The value of }f(0)\text{ is `0',
i.e.~}f(0)  &  =0\text{.\textquotedblright}\\
\text{(B)}\quad\text{\textquotedblleft The value of }f(1)\text{ is `0',
i.e.~}f(1)  &  =0\text{.\textquotedblright}%
\end{align*}
These two propositions are \textit{independent}. Knowing the truth value of
one of them does not allow to infer the truth value of the other. Ascribing
truth values to both propositions requires two bits of information. If one
postulates only proposition (A) to be true, i.e.\ if we choose (A) as an
\textquotedblleft axiom\textquotedblright, then it is impossible to prove
proposition (B) from (A). Having only axiom (A), i.e.\ only one bit of
information, there is not enough information to know also the truth value of
(B). Hence, proposition (B) is \textit{mathematically undecidable} within the
system containing the single axiom (A). Another example of an undecidable
proposition within the same axiomatic system is:%
\[
\text{(C)}\quad\text{\textquotedblleft The function is constant,
i.e.~}f(0)=f(1)\text{.\textquotedblright}%
\]
Again, this statement cannot be proved or disproved from the axiom (A) alone
because (C) is independent of (A) as it involves $f(1)$.

We refer to independent propositions to which one cannot simultaneously
ascribe definite truth values---given a limited amount of information
resources---as \textit{logically complementary propositions}. Knowing the
truth value of one of them (i.e.\ having the proposition itself or its
negation as an axiom) precludes any knowledge about the others. For example,
given the limitation to one bit of information, the three propositions (A),
(B) and (C) are logically complementary.

When the information content of the axioms and the number of independent
propositions increase, more possibilities arise. Already the case of two bits
as the information content is instructive. Consider two independent Boolean
functions $f_{1}(x)$ and $f_{2}(x)$ of a binary argument. The two bits can be
used to define truth values of properties of the individual functions or they
can define joint features of the functions. An example of the first type is
the following two-bit proposition:%
\begin{align*}
\text{(D)}\quad\text{``The value of }f_{1}(0)\text{ is `0', i.e.~}f_{1}(0)  &
=0\text{.''}\\
\text{``The value of }f_{2}(1)\text{ is `0', i.e.~}f_{2}(1)  &  =0\text{.''}%
\end{align*}
An example of the second type is:%
\begin{align*}
\text{(E)}\quad\text{``The functions have the same values for argument `0',
i.e.~}f_{1}(0)  &  =f_{2}(0)\text{.''}\\
\text{``The functions have the same values for argument `1', i.e.~}f_{1}(1)
&  =f_{2}(1)\text{.''}%
\end{align*}
Both (D) and (E) consist of two elementary (binary) propositions. Their truth
values are of the form of ``vectors''\ with two components being the truth
values of their elementary propositions. The propositions (D) and (E) are
logically complementary. Given (E) as a two-bit axiom, all the
\textit{individual} function values remain undefined and thus one can
determine neither of the two truth values of (D).

The new aspect of multi-bit axioms is the existence of ``partially''
undecidable propositions, containing more than one elementary proposition only
some of which are undecidable. An example of such a partially undecidable
proposition within the system consisting of the two-bit axiom (D) is:%
\begin{align*}
\text{(F)}\quad\text{``The value of }f_{1}(0)\text{ is `0', i.e.~}f_{1}(0)  &
=0\text{.''}\\
\text{``The value of }f_{2}(0)\text{ is `0', i.e.~}f_{2}(0)  &  =0\text{.''}%
\end{align*}
The first elementary proposition is the same as in (D) and thus it is
definitely true. The impossibility to decide the second elementary proposition
leads to (partial) undecidability of the proposition (F). In a similar way,
proposition (F) is partially undecidable within the axiomatic system of (E).

\section{Physical implementation and quantum randomness}

The discussion so far was purely \textit{mathematical}. We have described
axiomatic systems (of limited information content) using properties of Boolean
functions. Now we show that the undecidability of mathematical propositions
can be tested in quantum experiments. To this end we introduce a
\textit{physical} ``black box'' whose internal configuration encodes Boolean
functions. The black box hence forms a bridge between mathematics and physics.
Quantum systems enter it and the properties of the functions are written onto
the quantum states of the systems. Finally, measurements performed on the
systems extract information about the properties of the configuration of the
black box and thus about the properties of the functions.\begin{figure}[t]
\begin{center}
\includegraphics[width=.55\textwidth]{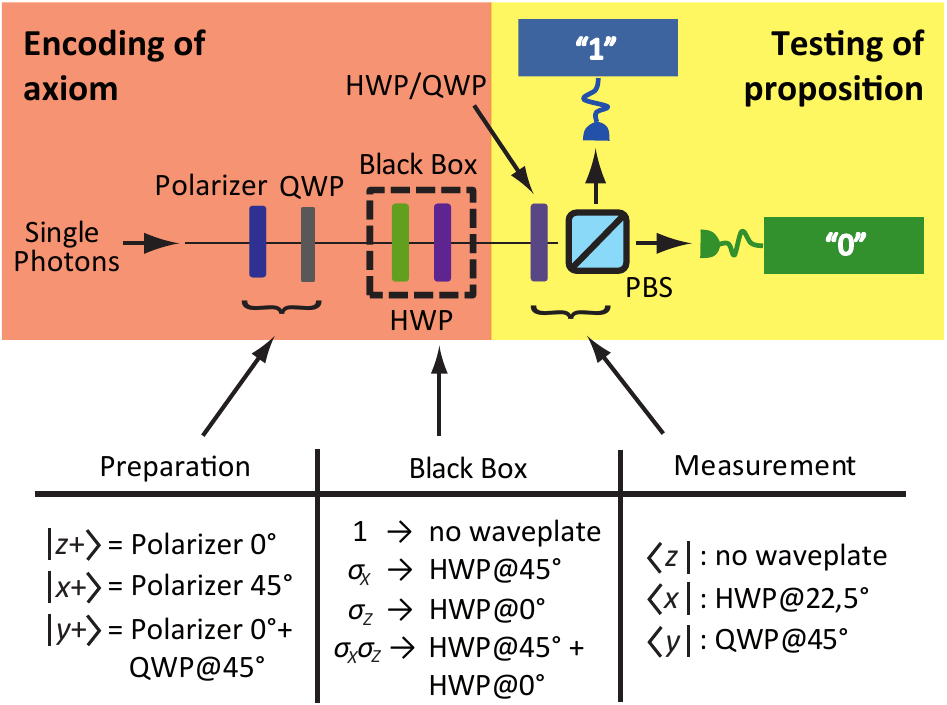}
\end{center}
\par
\vspace{-0.25cm}\caption{The experimental setup. We use the polarization of
single photons as information carriers of binary properties encoded by the
configuration in the ``black box''. The photons are initialized in a definite
polarization state, belonging to one of the three complementary bases $z,x,y$
(Preparation). The Boolean functions are realized within the black box by
inserting half-wave plates (HWPs) which implement the product of Pauli
operators $\hat{\sigma}_{x}^{f(0)}\,\hat{\sigma}_{z}^{f(1)}$, eq.\ (\ref{eq U}%
). The measurement apparatus consists of a quarter-wave plate (QWP) or HWP,
followed by a polarizing beam-splitter (PBS) and two fibre-coupled
single-photon detector modules. In this way, measurements in all complementary
bases ($z,x,y$) can be realized.}%
\label{Figure_Undecidability_schematic}%
\end{figure}

We begin with the simplest case of a spin-$\frac{1}{2}$ particle (qubit)
entering the black box (see Figure~\ref{Figure_Undecidability_schematic} for
the experimental implementation) and a single bit-to-bit function $f(x)$
implemented by the black box. Inside the black box two subsequent operations
alter the state of the input spin. The first operation encodes the value of
$f(1)$ via application of $\hat{\sigma}_{z}^{f(1)}$, i.e.\ the Pauli
$z$-operator taken to the power of $f(1)$. The second operation encodes $f(0)$
with $\hat{\sigma}_{x}^{f(0)}$, i.e.\ the Pauli $x$-operator taken to the
power of $f(0)$. The total action of the black box is%
\begin{equation}
\hat{U}=\hat{\sigma}_{x}^{f(0)}\,\hat{\sigma}_{z}^{f(1)}\,. \label{eq U}%
\end{equation}

Consider the input spin to be in one of the eigenstates of the Pauli operator
i$^{mn}\,\hat{\sigma}_{x}^{m}\,\hat{\sigma}_{z}^{n}$ (with i the imaginary
unit). The three particular choices $(m,n)=(0,1)$, $(1,0)$, or $(1,1)$
correspond to the three spin operators along orthogonal directions
$\hat{\sigma}_{z}$, $\hat{\sigma}_{x}$, or $\hat{\sigma}_{y}=\;$%
i$\,\hat{\sigma}_{x}\,\hat{\sigma}_{z}$, respectively. The input density
matrix reads%
\begin{equation}
\hat{\rho}=\tfrac{1}{2}%
\,[\leavevmode\hbox{\small1\kern-3.3pt\normalsize1}+\lambda_{mn}%
\,\text{i}^{mn}\,\hat{\sigma}_{x}^{m}\,\hat{\sigma}_{z}^{n}]\,,
\end{equation}
with $\lambda_{mn}=\pm1$ and
$\leavevmode\hbox{\small1\kern-3.3pt\normalsize1}$ the identity operator. It
evolves under the action of the black box to%
\begin{equation}
\hat{U}\hat{\rho}\,\hat{U}^{\dag}=\tfrac{1}{2}%
\,[\leavevmode\hbox{\small1\kern-3.3pt\normalsize1}+\lambda_{mn}%
\,(-1)^{nf(0)+mf(1)}\,\text{i}^{mn}\,\hat{\sigma}_{x}^{m}\,\hat{\sigma}%
_{z}^{n}]\,.
\end{equation}

Depending on the value of $n\,f(0)+m\,f(1)$ (in this chapter all sums are
taken modulo 2), the state after the black box is either the same or
orthogonal to the initial one. If one now performs a measurement in the basis
of the initial state, i.e.\ the eigenbasis of the operator i$^{mn}%
\,\hat{\sigma}_{x}^{m}\,\hat{\sigma}_{z}^{n}$, the outcome reveals the value
of $n\,f(0)+m\,f(1)$ and hence the measurement can be considered as
\textit{checking the truth value of the proposition}%
\[
\text{(G)}\quad\text{\textquotedblleft}%
n\,f(0)+m\,f(1)=0\text{.\textquotedblright}%
\]

Each of the three \textit{quantum complementary} measurements $\hat{\sigma
}_{z}$, $\hat{\sigma}_{x}$, or $\hat{\sigma}_{y}$ reveals the truth value of
one of the independent propositions (A), (B), or (C), respectively.

\textit{Independent} of the initial state, we now identify the quantum
measurement $(m,n)$ with the question about the truth value of the
corresponding mathematical proposition (G). The states that give a definite
answer in the quantum experiment encode (G) or its negation as an axiom. A
measurement \textit{quantum physically complementary} to the one identified
with (G) gives random results for those states, and the corresponding
\textit{logically complementary} proposition is undecidable within the one-bit axiom.

\begin{figure}[t]
\begin{center}
\includegraphics{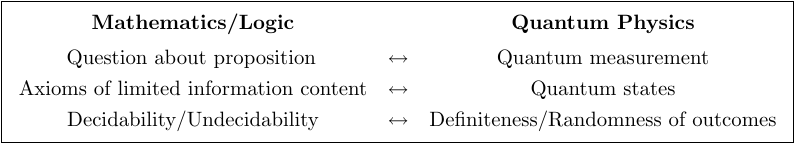}
\end{center}
\par
\vspace{-0.25cm}\caption{The link between mathematical undecidability and
quantum randomness.}%
\label{Figure_Undecidability_table}%
\end{figure}

We will show in general that \textit{whenever} the proposition identified with
the measurement is decidable (in the axiomatic system encoded by the state
after the black box), the measurement outcome is definite, and whenever it is
undecidable, the measurement outcome is random. This links mathematical
undecidability and quantum randomness and allows to experimentally find out
whether a proposition is decidable or not. The essence of this chapter is
summarized in Figure~\ref{Figure_Undecidability_table}.

A natural explanation of \textit{irreducible} (objective) randomness of
individual quantum outcomes arises. After leaving the black box the spin's
quantum state encodes exactly one bit of information about $f(x)$, namely the
truth value of the proposition (G). \textit{In mathematical language, the
system encodes a one-bit axiom.} One bit is the maximum amount of information
which can be carried by a single spin-$\frac{1}{2}$
particle~\cite{Hole1973,Zeil1999}. Thus, if this bit represents the truth
value of, say, proposition (A), therefore defining (A) or its negation as an
axiom, the other two logically complementary propositions, (B) and (C), are
undecidable. This is because there is no information left for specifying their
truth values. However, the spin can nevertheless be measured in the bases
corresponding to (B) or (C) and---as in any measurement---will inevitably give
an outcome, e.g.\ a click in a detector. The clicks must not contain any
information whatsoever about the truth of the undecidable proposition.
Therefore, the individual quantum outcome must be random, reconciling
mathematical undecidability with the fact that a quantum system always gives
an ``answer'' when ``asked'' in an experiment. This provides an intuitive
understanding of quantum randomness, a key quantum feature, using mathematical reasoning.

To find out whether a proposition is undecidable, it is necessary to repeat an
experiment sufficiently many times, such that---even in the presence of
unavoidable experimental imperfections---the two possible different outcomes
occur significantly often. To reveal decidability, only one outcome has to
appear again and again, but sufficiently many repetitions of the experiment
are needed, because in practice the other outcome sometimes occurs due to
experimental errors.\begin{figure}[t]
\begin{center}
\includegraphics[width=.45\textwidth]{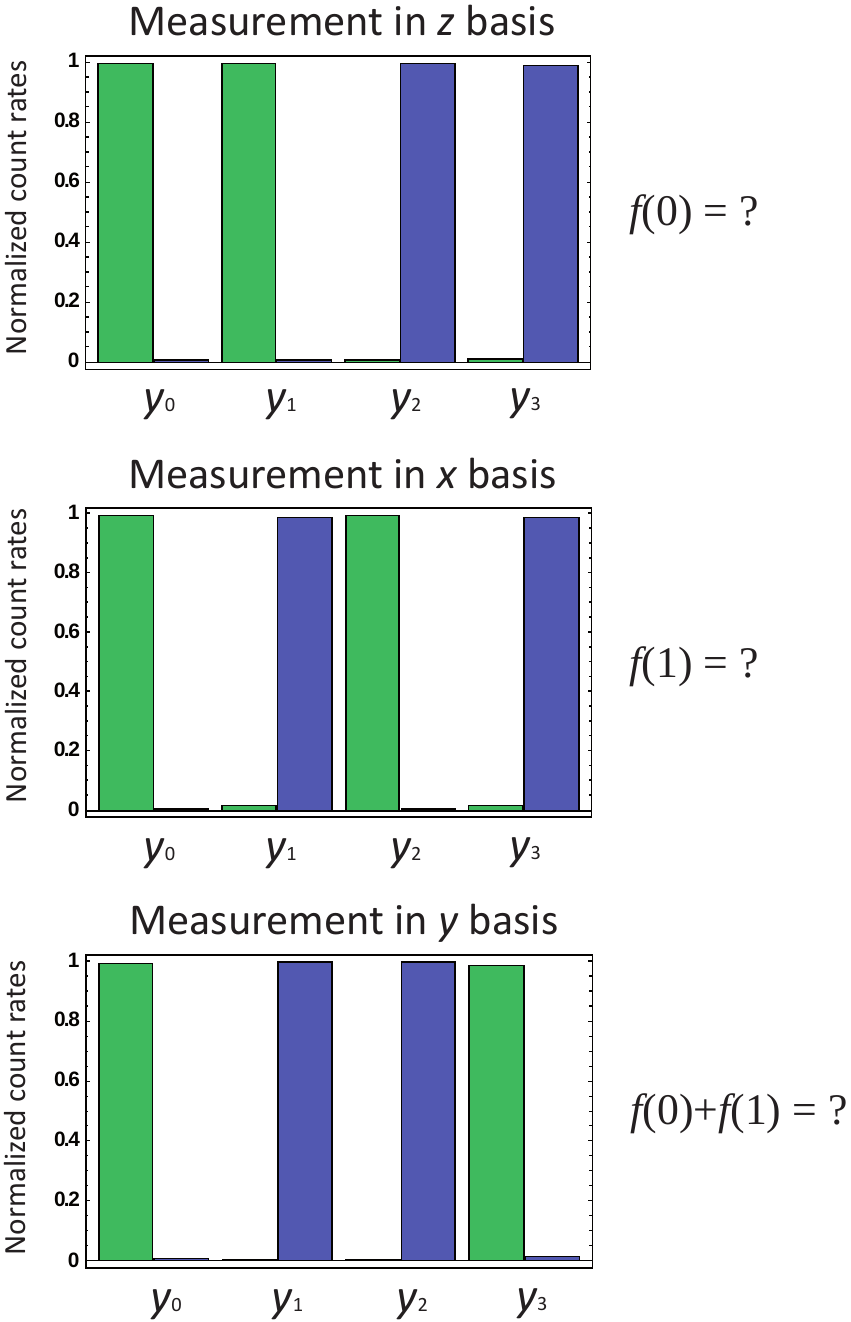}
\end{center}
\par
\vspace{-0.25cm}\caption{We confirm that a measurement in the $z$ basis gives
the value of $f(0)$ and similarly, measurements in $x$ and $y$ bases give the
value of $f(1)$ and $f(0)+f(1)$, respectively. Shown are the experimental
results for the cases where we prepared and measured the qubit in the
\textit{same} basis. Here, green (blue) bars represent counts in detector
``0'' (detector ``1''), see Figure~\ref{Figure_Undecidability_schematic}. The
top graph, e.g., shows the measurement in the $z$ basis: $f(0)$ is `0' (green)
for the functions $y_{0}$ and $y_{1}$, and `1' (blue) for $y_{2}$ and $y_{3}%
$.}%
\label{Figure_Undecidability_confirmation}%
\end{figure}\begin{figure}[tt]
\begin{center}
\includegraphics[width=.45\textwidth]{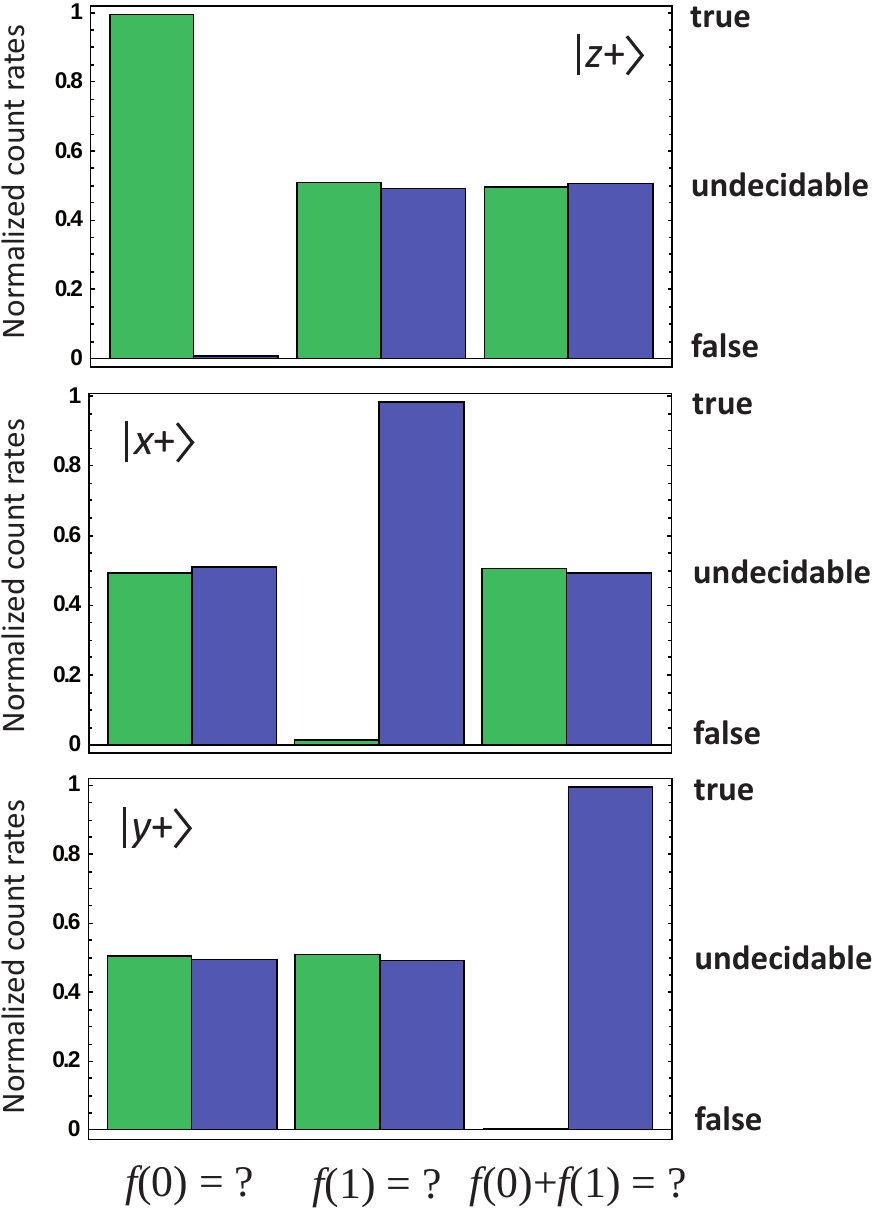}
\end{center}
\par
\vspace{-0.25cm}\caption{Using the setup of
Figure~\ref{Figure_Undecidability_schematic}, we input the qubit in a
well-defined Pauli operator eigenstate $\left\vert z+\right\rangle $,
$\left\vert x+\right\rangle $, or $\left\vert y+\right\rangle $ into the black
box, shown from top to bottom. The black box encodes two classical bits,
$f(0)$ and $f(1)$, and the choice of the initial state determines which single
bit, $f(0)$, $f(1)$, or $f(0)+f(1)$, is read out. For every input state we
measure in all three complementary bases, $z$ [asking for $f(0)$], $x$
[$f(1)$], and $y$ [$f(0)+f(1)$], shown from left to right. The three
measurements are related to three logically complementary questions (A), (B),
(C) of the main text as indicated by the labels. This particular plot is the
experimentally obtained data for the black box realizing the function $y_{1}$.
Similar results were obtained for the other black box configurations $y_{0}$,
$y_{2}$, and $y_{3}$. Again, green (blue) bars represent counts in detector
``0'' (detector ``1'') giving the answer to the corresponding question. Each
input state, after leaving the black box, reveals the truth value of one and
only one of the propositions, i.e.\ it encodes a one-bit axiom. Given this
axiom, the remaining two logically complementary propositions are undecidable.
This undecidability is revealed by complete randomness of the outcomes in the
other two measurement bases.}%
\label{Figure_Undecidability_Q1}%
\end{figure}

We illustrate the link between mathematical undecidability and quantum
randomness in an experiment. The qubit is realized by two orthogonal
polarization states of a single photon generated in the process of spontaneous
parametric down-conversion~\cite{Kwia1995} (SPDC). This system is formally
equivalent to a spin-$\frac{1}{2}$ particle. The horizontal/vertical, $+45%
{{}^\circ}%
$/$-45%
{{}^\circ}%
$, right/left circular polarization of the photon corresponds to eigenstates
$\left\vert z\pm\right\rangle $, $\left\vert x\pm\right\rangle $, and
$\left\vert y\pm\right\rangle $ of the spin-$\frac{1}{2}$ particle,
respectively. We start by initializing the qubit in a definite polarization
state by inserting a linear polarizer in the beam path (see
Figure~\ref{Figure_Undecidability_schematic}). The qubit then propagates
through the black box in which the Boolean functions are encoded with the help
of half-wave plates (HWP). The desired unitary transformation $\hat{\sigma
}_{z}$ ($\hat{\sigma}_{x}$) on the polarization states is implemented by a HWP
at an angle 0%
${{}^\circ}$
(45%
${{}^\circ}$%
) with respect to the $z$-basis. Subsequently, measurements of $\hat{\sigma
}_{z}$, $\hat{\sigma}_{x}$, and $\hat{\sigma}_{y}$, which test the truth value
of a specific proposition, are performed as projective measurements in the
corresponding polarization basis. Specifically, we use a polarizing
beam-splitter (PBS) performing $\hat{\sigma}_{z}$ measurements whose output
modes are fiber-coupled to single-photon detector modules and use wave plates
in front of the PBS to rotate the measurement basis (see
Figure~\ref{Figure_Undecidability_schematic}). The truth value of the
proposition now corresponds to photon detection in one of the two output modes
of the PBS.

First, we confirm that complementary quantum measurements indeed reveal truth
values of logically complementary propositions. To this aim we prepare the
system in a state belonging to the basis in which we finally measure. The
results are presented in Figure~\ref{Figure_Undecidability_confirmation}.

Next, for each of the three choices of the initial state, we ask\ all three
logically complementary questions by measuring in all three different
complementary bases. In Figure~\ref{Figure_Undecidability_Q1}, we plot the
count rate of photons measured after the PBS for a fixed configuration in the
black box that encodes function $y_{1}$. Similar results are obtained for
other black box configurations (not shown).
Figure~\ref{Figure_Undecidability_Q1} shows that for every input state one and
only one question has a definite answer, as already indicated in
Figure~\ref{Figure_Undecidability_confirmation}. This is the axiom encoded in
the system after it leaves the black box. The remaining propositions are
undecidable given that axiom, and the corresponding measurement outcomes are
completely random, i.e.\ evenly distributed.

\section{Generalization to many qubits}

We now generalize the above reasoning using multiple qubits. Consider a black
box with $N$ input and $N$ output ports, one for each qubit. It encodes $N$
Boolean functions $f_{j}(x)$ numbered by $j=1,\ldots,N$ by applying the
operation%
\begin{equation}
\hat{U}_{N}=\hat{\sigma}_{x}^{f_{1}(0)}\,\hat{\sigma}_{z}^{f_{1}(1)}%
\otimes\cdots\otimes\hat{\sigma}_{x}^{f_{N}(0)}\,\hat{\sigma}_{z}^{f_{N}%
(1)}\,.
\end{equation}
The initial $N$-qubit state is chosen to be a particular one of the $2^{N}$
eigenstates of \textit{independent and mutually commuting} tensor products of
Pauli operators, numbered by $p=1,...,N$:%
\begin{equation}
\hat{\Omega}_{p}\equiv\text{i}^{m_{1}(p)n_{1}(p)}\,\hat{\sigma}_{x}^{m_{1}%
(p)}\,\hat{\sigma}_{z}^{n_{1}(p)}\otimes\cdots\otimes\text{i}^{m_{N}%
(p)n_{N}(p)}\,\hat{\sigma}_{x}^{m_{N}(p)}\,\hat{\sigma}_{z}^{n_{N}(p)}\,,
\end{equation}
with $m_{j}(p),n_{j}(p)\in\{0,1\}$. A broad family of such states is the
family of graph states~\cite{Raus2003}. As before, the qubits propagate
through the black box. After leaving it, their state encodes the truth values
of the following $N$ independent binary propositions (negating the false
propositions, one has $N$ true ones which serve as axioms):%
\[
\text{(H}_{p}\text{)}\quad\text{``}%
{\textstyle\sum\nolimits_{j=1}^{N}}
[n_{j}(p)\,f_{j}(0)+m_{j}(p)\,f_{j}(1)]=0\text{.''}%
\]
with $p=1,\ldots,N$. Mathematically, $N$ binary arguments being the $N$ truth
values of the (H$_{p}$) allow to construct $2^{2^{N}}$ different Boolean
functions. The statements about the values of these functions form all
possible decidable (not independent) binary propositions. In suitable
measurements quantum mechanics provides a way to test whether a given
proposition is decidable or not. If one measures the operator%
\begin{equation}
\hat{\Theta}\equiv\text{i}^{\alpha_{1}\beta_{1}}\,\hat{\sigma}_{x}^{\alpha
_{1}}\,\hat{\sigma}_{z}^{\beta_{1}}\otimes\cdots\otimes\text{i}^{\alpha
_{N}\beta_{N}}\,\hat{\sigma}_{x}^{\alpha_{N}}\,\hat{\sigma}_{z}^{\beta_{N}}\,,
\end{equation}
with $\alpha_{j},\beta_{j}\in\{0,1\}$, one tests whether the proposition%
\[
\text{(J)}\quad\text{``}%
{\textstyle\sum\nolimits_{j=1}^{N}}
[\beta_{j}\,f_{j}(0)+\alpha_{j}\,f_{j}(1)]=0\text{.''}%
\]
is decidable or not. In case $\hat{\Theta}$ can be written as a product
$\hat{\Omega}_{1}^{k_{1}}\cdots\hat{\Omega}_{N}^{k_{N}}$ with $k_{p}%
\in\{0,1\}$, it commutes with all the $\hat{\Omega}_{p}$'s and consequently
the measurement of $\hat{\Theta}$ has a definite outcome. On the other hand,
proposition (J) is decidable within the set of axioms (H$_{p}$). Its truth
value can be (logically) derived from the axioms in the sense that the sum in
(J), $%
{\textstyle\sum\nolimits_{j=1}^{N}}
[\beta_{j}\,f_{j}(0)+\alpha_{j}\,f_{j}(1)]$, is a linear combination of the
sums in (H$_{p}$) with binary coefficients $k_{p}$. There are $2^{N}$ such
linear combinations and therefore $2^{N}$ such $\hat{\Theta}$'s. Since in
general $4^{N}$ different $\hat{\Theta}$'s exist, the remaining $4^{N}%
-2^{N}=2^{N}(2^{N}-1)$ operators are linked with undecidable propositions and
their measurement outcomes are random. Hence, there are much more undecidable
propositions of the form (J) than decidable ones. The ratio between their
numbers increases exponentially with the number of qubits, i.e.\ $\frac
{2^{N}(2^{N}-1)}{2^{N}}=O(2^{N})$.

Interestingly, in case of (J) being decidable, its truth value imposed by
(classical) logic is not necessarily the same as found in the quantum
measurement. This can be demonstrated for three qubits initially in the
Greenberger-Horne-Zeilinger (GHZ) state~\cite{Gree1989}%
\begin{equation}
\left\vert \text{GHZ}\right\rangle =\tfrac{1}{\sqrt{2}}\left(  \left\vert
z+\right\rangle _{1}\left\vert z+\right\rangle _{2}\left\vert z+\right\rangle
_{3}+\left\vert z-\right\rangle _{1}\left\vert z-\right\rangle _{2}\left\vert
z-\right\rangle _{3}\right)  ,
\end{equation}
where e.g.\ $\left\vert z\pm\right\rangle _{1}$ denotes the eigenstate with
the eigenvalue $\pm1$ of $\hat{\sigma}_{z}$ for the first qubit. We choose as
axioms the propositions%
\begin{align*}
\text{(K}_{1}\text{)}\quad\text{``}f_{1}(0)+f_{1}(1)+f_{2}(0)+f_{2}%
(1)\quad\quad\quad\;\,+f_{3}(1)  &  =1\text{.''}\\
\text{(K}_{2}\text{)}\quad\text{``}f_{1}(0)+f_{1}(1)\quad\quad\quad
\;\,+f_{2}(1)+f_{3}(0)+f_{3}(1)  &  =1\text{.''}\\
\text{(K}_{3}\text{)}\quad\text{``}\quad\quad\quad\;\,f_{1}(1)+f_{2}%
(0)+f_{2}(1)+f_{3}(0)+f_{3}(1)  &  =1\text{.''}%
\end{align*}
linked with the operators $\hat{\sigma}_{y}\otimes\hat{\sigma}_{y}\otimes
\hat{\sigma}_{x}$, $\hat{\sigma}_{y}\otimes\hat{\sigma}_{x}\otimes\hat{\sigma
}_{y}$, and $\hat{\sigma}_{x}\otimes\hat{\sigma}_{y}\otimes\hat{\sigma}_{y}$,
respectively. One can \textit{logically} derive from (K$_{1}$) to (K$_{3}$)
the true proposition%
\[
\text{(L)}\quad\text{``}f_{1}(1)+f_{2}(1)+f_{3}(1)=1\text{.''}%
\]
On the other hand, the proposition (L) is identified with the measurement of
$\hat{\sigma}_{x}\otimes\hat{\sigma}_{x}\otimes\hat{\sigma}_{x}$. But the
result imposed by quantum mechanics corresponds to the \textit{negation} of
(L), namely: ``$f_{1}(1)+f_{2}(1)+f_{3}(1)=0$.'' This is the heart of the
experimentally confirmed GHZ argument against local
realism~\cite{Gree1989,Merm1990,Pan2000}. In the (standard logical) derivation
of (L) the individual function values are well defined and the same
independently of the axiom in which they appear. Since this is equivalent to
the assumptions of local realism, the truth values of decidable propositions
found in quantum experiments do not necessarily have to be the same as the
ones derived by logic. However, this does not change the connection between
decidability (undecidability) of propositions and definiteness (randomness) of
measurement outcomes.

The question arises whether one can construct a classical device to reveal the
undecidability of propositions. This is possible, provided one uses $2N$
classical bits to simulate quantum complementary measurements on $N$
qubits~\cite{Spek2007}. Since such a device satisfies local realism, it gives
the truth values of decidable propositions according to classical logic. On
the level of elementary physical systems, however, the world is known to be
quantum. It is intriguing that nature supplies us with quantum systems which
can reveal decidability but cannot be used to learn the classical truth values.

\begin{figure}[t]
\begin{center}
\includegraphics[width=.99\textwidth]{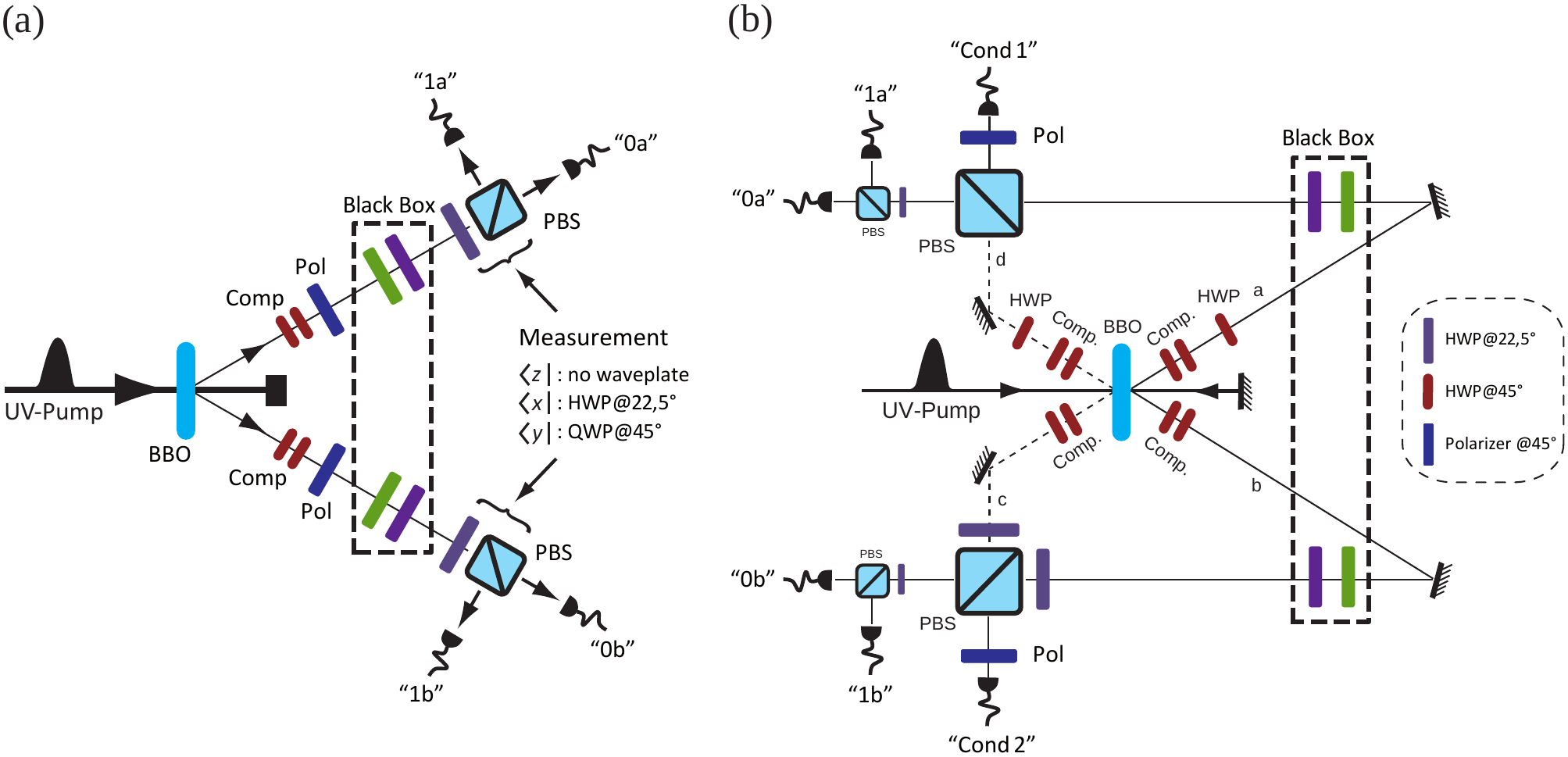}
\end{center}
\par
\vspace{-0.25cm}\caption{The setup in (a) allows for separable two-qubit
measurements on product as well as on entangled input states. Joint two-qubit
measurements require a so-called Bell State Analyzer (BSA), whose experimental
realization is depicted in (b). In both cases, an ultra-violet laser pulse
passes through a non-linear crystal (BBO) to produce polarization-entangled
photon pairs in the process of spontaneous parametric down-conversion.
Compensators (Comp), made up of half-wave plates (HWP) and BBO crystals, are
used to counter walk-off effects in the down-conversion crystal. They are set
such that $\left\vert \Phi^{+}\right\rangle $ states are emitted. In (a),
local measurements in all complementary bases are performed with the help of
HWP, QWP and PBSs. Measurements in the Bell basis, as depicted in (b), require
an ancillary entangled Bell state $\left\vert \Phi^{+}\right\rangle $.
Therefore the UV-laser passes twice through the crystal, emitting the input
Bell state in the forward direction (modes a \& b), and the ancilla pair in
the backward direction (modes c \& d). Coherently combining these photons on
PBSs, allows the identification of all four Bell states whenever there is one
photon in each output mode of the PBSs. Then, conditioned on the detection of
one photon with +45${{}^{\circ}}$ polarization in both detectors,
\textquotedblleft Cond 1\textquotedblright\ and \textquotedblleft Cond
2\textquotedblright, the Bell state can be identified by analyzing the
remaining two photons in the $+45{{}^{\circ}}$/$-45{{}^{\circ}}$ basis (see
Reference~\cite{Walt2005b} for details).}%
\label{Figure_Undecidability_setup}%
\end{figure}In order to illustrate the concept of partial undecidability as
introduced above, we performed a two-qubit experiment. The two bits of
proposition (E) described above correspond to the set of independent commuting
operators $\hat{\Omega}_{1}=\hat{\sigma}_{z}\otimes\hat{\sigma}_{z}$ and
$\hat{\Omega}_{2}=\hat{\sigma}_{x}\otimes\hat{\sigma}_{x}$. The common
eigenbasis of these operators is spanned by the maximally entangled Bell
states (basis $b_{\text{E}}$): $\left\vert \Phi^{\pm}\right\rangle =\tfrac
{1}{\sqrt{2}}\left(  \left\vert z+\right\rangle _{1}\left\vert z+\right\rangle
_{2}\pm\left\vert z-\right\rangle _{1}\left\vert z-\right\rangle _{2}\right)
$, $\left\vert \Psi^{\pm}\right\rangle =\tfrac{1}{\sqrt{2}}\left(  \left\vert
z+\right\rangle _{1}\left\vert z-\right\rangle _{2}\pm\left\vert
z-\right\rangle _{1}\left\vert z+\right\rangle _{2}\right)  $. Thus, after the
black box the four Bell states encode the four possible truth values of the
elementary propositions in (E) and a so-called Bell State Analyzer (i.e.\ an
apparatus that measures in the Bell basis) reveals these values. In the same
way, the truth values of the elementary propositions in (F) are encoded in the
eigenstates of local $\hat{\sigma}_{z}$ bases, i.e.\ by the four states
$\left\vert z\pm\right\rangle _{1}\left\vert z\pm\right\rangle _{2}$ (basis
$b_{\text{F}}$). Finally, the elementary propositions in (D) are linked with
the product states $\left\vert z\pm\right\rangle _{1}\left\vert x\pm
\right\rangle _{2}$ (basis $b_{\text{D}}$).

The setup that was employed to generate maximally entangled Bell states via
type-II SPDC is shown in Figure~\ref{Figure_Undecidability_setup}. We also
prepared separable two-qubit states by inserting polarizers into the optical
paths just before the black box. The encoding of functions within the black
box as well as the polarization measurements on the individual qubits were
done similarly to the single-qubit case. However, this only allows for
\textit{independent} single-qubit measurements. \textit{Joint} measurements on
both qubits in the Bell basis require conditional operations on two qubits.
The heart of such a Bell State Analyzer~\cite{Wein1994,Mich1996,Walt2005b}
(BSA) is a controlled-NOT gate~\cite{Bare1995,Pitt2002,OBri2003,Gasp2004}. In
Figure~\ref{Figure_Undecidability_setup} we depict the experimental setup
implementing the BSA. Details of its working can be found in
Reference~\cite{Walt2005b}.

\begin{figure}[t]
\begin{center}
\includegraphics[width=.7\textwidth]{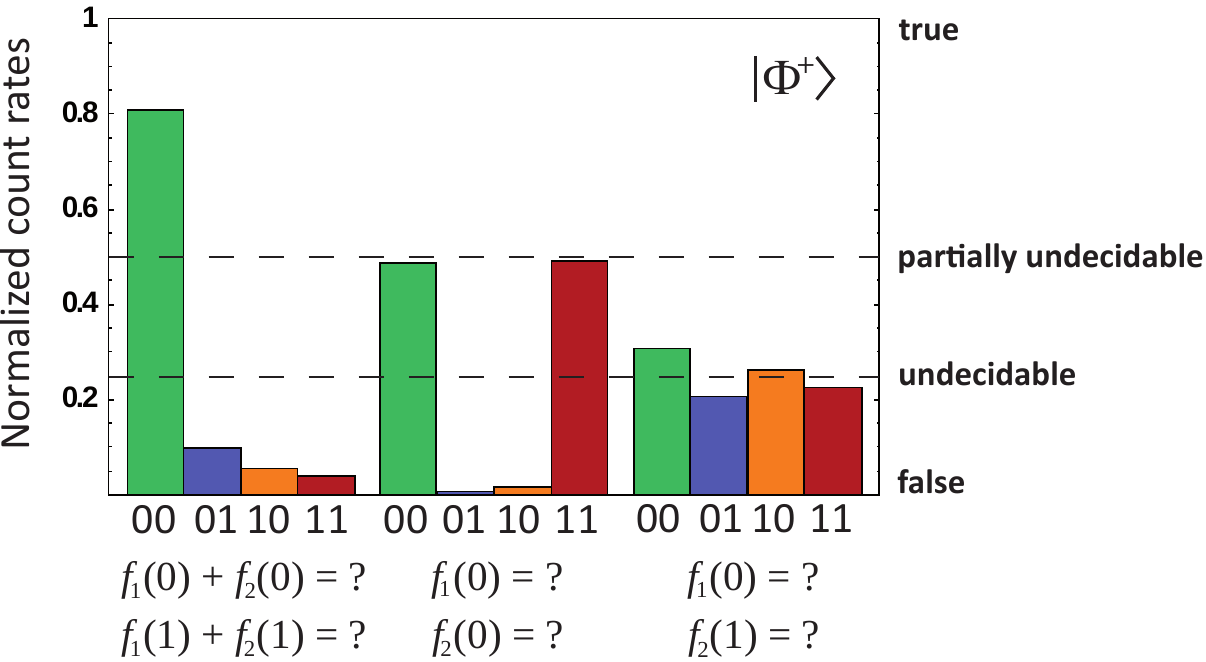}
\end{center}
\par
\vspace{-0.25cm}\caption{In this two-qubit experiment a $\left\vert \Phi
^{+}\right\rangle $ Bell state is measured in three different bases (the Bell
basis as well as $\left\vert z\pm\right\rangle _{1}\left\vert z\pm
\right\rangle _{2}$ and $\left\vert z\pm\right\rangle _{1}\left\vert
x\pm\right\rangle _{2}$, shown from left to right) using the setups depicted
in Figure~\ref{Figure_Undecidability_setup}. Plotted are the count rates
associated with different detector combinations for the black box encoding
function $y_{2}$ on both photons. Label ``00'' corresponds to a coincidence
event in detectors ``0a'' and ``0b'', and similarly for the other labels. The
first (second) label gives the answer to the upper (lower) question. The
measurements performed in the Bell basis show that the two qubits encode
proposition (E) of the main text. The data in the middle plot reveal partial
undecidability of proposition (F), given (E) as an axiom, as indicated by the
random outcomes in two out of four detector combinations. In contrast, the
right plot presents the data corresponding to the fully undecidable
proposition (D), where the outcomes are completely random. Similar results for
propositions (E), (F) and (D) were obtained for other black box encodings. The
reason for the fact that the Bell state in the left plot is not identified
with unit fidelity stems from imperfections in the experimental setup. Unequal
detector efficiencies explain the small bias in the right plot.}%
\label{Figure_Undecidability_Q2}%
\end{figure}We start our investigation by confirming that the truth values of
the (elementary) propositions in (E), (F) and (D) are revealed by measurements
performed in the bases $b_{\text{E}}$, $b_{\text{F}}$ and $b_{\text{D}}$,
respectively. As in the single-qubit case, preparation and measurement are in
the same basis. The obtained results (not shown) are similar to
Figure~\ref{Figure_Undecidability_confirmation}. Moreover, one can initially
prepare a Bell state and measure it in the bases $b_{\text{E}}$, $b_{\text{F}%
}$ and $b_{\text{D}}$. As can be seen in the left plot of
Figure~\ref{Figure_Undecidability_Q2}, measurements in the Bell basis,
$b_{\text{E}}$, prove that the entangled state indeed encodes joint properties
of the functions $f_{1}(x)$ and $f_{2}(x)$, i.e.\ information about (E).
Measurements in other bases can then be interpreted in terms of partial\ and
full\ undecidability. Proposition (D) is fully undecidable given (E) as an
axiom which is encoded by $\left\vert \Phi^{+}\right\rangle $ after the black
box. This can be seen from the right part in
Figure~\ref{Figure_Undecidability_Q2}, in which all four measurement outcomes
occur with equal probability. On the other hand, proposition (F) is partially
undecidable. This is experimentally revealed by the count distribution of the
middle part in Figure~\ref{Figure_Undecidability_Q2}. The partial
undecidability is uncovered by the randomness of the two occurring outcomes,
while the other two outcomes do not appear.

In conclusion, we have demonstrated how mathematical axioms can be encoded in
quantum states and how decidability of mathematical propositions can be
verified in quantum measurements. This extends the concept of
\textquotedblleft experimental mathematics\textquotedblright~\cite{Chai2007}
to a new domain and sheds new light on the (mathematical) origin of quantum
randomness. We have performed an experiment which showed for the first time
the link between mathematical undecidability and quantum randomness. There, we
found that whenever the quantum system is measured in a basis associated to an
undecidable proposition, it gives random outcomes. Our results can be extended
in many ways.

An interesting avenue for further research arises from the observation that in
mathematical logic one usually makes a difference only between statements that
are either decidable\ or undecidable\ within a formal system of axioms.
Quantum complementarity implies that losing certainty about one of the
propositions is followed by a corresponding gain of certainty in other,
logically complementary, propositions~\cite{Bruk1999,Bruk2004}. This opens up
the possibility to quantify the \textit{amount of undecidability} of a
proposition from \textquotedblleft impossible\textquotedblright\ to
\textquotedblleft necessary\textquotedblright, and to describe the continuous
\textit{trade-off} in gaining and losing knowledge about them.

\chapter*{\vspace{-3cm}Conclusions and outlook}

\addcontentsline{toc}{chapter}{Conclusions and outlook}

\pagestyle{fancy} \fancyhf{} \fancyhead[LE,RO]{\thepage}
\fancyhead[LO]{Conclusions and outlook}
\fancyhead[RE]{Conclusions and outlook} \renewcommand{\headrulewidth}{0.25pt}
\renewcommand{\footrulewidth}{0pt} \addtolength{\headheight}{0.5pt}
\fancypagestyle{plain}{\fancyhead{}
\renewcommand{\headrulewidth}{0pt}}

\vspace{0.25cm}This final part of the present work offers the room to point
out some possibilities for further research, including not only conservative
but also speculative thoughts.\vspace{0.5cm}

\noindent\textbf{Macroscopic realism and the quantum-to-classical transition:}

\begin{itemize}
\item We have put forward an approach to the quantum-to-classical transition,
resting solely on the idea that coarse-grained measurements give rise to the
emergence of macroscopic realism. However, to offer a complementary approach
to the decoherence program that also applies to isolated systems, it is
necessary to \textit{generalize the formalism to systems other than spins}.
E.g., for a particle in conventional phase space with position and momentum
and an arbitrary time evolution there is no quantum number that can be made
larger and larger (like spin length). But Reference~\cite{Pere1995} shows how
a small blurring of a particle's Wigner function makes it positive (by
convoluting it with the Wigner function of a coherent state). Again, it should
be possible to model coarse-grained measurements in such a way that they are
non-invasive at the level of a positive probability distribution. The latter
represents an ensemble of classical objects with definite positions and
momenta and must be sufficient to compute probabilities for coarse-grained
outcomes. It can be seen already now that the interesting issue of
non-classical Hamiltonians will arise again.

\item We have found that coarse-grained spin measurements are best modeled
with a positive operator value measure because coarse-grained von Neumann
measurements allow to distinguish microstates at two sides of a slot border.
Still, under all circumstances it is unavoidable that a quantum measurement
always introduces a slight disturbance of the state even on the level of the
$Q$-distribution. Under ideal experimental conditions and sufficiently many
runs, we should be able to see \textit{tiny deviations from classical physics}
even for classical Hamiltonians, macroscopic systems, and coarse-grained measurements.

\item We have demonstrated that every non-trivial Hamiltonian allows to
violate the Leggett-Garg inequality as long as sharp quantum measurements can
be performed. This might be seen as related to the \textit{quantum Zeno
effect}~\cite{Misr1977}: Quantum measurements not only are capable of freezing
the time evolution of a system but they also may influence it in such a way
that it cannot be described classically anymore. Note that a perfect quantum
Zeno effect itself can easily be simulated classically just by using no time
evolution at all.

\item We have seen that non-classical Hamiltonians can lead to a violation of
macrorealism even under the restriction of coarse-grained measurements.
Moreover, even environmental decoherence, though leading to macrorealism, does
not resolve the problem that in principle no classical laws of motion are able
to describe such time evolutions. We have tried to argue why non-classical
Hamiltonians do not appear in nature. One avenue of research would be to
investigate the relation between the interactions in our measurement
apparatuses and the ones governing the time evolutions of systems. The Hilbert
space has no notion of closeness or distance of orthogonal states. We expect
that under coarse-grained measurements which bunch together states by whatever
notion of closeness or distance \textit{quantumness can be seen only if the
system Hamiltonian and the apparatus Hamiltonian connect states differently}.

\item We may even conjecture that the \textit{three-dimensionality of our real
space} is a necessity under the postulate that an intersubjective classical
world arises out of the quantum realm merely due to coarse-graining. This is
supported by the fact that the Lie algebras SU(2) and SO(3) are isomorphic.
\end{itemize}

\noindent\textbf{Quantum randomness and mathematical undecidability:}

\begin{itemize}
\item In the spirit of the approach in Reference~\cite{Bruk2004}, one
important question is whether it is possible to \textit{derive Malus's law in
the framework of propositions}, i.e.~a very particular kind of continuous
trade-off between gaining and losing certainty about truth values like in the
quantum case. It might be that to this end the necessary number of independent
binary functions immediately becomes very large and is related to the number
of spatial directions that can be experimentally distinguished, or that one
cannot use binary functions any longer.

\item Our link between quantum physics and pure mathematics offers an
intuitive understanding of randomness in which the question of decidability is
more fundamental than about truth. Which other ingredients beyond mathematical
undecidability are necessary to fully \textit{reconstruct the quantum
formalism} of state vectors in a Hilbert space? Is it possible to derive from
this link any other physical phenomena that cannot be derived or understood
from the conventional quantum postulates?

\item Gödel's incompleteness theorem is known to be equivalent to Turing's
halting problem. The latter can be seen as a consequence of the fact that
there are more real numbers than natural numbers, or equivalently, not
uncountable many Turing machines for countably many possible inputs. Chaitin's
version rests on the intuitive observation that a theorem cannot be derived
from axioms if the information content of theorem plus axioms is larger than
the information content of the axioms itself. Thus, in both cases it is a
\textit{mismatch of \textquotedblleft resources for answers\textquotedblright%
\ and \textquotedblleft amount of possible questions\textquotedblright}
leading to undecidability---just like in quantum mechanics where randomness is
a consequence of the fact that we can measure more things than the state can
definitely define. Formalizing this link between Chaitin's and Gödel's
original version would strengthen the connection between quantum randomness
and mathematical undecidability.
\end{itemize}

\chapter*{\vspace{-3cm}Curriculum vitae}

\addcontentsline{toc}{chapter}{Curriculum vitae}

\pagestyle{fancy} \fancyhead{} \renewcommand{\headrulewidth}{0pt}
\addtolength{\headheight}{0pt} \fancyfoot[c]{\thepage}

\vspace{0.5cm}

\subsubsection*{Personal Information}

\vspace{0.33cm}
\begin{tabular}
[t]{p{3.8cm}p{10cm}}%
Name & Johannes Kofler\vspace{0.1cm}\\
Date and place of birth & June 16, 1980, Linz, Austria\vspace{0.1cm}\\
Citizenship & Austrian\vspace{0.1cm}%
\end{tabular}

\vspace{0.33cm}

\subsubsection*{Professional History}

\vspace{0.33cm}
\begin{tabular}
[t]{p{3.8cm}p{10cm}}%
Since 01/2007 & \textit{DOC Fellow}\vspace{0.cm}\\
& Austrian Academy of Sciences\vspace{0.1cm}\\
01/2005 -- 12/2006 & \textit{Research Assistant}\vspace{0.cm}\\
& Faculty of Physics, University of Vienna \& Institute for Quantum Optics and
Quantum Information (IQOQI), Austrian Academy of Sciences\vspace{0.1cm}\\
03/2004 -- 06/2004 & \textit{Scientific co-worker}\vspace{0.cm}\\
& Institute for Theoretical Physics \& Institute for Applied Physics,
University of Linz
\end{tabular}

\vspace{0.33cm}

\subsubsection*{Education}

\vspace{0.33cm}
\begin{tabular}
[t]{p{3.8cm}p{10cm}}%
Since 03/2005 & Doctoral studies of Natural Sciences (University of
Vienna)\vspace{0.1cm}\\
10/1999 -- 11/2004 & Diploma studies of Technical Physics (University of
Linz)\vspace{0.1cm}\\
07/1998 -- 02/1999 & Obligatory military service (Villach and
Klagenfurt)\vspace{0.1cm}\\
09/1990 -- 06/1998 & Grammar school (Linz and Klagenfurt)
\end{tabular}

\vspace{0.33cm}

\subsubsection*{Honors}

\vspace{0.33cm}
\begin{tabular}
[t]{p{3.8cm}p{10cm}}%
2007 & DOC Fellowship (Austrian Academy of Sciences)\vspace{0.1cm}\\
2004 & Wilhelm-Macke-Prize (University of Linz)\vspace{0.1cm}\\
2003 & Merit scholarship (University of Linz)\vspace{0.1cm}\\
2002 & Merit scholarship (University of Linz)\vspace{0.1cm}\\
2001 & Merit scholarship (University of Linz)
\end{tabular}

\end{document}